\documentclass[]{spie}  

\usepackage{amsmath,amsfonts,amssymb}
\usepackage{graphicx}
\usepackage[colorlinks=true, allcolors=blue]{hyperref}

\usepackage{float}
\usepackage{caption}
\usepackage{subcaption}
\usepackage{textcomp}
\usepackage{array}
\DeclareMathOperator{\sinc}{sinc}

\title{VIPER: A high-resolution multimode fiber-fed VIPA spectrograph concept for characterizing exoplanet atmospheric escape}

\author[a,b]{Matthew C. H. Leung}
\author[a,b]{David Charbonneau}
\author[a,b,c]{Andrew Szentgyorgyi}
\author[a,c]{Colby Jurgenson}
\author[a,b,c]{Morgan MacLeod}
\author[d]{Surangkhana Rukdee}
\author[e]{Shreyas Vissapragada}
\author[f]{Fabienne Nail}
\author[a,c]{Joseph~Zajac}
\author[a,c]{Andrea K. Dupree}

\affil[a]{\normalsize Center for Astrophysics \textbar{} Harvard \& Smithsonian, 60 Garden St, Cambridge, MA 02138, USA}
\affil[b]{Department of Astronomy, Harvard University, 60 Garden St, Cambridge, MA 02138, USA}
\affil[c]{Smithsonian Astrophysical Observatory, 100 Acorn Park Dr, Cambridge, MA 02140, USA}
\affil[d]{Max Planck Institute for Extraterrestrial Physics, Giessenbachstrasse, 85748, Garching, Germany}
\affil[e]{Carnegie Science Observatories, 813 Santa Barbara St, Pasadena, CA 91101, USA}
\affil[f]{Anton Pannekoek Institute for Astronomy, University of Amsterdam, 1090 GE Amsterdam, The Netherlands}

\authorinfo{Further author information: (Send correspondence to M.C.H. Leung)\\M.C.H. Leung: E-mail: matthew.leung (at) cfa.harvard.edu \\ D. Charbonneau: E-mail: dcharbonneau (at) cfa.harvard.edu \\ A. Szentgyorgyi: E-mail: saint (at) cfa.harvard.edu}

\begin{document} 
\maketitle

\begin{abstract}
An increasing number of applications in exoplanetary science require spectrographs with high resolution and high throughput without the need for a broad spectral range. Examples include the search for biosignatures through the detection of the oxygen A-band at 760 nm, and the study of atmospheric escape through the helium 1083~nm triplet. These applications align well with the capabilities of a spectrograph based on a Virtually Imaged Phased Array (VIPA), a high-throughput dispersive element that is essentially a modified Fabry-Perot etalon. We are developing VIPER, a high-resolution, narrowband, multimode fiber-fed VIPA spectrograph specifically designed to observe the helium 1083 nm triplet absorption line in the atmospheres of gaseous exoplanets. VIPER will achieve a resolving power of 300,000 over a wavelength range of 25 nm, and will be cross-dispersed by an echelle grating. VIPER is intended for operation on the 1.5~m Tillinghast Telescope and potentially on the 6.5~m MMT, both located at the Fred Lawrence Whipple Observatory (FLWO) on Mount Hopkins, Arizona, USA. In this paper, we present VIPER’s instrument requirements, derived from the primary science goal of detecting anisotropic atmospheric escape from exoplanets. We discuss the design methodology for VIPA-based spectrographs aimed at maximizing throughput and diffraction efficiency, and we derive a wave-optics-based end-to-end model of the spectrograph to simulate the intensity distribution at the detector. We present an optical design for VIPER and highlight the potential of VIPA-based spectrographs for advancing exoplanetary science.
\end{abstract}



\section{Introduction}\label{sec:intro}

One of the most striking discoveries in the past decade of exoplanetary science is that small planets are divided into two distinct groups: rocky terrestrial worlds, and larger worlds enveloped in hydrogen-helium atmospheres. This divide, known as the radius gap\cite{Fulton2017}, challenged more conventional planet formation theories at the time. Understanding the origin of this divide is now considered crucial to developing accurate models of how planets form and evolve. A family of processes known as atmospheric escape, by which a planet loses its gaseous envelope, has been proposed as a key driver of this divide \cite{Owen2019}. Atmospheric escape can be studied through transit spectroscopy using specific spectral lines. In recent years, the metastable helium triplet absorption line at 1083~nm has emerged as a particularly powerful probe for atmospheric escape\cite{Oklopi2018}. High-resolution spectra of this helium feature, taken during and beyond transit, provide vital insights into the 3D velocity and density distributions of the escaping gas\cite{Zhang2023}. This is crucial for constraining mechanisms that lead to anisotropic mass loss\cite{MacLeod2022}, such as planetary magnetic fields\cite{Schreyer2023} and temperature gradients\cite{Nail2024,Nail2025}, which result in different population statistics for exoplanets\cite{Owen2023}.

Applications such as this, requiring high-resolution observations of specific spectral lines, are becoming increasingly central to exoplanetary science. Another prominent example is the search for biosignatures through the detection of the oxygen A-band at 760 nm\cite{Rukdee2023}. These targeted applications require spectrographs with both high resolving power and high throughput, but do not necessarily require broad spectral coverage. Such applications align well with the capabilities of a spectrograph based on a Virtually Imaged Phased Array (VIPA\cite{Shirasaki1996}), a high-throughput dispersive element that is essentially a modified Fabry-Perot etalon. VIPA spectrographs have recently begun to gain traction in stellar and exoplanetary science, with a few groups starting to explore and develop such spectrographs in the optical and near-infrared\cite{Bourdarot2017,Bourdarot2018,Carlotti2022,Zhu2020,Zhu2023,Rukdee2024}. While most implementations to date rely on a single mode optical fiber feed due to technical challenges, recent demonstrations using multimode optical fibers\cite{Zhu2023} suggest a pathway toward higher throughput and relaxed injection requirements at the telescope focal plane.

We are developing VIPER (\underline{V}irtually \underline{I}maged \underline{P}hased Array Probe for \underline{E}xoplanetary \underline{R}esearch), a high-resolution, narrowband, multimode fiber-fed, cross-dispersed VIPA spectrograph specifically designed to observe the helium 1083 nm triplet absorption line. VIPER will achieve a resolving power of $\mathcal{R}$ = 300,000 over a wavelength range of 25 nm. Cross-dispersion will be provided by an echelle grating. VIPER is intended for operation on the 1.5 m Tillinghast Telescope and potentially on the 6.5 MMT, both located at the Fred Lawrence Whipple Observatory (FLWO) on Mount Hopkins, Arizona, USA. The primary science goal of VIPER is to detect anisotropic atmospheric escape in gaseous exoplanets. Additional applications involve studies of stellar activity\cite{Strader2015}, stellar mass loss\cite{Dupree1986}, and galactic chemical evolution\cite{Cooke2022}, also possible through the helium 1083 nm triplet.

An outline of this paper is as follows. In Section \ref{sec:science}, we discuss the science background for atmospheric escape and the instrument requirements of VIPER. Readers primary interested in the optical design process of VIPA-based spectrographs may wish to skip this section altogether and proceed directly to Section \ref{sec:VIPA_spec_overview}, where we provide a brief overview of how cross-dispersed VIPA spectrographs function. In Section \ref{sec:VIPA}, we discuss how a VIPA functions. In Section~\ref{sec:VIPA_inject}, we model the injection of light into the VIPA, which determines the VIPA diffraction efficiency. In Section~\ref{sec:cross_disp}, we model the cross-disperser, which is an echelle grating. In Section \ref{sec:spec_model}, we combine the results of Sections \ref{sec:VIPA_spec_overview}--\ref{sec:cross_disp} and present an end-to-end model of a cross-dispersed VIPA spectrograph. In Section \ref{sec:spec_design}, we discuss the optical design of VIPER and demonstrate its expected performance.
\section{Science Goal and Instrument Requirements}\label{sec:science}

\subsection{Science Background}

\subsubsection{Radius Gap, Neptune Desert, and Atmospheric Escape}

\begin{figure}[h]
    \centering
    \begin{subfigure}{0.49\textwidth}
        \centering
        \includegraphics[width=\textwidth]{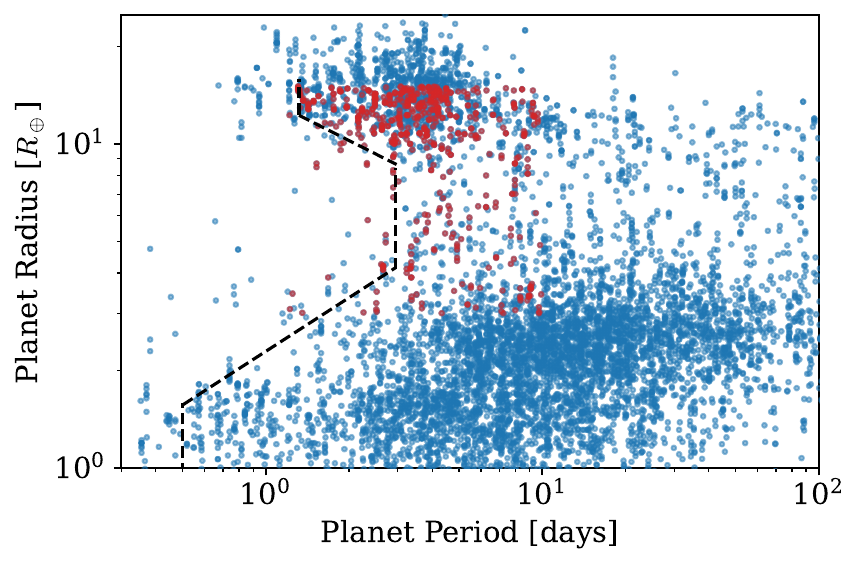}
        \caption{Exoplanet radius-period diagram}
        \label{fig:radius_VS_period}
    \end{subfigure}
    \begin{subfigure}{0.49\textwidth}
        \centering
        \includegraphics[width=\textwidth]{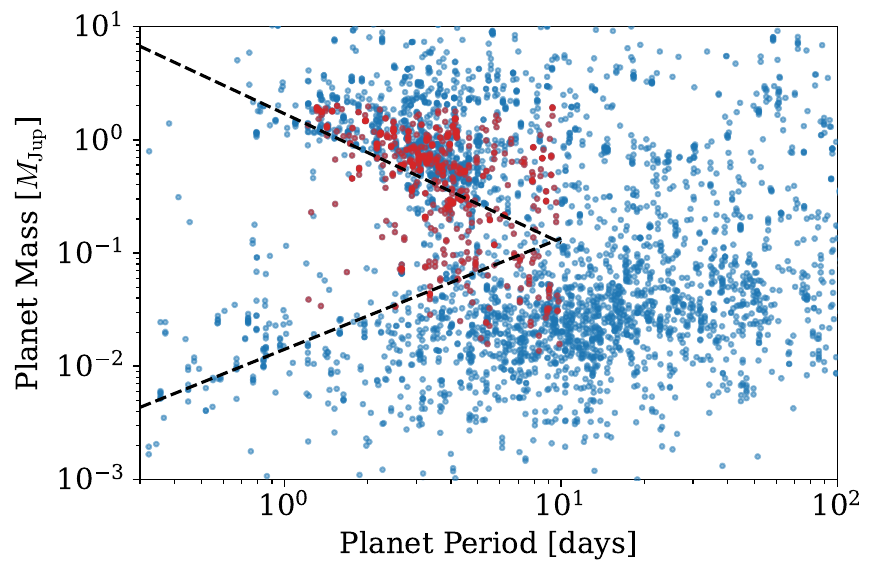}
        \caption{Exoplanet mass-period diagram}
        \label{fig:mass_VS_period}
    \end{subfigure}
    \caption{Radius-period diagram and mass-period diagram for all confirmed exoplanets from the NASA Exoplanet Archive\cite{NASAExoArchive} as of July 2025, shown by the blue points. The red points are the planets that VIPER might probe. The Neptune desert is left of the black dashed lines in both diagrams. The plotted Neptune desert boundaries for the radius-period diagram and mass-period diagram are from Ref.~\citenum{CastroGonzlez2024} and Ref.~\citenum{Mazeh2016} respectively.}
    \label{fig:mass_radius_period}
\end{figure}

Figure \ref{fig:mass_radius_period} shows a radius-period diagram (Figure \ref{fig:radius_VS_period}) and a mass-period diagram (Figure \ref{fig:mass_VS_period}) for all confirmed exoplanets in the NASA Exoplanet Archive\cite{NASAExoArchive} as of July 2025. These diagrams contain some notable demographic features. In radius-period space, there is a bimodal distribution in exoplanet radii, divided by the radius gap (also called the ``radius valley''). In mass-period space, there is a lack of Neptune-mass planets orbiting close to their host stars (with period $\lesssim$ 3 days\cite{CastroGonzlez2024}); this is known as the ``Neptune desert''\cite{Mazeh2016}, which occupies the region of the mass-period diagram left of the black dashed lines. This Neptune desert feature can also be seen in radius-period space, indicated by the region left of the black dashed lines\cite{CastroGonzlez2024} in the radius-period diagram. Deciphering the origins of the radius gap and the Neptune desert is a central challenge in contemporary exoplanetary science, crucial for constraining models of planet formation, evolution, and migration.

Atmospheric escape is believed to be an important contributor to carving out both the radius valley and the Neptune desert\cite{Owen2022}. Planet evolution models invoking atmospheric escape have been able to partially recover the observed exoplanet demographics\cite{Owen2022,Owen2017,Rogers2021,Gupta2020,Ginzburg2018}. 
Atmospheric escape might be driven by several physical processes, and two competing classes of atmospheric escape models have been proposed: photoevaporation\cite{Owen2017} and core-powered mass loss\cite{Ginzburg2018}.

In the photoevaporation model, intense stellar XUV and X-ray irradiation strips a close-in planet of its gaseous envelope, leaving behind a bare core or potentially leading to the planet's destruction by stellar tidal forces\cite{Owen2017,Owen2019}. This results in a bimodal distribution of exoplanet radii, where the smaller planets are the stripped bare cores and the larger planets are the ones that survived the photoevaporation process\cite{Rogers2021}. This also explains the lower part\cite{Ionov2018} of the Neptune desert, since less-massive Neptune-like planets originally inside the desert do not have enough gravitational potential to retain their atmospheres, becoming bare cores.

On the other hand, in the core-powered mass loss model, a planet's cooling rocky core heats the planet's atmosphere, causing it to be reduced on a timescale of millions of years\cite{Ginzburg2018,Ginzburg2016}. Planets with more massive envelopes are able to retain them, while planets with lighter envelopes can lose them completely\cite{Ginzburg2018}, creating the radius gap. Core-powered mass loss occurs on a longer timescale compared to photoevaporation, and does not require high-energy stellar irradiation. The planet's primordial energy from its formation (gravitational and thermal), stored in its core, can be greater than its atmosphere's gravitational binding energy\cite{Gupta2020}, allowing for this process to be possible.

Although atmospheric escape models can reproduce the radius valley and Neptune desert, there are models that do not invoke atmospheric escape and can still explain the observed exoplanet demographics\cite{Krishnamurthy2024}. For example, it is possible that the radius gap is a result of the physics of gas accretion during plant formation, where the accretion rate is limited in lighter cores, producing a divide in exoplanet radii\cite{Lee2022}. The radius gap could also be explained as a result of planets having inherently different densities (``water worlds'' hypothesis\cite{Luque2022}). There are also alternate explanations for the Neptune desert. The Neptune desert could be formed by high-eccentricity migration of Neptune-mass planets. If a Neptune-mass planet forms far away from its host star and then migrates inward over time, its orbit will be highly eccentric, and so it may pass through the star's tidal disruption radius and be destroyed\cite{Owen2018,Vissapragada2022}. In contrast, more-massive Jupiter-like planets will have survivable orbits as a result of tidal circularization\cite{Matsakos2016}. In fact, studies have shown that photoevaporation is insufficient to explain the upper edge of the Neptune desert\cite{Ionov2018,Vissapragada2022}. Clearly, there are many different explanations for the radius gap and the Neptune desert, and more investigation is required into their origins.


\subsubsection{Helium 1083 nm Triplet}

To distinguish between these competing explanations for the radius gap and Neptune desert, reliable observations of atmospheric escape are required. There are several tracers for atmospheric escape, including Lyman-$\alpha$ 121.5~nm, Ca {\footnotesize II} 393.3 nm, Na {\footnotesize I} 589 nm, H$\alpha$ 656.3 nm, and He~{\footnotesize I} 1083 nm\cite{Oklopi2018,Cauley2018,Seager2000}. Out of these, the helium 1083 nm triplet has clear advantages as a tracer. Compared to Lyman-$\alpha$, helium 1083 nm is less susceptible to interstellar medium absorption and geocoronal emission \cite{Oklopi2018}. Compared to H$\alpha$, Ca {\footnotesize II}, and Na {\footnotesize I}, helium 1083 nm is less affected by contamination from stellar activity such as starspots \cite{Cauley2018}. Compared to UV lines, helium 1083 nm is more accessible to ground-based observatories. This feature is also extremely weak in inactive host stars\cite{Seager2000}. For these reasons, the helium 1083~nm triplet has been preferred for use as a tracer for atmospheric escape in recent years.

A helium atom can be in a singlet state or a triplet state, which happens when its two electrons have opposite spin and parallel spin respectively. The ground state $1^1S$ (electron configuration $1s^2$) is a singlet state. The excited state with the lowest energy is $2^3S$ (electron configuration $1s^1\,2s^1$), which is a triplet state. Since these two states have different total spin, the transition from $2^3S$ to $1^1S$ is spin-forbidden, and so radiative transitions are suppressed between these two states\cite{Oklopi2018}. However, through a single-photon magnetic dipole process, the $2^3S$ state can radiatively decay to the $1^1S$ ground state at a rate of $1.27\times10^{-4}$ s$^{-1}$, or in other words, one decay over $2.18$ hours\cite{Drake1971}. This makes the $2^3S$ state ``metastable''. The next excited triplet state with energy above that of the $2^3S$ state is the $2^3P$ triplet state (electron configuration $1s^1\,2p^1$). This $2^3P$ state has fine structure, with three levels with different total angular momentum\cite{Corsaro1973}. Hence the transition from the $2^3S$ state to the $2^3P$ state via resonance scattering\cite{Oklopi2018} will produce three absorption lines at 1082.909 nm, 1083.025 nm, and 1083.034~nm\cite{NIST_ASD}. This is the metstable helium 1083 nm triplet line. The metastable nature of the $2^3S$ state has made it a great source of absorption lines. Consequently, the helium 1083 nm triplet has been used in a variety of applications in astrophysics, including exoplanet atmospheres, stellar wind dynamics\cite{Strader2015,Dupree2009}, galactic chemical evolution\cite{Cooke2022}, and supernovae.


\subsubsection{Out-of-Transit Observations}

\begin{figure}[h]
    \centering
    \begin{subfigure}{0.49\textwidth}
        \centering
        \includegraphics[width=0.87\textwidth]{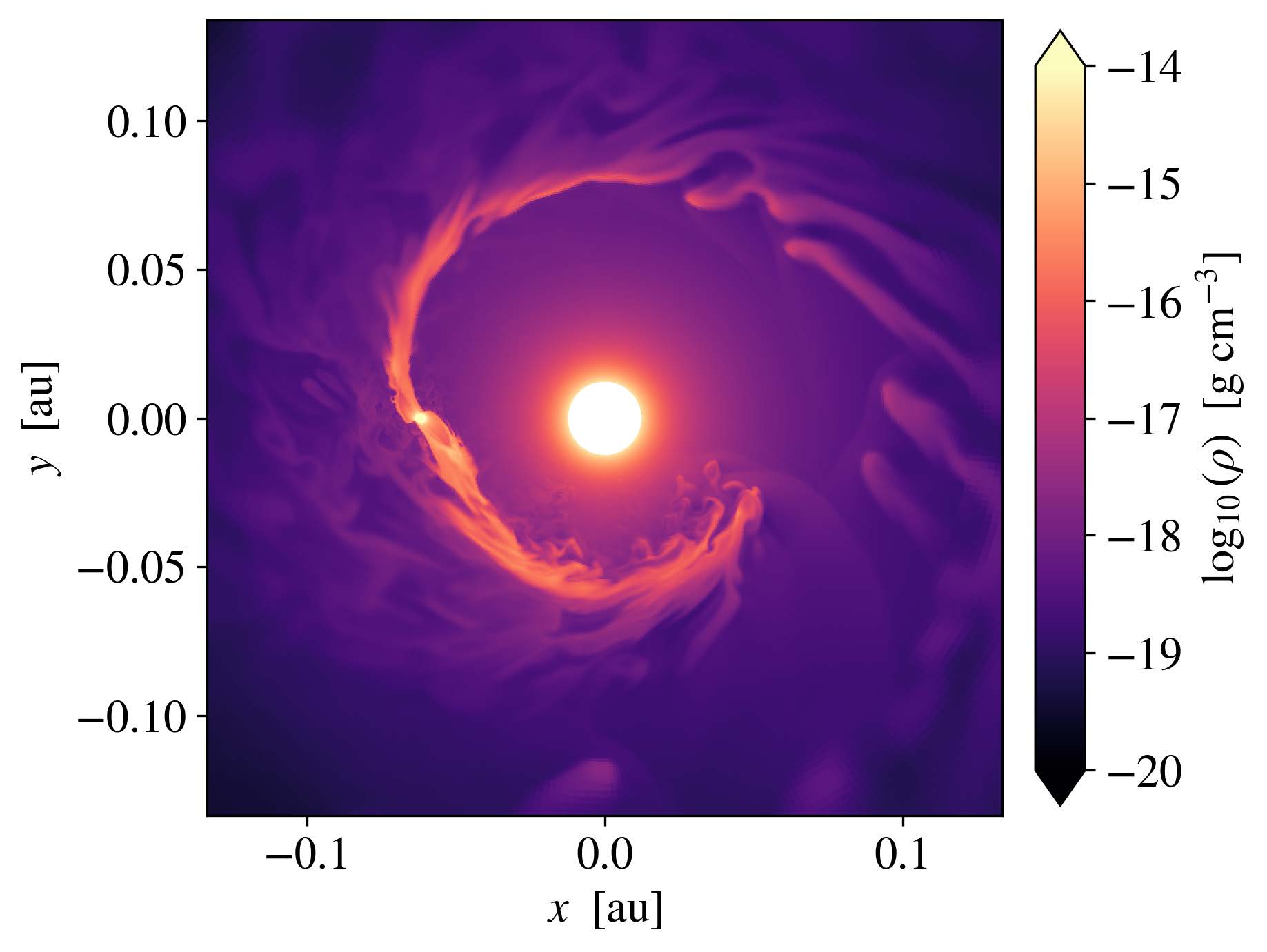}
        \caption{Hydrodynamic simulation of atmospheric escape}
        \label{fig:Nail2025_figA1}
    \end{subfigure}
    \begin{subfigure}{0.49\textwidth}
        \centering
        \includegraphics[width=0.87\textwidth]{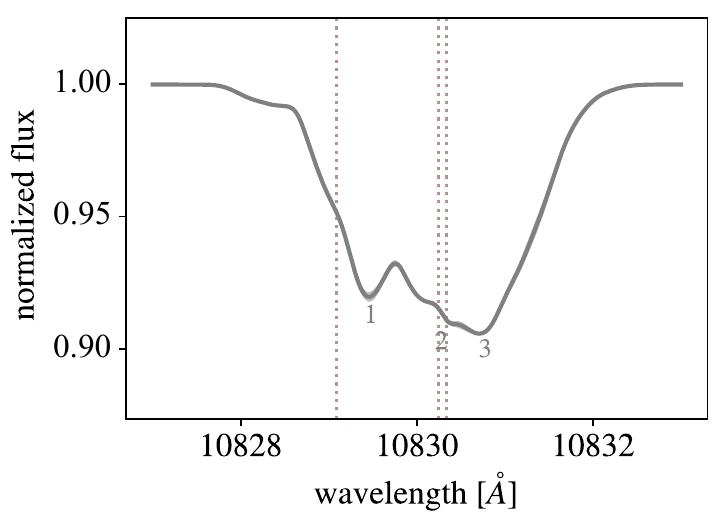}
        \caption{Corresponding spectrum at mid-transit}
        \label{fig:Nail2025_fig5}
    \end{subfigure}
    \caption{3D hydrodynamic simulation of atmospheric escape from Ref.~\citenum{Nail2025}, for a system similar to HAT-P-67 b. Figure~\ref{fig:Nail2025_figA1} shows the gas density. The star is in the center of the plot and the observer is looking from the left. Figure \ref{fig:Nail2025_fig5} shows a mid-transit spectrum. Three components can be seen: (1) a blueshifted trailing stream, (2) the planet terminator with minimal velocity shift, and (3) a redshifted leading stream.}
    \label{fig:Nail2025}
\end{figure}

The helium 1083 nm triplet was first used to conclusively detect atmospheric escape during the transit of planet WASP-107b, observed by the Hubble Space Telescope in 2017\cite{Spake2018}. Since then, over 50 exoplanets have been probed for the helium 1083~nm feature\cite{Krishnamurthy2024}. Most studies to date have focused on observing the helium 1083~nm triplet during or near the planet's transit just to detect a signature of atmospheric escape\cite{Allart2023,Guilluy2024,Masson2024}. However, extending observations beyond transit to cover the full orbital phase can provide deeper insights into the escaping gas' velocity and density distributions, offering a 3D time-evolved picture of the system's dynamics. Two prominent examples are the hot Jupiter HAT-P-32 b\cite{Zhang2023} and the hot Saturn HAT-P-67 b\cite{GullySantiago2024}. Extended out-of-transit observations have revealed giant tails of gas anisotropically outflowing from these two planets, spanning hundreds of planetary radii\cite{GullySantiago2024,Zhang2023}. 3D hydrodynamic simulations have been used to model the outflows in such systems, showing complex gas kinematics\cite{Zhang2023,Nail2025}. Figure \ref{fig:Nail2025_figA1} shows a 3D hydrodynamic simulation by Ref.~\citenum{Nail2025} of atmospheric escape in a system similar to HAT-P-67 b. This simulation includes the effects of stellar winds\cite{MacLeod2022} and planetary day-night temperature gradients\cite{Nail2024}, and was able to reproduce the giant tails inferred from the time-series spectra of HAT-P-67 b.

Time-series spectra offer valuable insights into the underlying gas velocity and density distributions seen in Figure \ref{fig:Nail2025_figA1}. The gas kinematics can be inferred from the absorption line profile of the helium 1083 nm triplet. For example, Figure \ref{fig:Nail2025_fig5} shows a simulated observed spectrum at mid-transit, which corresponds to the simulation in Figure \ref{fig:Nail2025_figA1}. Ref.~\citenum{Nail2025} remarks that there are three components in this spectrum, which correspond to (1) the blueshifted trailing stream launched from the planet's nightside, (2) a component with little velocity shift which corresponds to gas at the planet terminator, and (3) the redshifted leading stream launched from the planet's dayside. From this spectrum alone, we can deduce the kinematics of the escaping atmosphere and the underlying physical processes driving escape. This multiple-component absorption line profile can also be seen in spectra outside of transit since the outflows are extended.

Hydrodynamic simulations have shown that these extended streams are created when the outflow is relatively cool (a few thousand kelvin); the line profiles are not thermally broadened, but kinematically broadened, resulting in irregular shapes\cite{Nail2025}. This may suggest that higher resolution spectra could reveal even more details about the kinematics of the escaping gas. This would provide stronger constraints on the underlying physical mechanisms driving atmospheric escape. To this end, we are developing VIPER -- a spectrograph with a resolving power of $\mathcal{R}$ = 300,000, which is notably greater than the typical 50,000--100,000 resolving powers of near-infrared spectrographs used to observe the line profile of the helium 1083~nm triplet. The remainder of this section will discuss the instrument requirements of VIPER, derived from the primary science goal of observing anisotropic atmospheric escape in gaseous planets.


\subsection{Spectrograph Resolving Power}\label{sec:req_R}

The resolving power of an astronomical spectrograph is one of the most important instrument requirements, as it informs the choice of the primary disperser and hence the spectrograph design. The resolving power $\mathcal{R}$ at a particular wavelength $\lambda_0$ is defined as:
\begin{equation}
    \mathcal{R}\equiv\frac{\lambda_0}{\Delta\lambda}
\end{equation}
where $\Delta\lambda$ is the resolution element. To determine the required resolving power for VIPER, we first created a spectrum simulator that produces realistic astronomical spectra from ideal, infinitely-resolved, synthetic spectra of helium 1083 nm observations, given parameters such as the target apparent magnitude, telescope diameter, exposure time, and resolving power. We then passed synthetic spectra of anisotropic atmospheric escape through our spectrum simulator, and then computed the detection significance of anisotropic atmospheric escape for different resolving powers.


\subsubsection{Spectrum Simulator}

Let $S(\lambda)$ be an ideal, infinitely-resolved, normalized, synthetic spectrum. Given a 2MASS $J$-band apparent magnitude $m_J$, telescope diameter $D_T$, exposure time $t_\mathrm{exp}$, and instrument efficiency $\epsilon$, we can compute an energy density $E_\lambda$ at the telescope, with units of J/\AA:
\begin{align}
    E_\lambda= \left(F_{\mathrm{ZP},\lambda} 10^{-m_J/2.5}\right) \times \pi\left(\frac{D_T}{2}\right)^2 \times t_\mathrm{exp} \times \epsilon \:\:\:\:\:\:\:\:\:\:\:\: \textrm{[J/\AA]}
\end{align}
where $F_{\mathrm{ZP},\lambda}$ is the $J$-band zero point flux density from Ref.~\citenum{Cohen2003}. The effect of the instrument can be modeled as a convolution of $E_\lambda \,S(\lambda)$ with the instrument line spread function, which we model as a normal function:
\begin{equation}
    \mathcal{N}(\lambda,\sigma,\mu) = \frac{1}{\sqrt{2\pi}\sigma} \exp{\left(-\frac{(\lambda-\mu)^2}{2\sigma^2}\right)}
\end{equation}
where we take $\sigma=\frac{\lambda_0/\mathcal{R}}{2\sqrt{2\ln{2}}}$ and $\lambda_0=10830$ {\AA}. The convolved energy density is then:
\begin{equation}
    \hat{E}_\lambda=\int_{-\infty}^{\infty} E_{\lambda'} \,S(\lambda') \,\mathcal{N}\left(\lambda-\lambda',\frac{\lambda_0/\mathcal{R}}{2\sqrt{2\ln{2}}},0\right) \,d\lambda' \:\:\:\:\:\:\:\:\:\:\:\: \textrm{[J/\AA]}
\end{equation}
which still has units of J/{\AA} because $\int_{-\infty}^{\infty} \mathcal{N}(\lambda,\sigma,\mu)\,d\lambda=1$. We then compute a count density $C_\lambda$ at the detector, which has units of counts/{\AA}, through division by the photon energy:
\begin{equation}
    C_\lambda = \hat{E}_\lambda\left(\frac{hc}{\lambda}\right)^{-1} \:\:\:\:\:\:\:\:\:\:\:\: \textrm{[counts/\AA]}
\end{equation}
where $h$ is the Planck constant and $c$ is the speed of light. Next, we bin and pixelize the spectrum by integrating $C_\lambda$ over the resolution element $\Delta\lambda_\mathrm{pix}$ corresponding to one pixel:
\begin{equation}
    C_\mathrm{counts}(\lambda) =\int_{\lambda-\Delta\lambda_\mathrm{pix}/2}^{\lambda+\Delta\lambda_\mathrm{pix}/2} C_{\lambda'} \,d\lambda' \:\:\:\:\:\:\:\:\:\:\:\: \textrm{[counts]}
\end{equation}
This integral can be evaluated numerically, by convolving $C_\lambda$ with a rectangle function of width $\Delta\lambda_\mathrm{pix}$. We take $\Delta\lambda_\mathrm{pix}=\Delta\lambda/3$. Note that Nyquist sampling demands that $\Delta\lambda\geq2\Delta\lambda_\mathrm{pix}$. Finally, we add Poisson noise to $C_\mathrm{counts}(\lambda)$ and hence obtain a realistic, noisy, simulated spectrum. Figure \ref{fig:specsim_res} shows some examples of simulated spectra with several different resolving powers.


\subsubsection{Synthetic Spectra}

The helium 1083 nm triplet absorption feature can be modeled as three Gaussian components due to thermal broadening:
\begin{align}
    f_\mathrm{He}(\lambda,T) &= \sum_{i=1}^3 A_\mathrm{He,i} \exp{\left(-\frac{(\lambda-\mu_\mathrm{He,i})^2}{2[\sigma(T)]^2} \right)} \\
    \sigma(T) &= \frac{\lambda}{c} \sqrt{\frac{k_\mathrm{B} T}{m_\mathrm{He}}}
\end{align}
where $\mu_\mathrm{He,1}=10829.09115$ \AA, $\mu_\mathrm{He,2}=10830.25011$ \AA, and $\mu_\mathrm{He,3} = 10830.33978$ {\AA} are the locations of the triplet lines, and $A_\mathrm{He,1}=300$, $A_\mathrm{He,2}=1000$ and $A_\mathrm{He,3}=2000$ are the relative amplitudes of the lines, from Ref.~\citenum{NIST_ASD}. $k_\mathrm{B}$~is the Boltzmann constant and $m_\mathrm{He}$ is the helium mass.

\begin{figure}[H]
    \centering
    \begin{subfigure}{0.49\textwidth}
        \centering
        \includegraphics[width=0.9\textwidth]{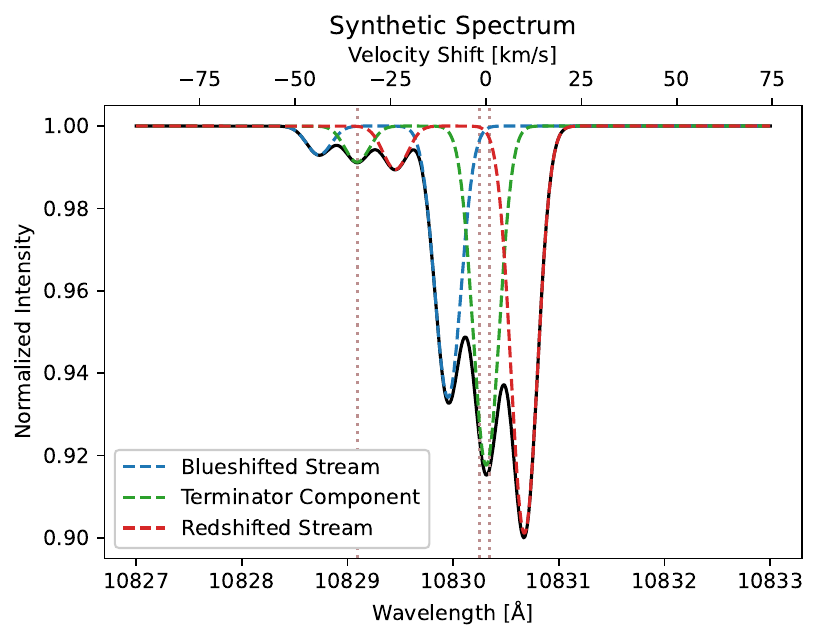}
        \caption{Synthetic spectrum from our toy model, Eq.~(\ref{eq:toymodel})}
        \label{fig:specsim_synthetic}
    \end{subfigure}
    \begin{subfigure}{0.49\textwidth}
        \centering
        \includegraphics[width=0.9\textwidth]{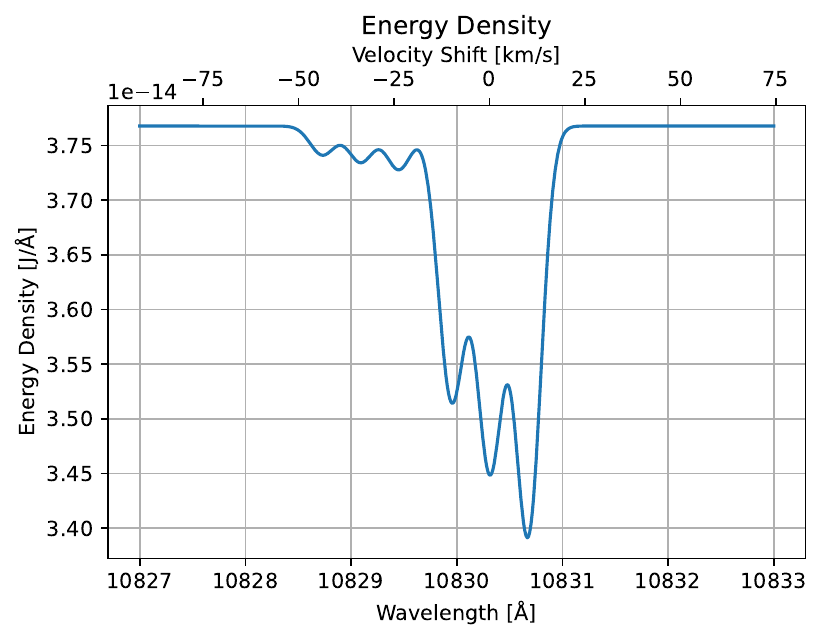}
        \caption{Corresponding energy density}
        \label{fig:specsim_energyden}
    \end{subfigure}
    \caption{The left figure shows a synthetic spectrum generated using our three-component toy model, Eq.~(\ref{eq:toymodel}). The three components correspond to the three components in Figure \ref{fig:Nail2025_fig5}. Here, we used $T_1=T_2=T_3=5000$ K. The three components are each separated by $\Delta v=10$ km/s. The right figure shows the corresponding energy density at the telescope, for $m_J=8$, $D_T=1.5$~m, $t_\mathrm{exp}=360$ s, and $\epsilon=0.3$.}
    \label{fig:specsim_1}
\end{figure}

\begin{figure}[H]
    \centering
    \begin{subfigure}{0.325\textwidth}
        \centering
        \includegraphics[width=\textwidth]{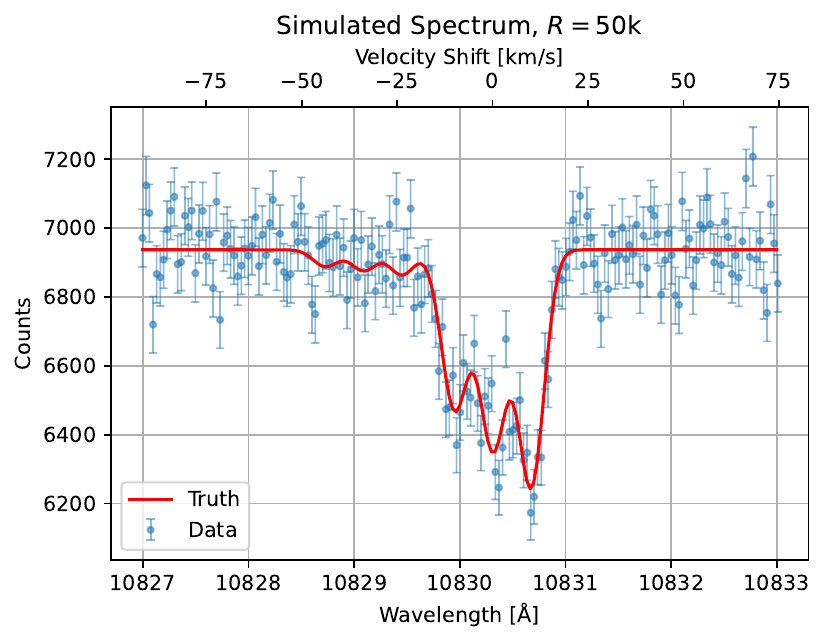}
        \caption{$\mathcal{R}=50,000$}
    \end{subfigure}
    \begin{subfigure}{0.325\textwidth}
        \centering
        \includegraphics[width=\textwidth]{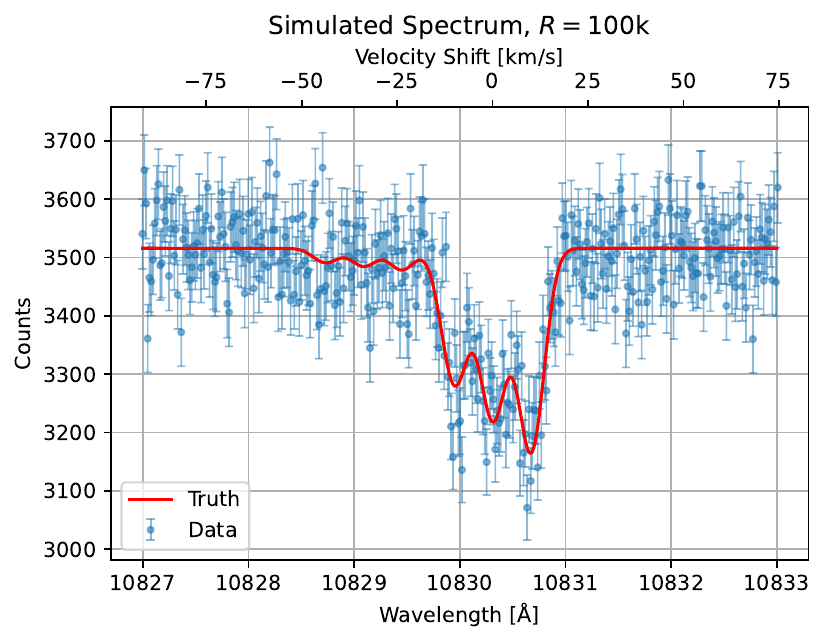}
        \caption{$\mathcal{R}=100,000$}
    \end{subfigure}
    \begin{subfigure}{0.325\textwidth}
        \centering
        \includegraphics[width=\textwidth]{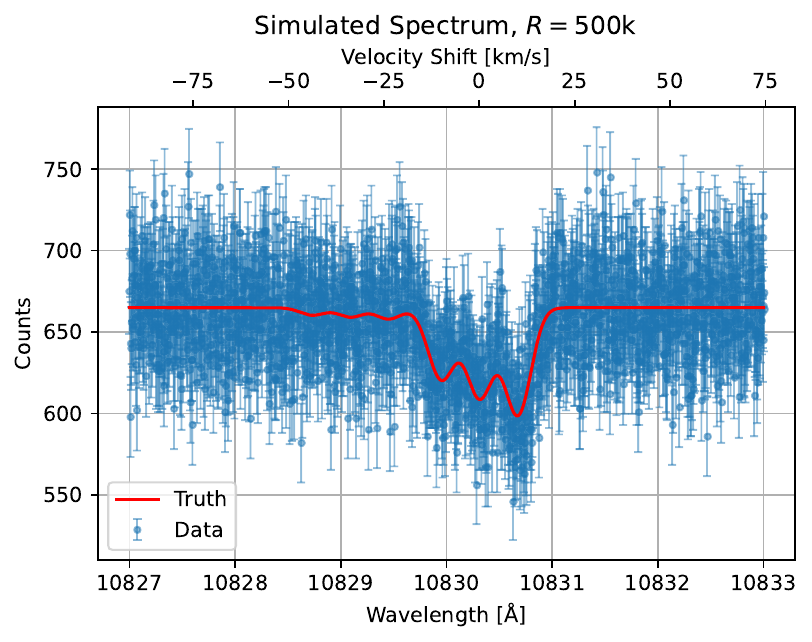}
        \caption{$\mathcal{R}=500,000$}
        \label{fig:specsim_res_500k}
    \end{subfigure}
    \caption{Simulated spectra from our spectrum simulator, with the synthetic spectrum in Figure \ref{fig:specsim_synthetic} used as the input. As the resolving power increases, the number of continuum counts decreases, as expected. The parameters used were $m_J=8$, $D_T=1.5$~m, $t_\mathrm{exp}=360$ s, and $\epsilon=0.3$.}
    \label{fig:specsim_res}
\end{figure}

To model anisotropic atmospheric escape, we used a toy model with three helium gas components, representing the three gas components seen in spectra in Ref.~\citenum{Nail2025} (see Figure \ref{fig:Nail2025_fig5}). This is a first-order approximation to the more complex 3D hydrodynamic simulations. Our toy model is:
\begin{equation}\label{eq:toymodel}
    f_\mathrm{M_1}(\lambda,T_1,T_2,T_3,A_1,A_2,A_3,\mu_1,\mu_2,\mu_3) = \sum_{i=1}^3 A_i f_\mathrm{He}(\lambda-\mu_i,T_i)
\end{equation}
where the free parameters are $T_1,T_2,T_3,A_1,A_2,A_3,\mu_1,\mu_2,\mu_3$. Figure \ref{fig:specsim_synthetic} shows an example synthetic spectrum generated using this model.

\subsubsection{Hypothesis Testing}\label{sec:science_hypo_test}

With the spectrum simulator and a synthetic spectra toy model, we can conduct hypothesis testing to see what combination of parameters will result in the detection of anisotropic atmospheric escape. The null hypothesis is that there is no anisotropic atmospheric escape, which can be described by the one-component model:
\begin{equation}\label{eq:null_hyp_model}
    f_\mathrm{M_0}(\lambda,T,A,\mu) = A f_\mathrm{He}(\lambda-\mu,T)
\end{equation}
where the free parameters are $T,A,\mu$. The alternative hypothesis is that there is anisotropic atmospheric escape, modeled by Eq.~(\ref{eq:toymodel}).

\begin{figure}[h]
    \centering
    \includegraphics[width=0.5\textwidth]{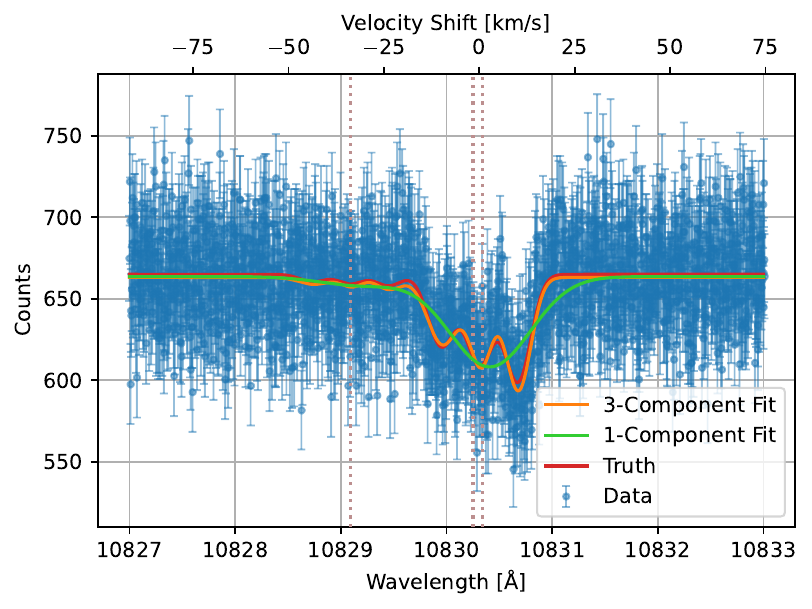}
    \caption{$9.5\sigma$ detection of anisotropic atmospheric escape in a $\mathcal{R}$ = 500,000 simulated spectrum from Figure \ref{fig:specsim_res_500k}. The 3-component fit is using the model from Eq.~(\ref{eq:toymodel}), and the 1-component fit is using the null hypothesis model from Eq.~(\ref{eq:null_hyp_model}).}
    \label{fig:hypo_test_example}
\end{figure}

For a particular simulated spectrum (e.g., those in Figure \ref{fig:specsim_res}), we fit Eqs.~(\ref{eq:null_hyp_model}) and (\ref{eq:toymodel}) to the data using nonlinear least squares regression, and computed the chi-squared of the fits, which we call $\chi^2_0$ and $\chi^2_1$ respectively for the null hypothesis model and the alternative hypothesis model. Let $\hat{\mathcal{L}}_0$ and $\hat{\mathcal{L}}_1$ be the maximum likelihood ratios of the data under the null hypothesis model and the alternative hypothesis model respectively. Then we computed the likelihood ratio test statistic:
\begin{equation}
    \lambda_\mathrm{LR} = -2\ln{\left(\frac{\hat{\mathcal{L}}_0}{\hat{\mathcal{L}}_1}\right)} \approx \chi^2_0 -\chi^2_1 \equiv\Delta\chi^2
\end{equation}
If there is a large sample size, then the likelihood ratio test statistic follows a chi-squared distribution $\Delta \chi^2$ with the number of degrees of freedom being equal to number of parameters in the alternative hypothesis model minus the number of parameters in the null hypothesis model. From the survival function of this chi-squared distribution, we then computed a two-tailed $p$-value. From the $p$-value, we computed a $z$-score, which is the detection significance of anisotropic atmospheric escape. Figure \ref{fig:hypo_test_example} shows an example of a $9.5\sigma$ detection of anisotropic atmospheric escape, with the detection significance computed following this described procedure.


\subsubsection{Detection Significance on a Grid of Parameters}

We repeated this procedure on a grid of simulated spectra with different parameters, varying the stream velocity shift $\Delta v$, stream temperature $T$, target $J$-band apparent magnitude $m_J$, and spectrograph resolving power~$\mathcal{R}$. The goal was to find what resolving power would result in at least $5\sigma$ detection of anisotropic atmospheric escape.

In this grid, we varied the velocity shift of the blueshifted and redshifted streams from 1 km/s to 20 km/s, which is the typical velocity for gas undergoing atmospheric escape when it leaves the planet's Hill sphere\cite{Owen2022}. We varied the temperature of the gas from 500 K to 5000 K. In Ref.~\citenum{Nail2025}, a 4700 K stream temperature was found to model the HAT-P-67 b system well, but in theory the temperature can be lower. The line broadening is mainly due to kinematic effects instead of thermal broadening. Hence, we chose 5000~K as the upper limit. We chose 500~K as the lower limit because this is around the lowest equilibrium temperature for the planets that we are trying to observe (see Figure \ref{fig:planet_eq}). We varied the $J$-band apparent magnitude from 8.5 to 10.5, and the resolving power from 25,000 to 500,000. For each quadruple of $(\Delta v, T, m_J, \mathcal{R})$, we created 15 spectra using different random seeds so that the Poisson noise and fitting process are different. By averaging over these 15 spectra, we reduce the spread in the resulting metrics.

In all of the synthetic spectra, we set $D_T=1.5$ m for the Tillinghast Telescope, $t_\mathrm{exp}=360$ s, and $\epsilon=0.3$. The exposure time $t_\mathrm{exp}$ was selected as 360 s because typical helium 1083 nm atmospheric escape observations in the literature had individual exposure times in the hundreds of seconds\cite{Allart2023,Guilluy2024,Masson2024}. The relative amplitudes of the three stream components were set as 0.8 to 1 to 1.2, inspired by Ref.~\citenum{Nail2025}. The maximum depth of the absorption feature was scaled in each spectra such that at 4700 K, the normalized depth would be 0.1, similar to Ref.~\citenum{Nail2025}.

\begin{figure}[h]
    \centering
    \begin{subfigure}{0.49\textwidth}
        \centering
        \includegraphics[width=0.9\textwidth]{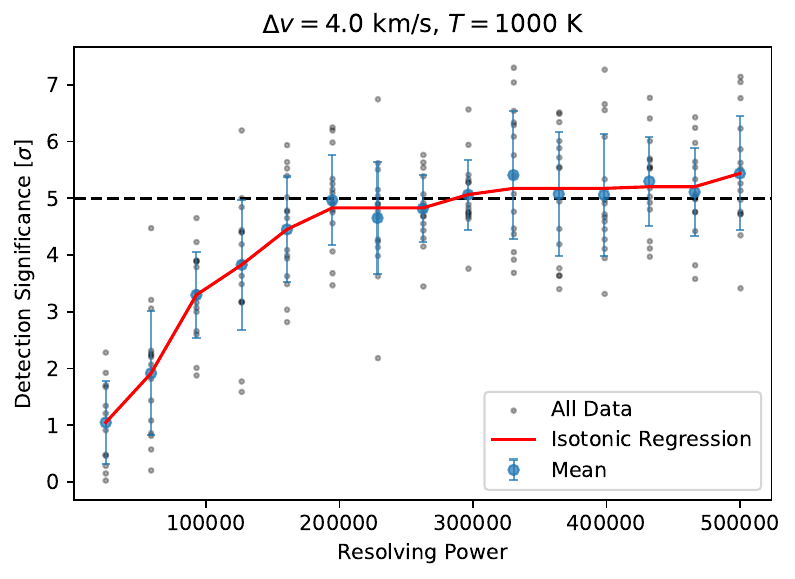}
        \caption{$\Delta v=4$ km/s, $T=1000$ K}
    \end{subfigure}
    \begin{subfigure}{0.49\textwidth}
        \centering
        \includegraphics[width=0.9\textwidth]{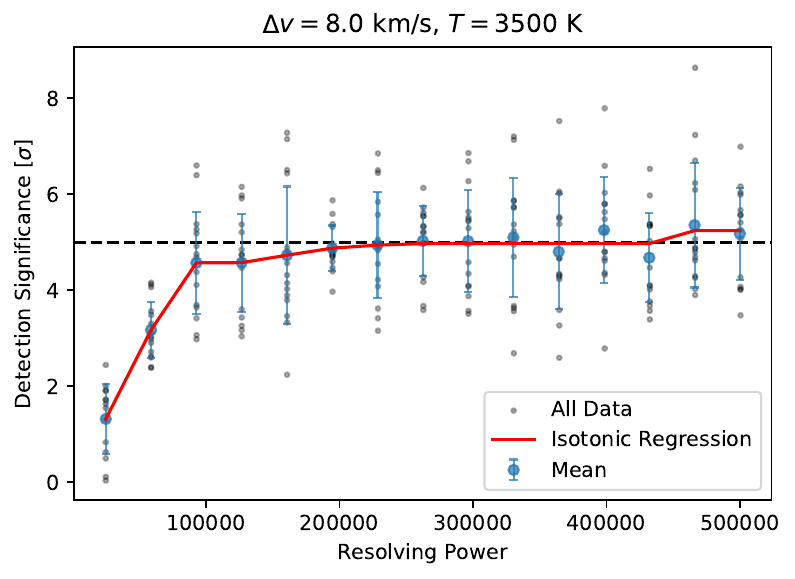}
        \caption{$\Delta v=8$ km/s, $T=3500$ K}
    \end{subfigure}
    \caption{Resolving power VS detection significance for anisotropic atmopsheric escape, for $m_J=9$, $D_T=1.5$ m, $t_\mathrm{exp}=360$ s, $\epsilon=0.3$. Each black point in this figure represents one simulated spectrum and fit (e.g., one Figure \ref{fig:hypo_test_example}).}
    \label{fig:R_VS_sigma}
\end{figure}

\begin{figure}[H]
    \centering
    \begin{subfigure}{0.49\textwidth}
        \centering
        \includegraphics[width=\textwidth]{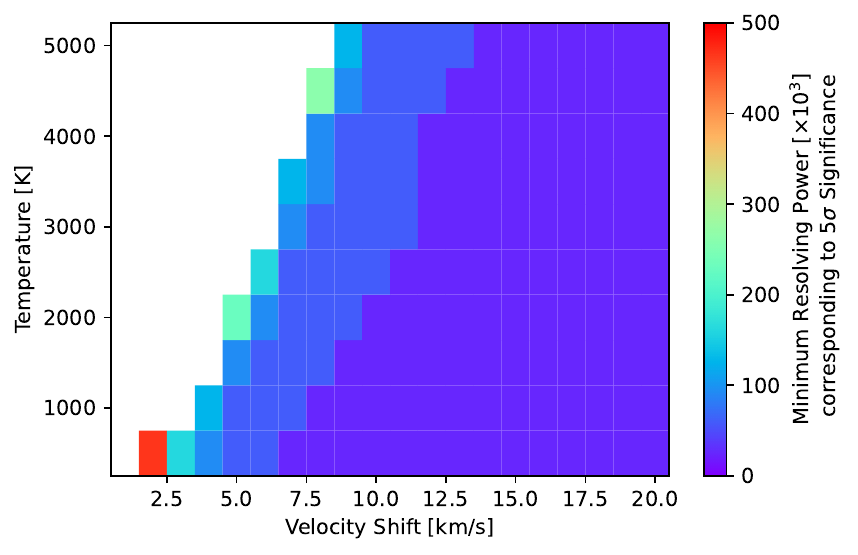}
        \caption{$m_J=8.5$}
    \end{subfigure}
    \begin{subfigure}{0.49\textwidth}
        \centering
        \includegraphics[width=\textwidth]{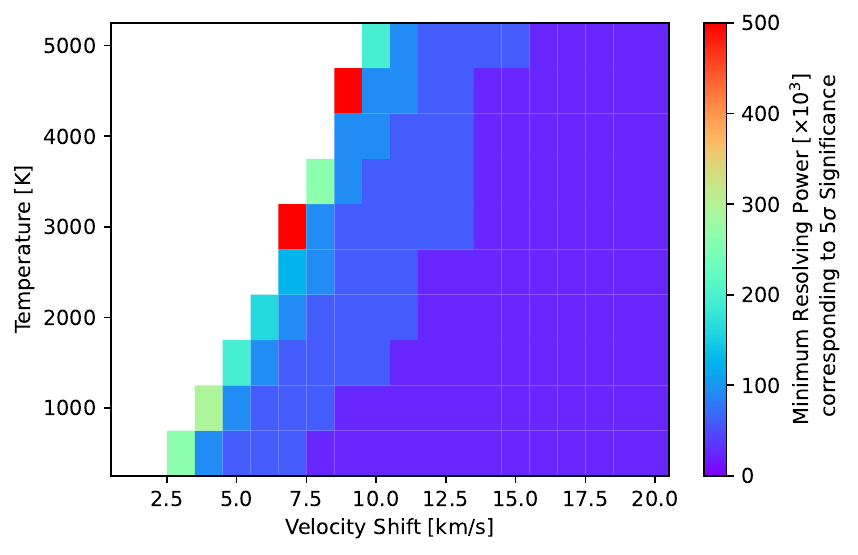}
        \caption{$m_J=9$}
    \end{subfigure}
    \begin{subfigure}{0.49\textwidth}
        \centering
        \includegraphics[width=\textwidth]{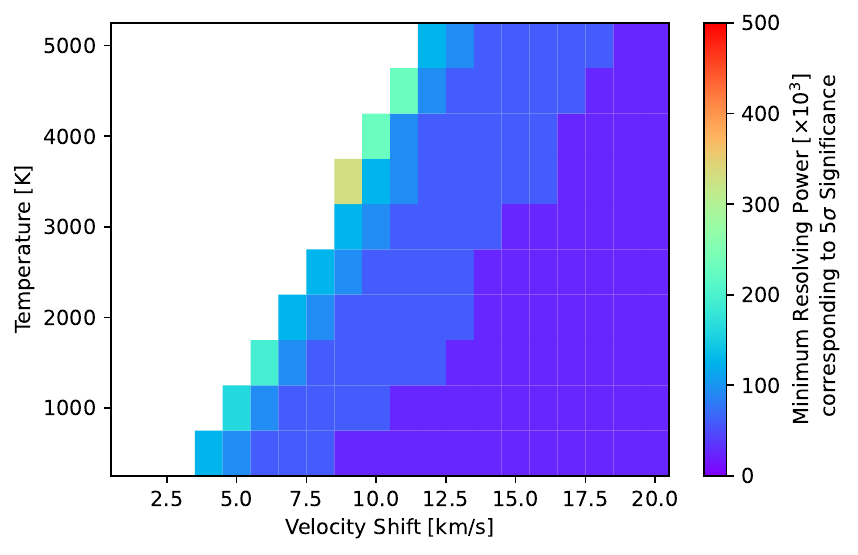}
        \caption{$m_J=9.5$}
    \end{subfigure}
    \begin{subfigure}{0.49\textwidth}
        \centering
        \includegraphics[width=\textwidth]{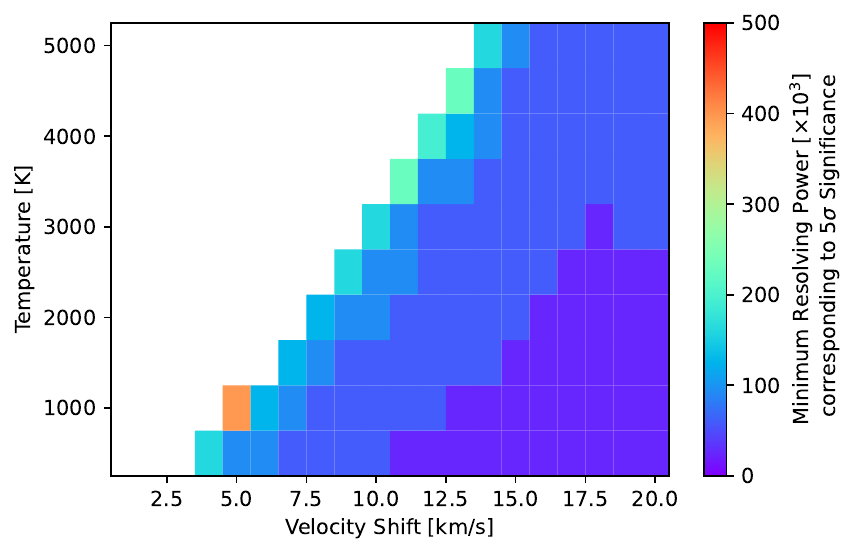}
        \caption{$m_J=10$}
    \end{subfigure}
    \begin{subfigure}{0.49\textwidth}
        \centering
        \includegraphics[width=\textwidth]{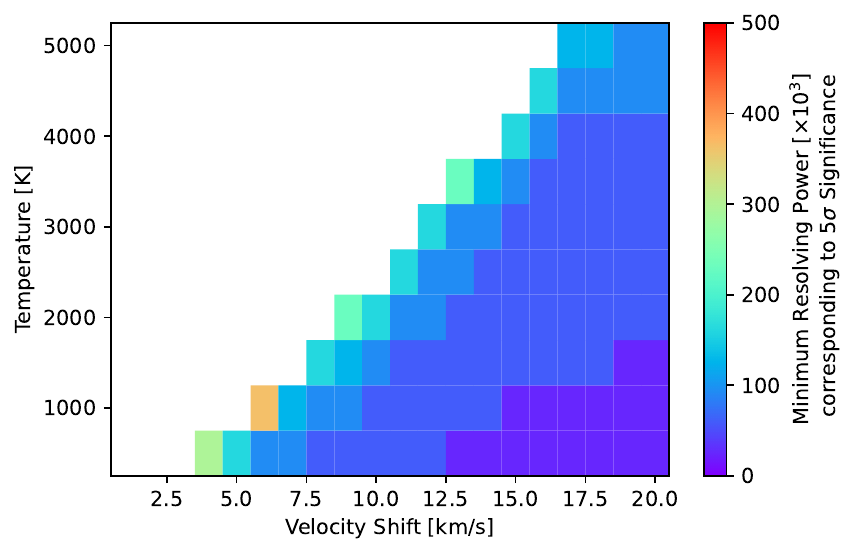}
        \caption{$m_J=10.5$}
    \end{subfigure}
    \caption{Resolving power corresponding to $5\sigma$ detection of anisotropic atmospheric escape, for several $m_J$. Here, we use $D_T=1.5$ m, $t_\mathrm{exp}=360$ s, and $\epsilon=0.3$. The white cells represent a ($\Delta v, T, m_J$) combination where the maximum possible detection significance, within the $\mathcal{R}$ values we investigate, is under $5\sigma$.}
    \label{fig:R_heatmap}
\end{figure}

Figure \ref{fig:R_VS_sigma} plots the resolving power versus the detection significance for anisotropic atmospheric escape, for $m_J=9$ and two specific $(\Delta v,T)$ pairs. Each black dot represents a detection significance computed from the process described in Section \ref{sec:science_hypo_test} applied to one spectrum, like in Figure \ref{fig:hypo_test_example}. The blue points show the mean detection significance for a fixed $(\Delta v, T, \mathcal{R})$ triplet, averaged over the black points at fixed $\mathcal{R}$. We then used isotonic regression to fit the blue points, which should show a monotonically increasing trend with $\mathcal{R}$. From this, we identified the minimum resolving power which corresponds to $5\sigma$ detection significance. These resolving powers are plotted in Figure~\ref{fig:R_heatmap}. Each grid cell in Figure \ref{fig:R_heatmap} represents the result of one resolving power VS detection significance curve. The white cells represent a ($\Delta v, T, m_J$) combination where the maximum possible detection significance, within the $\mathcal{R}$ values we investigate, is under $5\sigma$.

It should be noted that the trend in Figure \ref{fig:R_VS_sigma} is that the detection significance monotonically increases with resolving power but plateaus at some point. Hence, Figure \ref{fig:R_heatmap} should be interpreted as showing the minimum $\mathcal{R}$ required to achieve $5\sigma$ detection significance. Resolving powers above the plotted minimum $\mathcal{R}$ may be able to achieve greater than $5\sigma$ detection significance. Figure \ref{fig:R_heatmap} shows that as the velocity shift $\Delta v$ increases and as the temperature $T$ decreases, the required resolving power for a $5\sigma$ detection decreases. This makes sense because less resolution is needed to resolve separated or broader spectral features. We can also see that as the apparent magnitude $m_J$ increases, there are fewer parameter combinations where $5\sigma$ detection is possible. This is because there will be fewer photons in the spectrum, and hence more photon noise, reducing the detection significance. From Figure \ref{fig:R_heatmap}, we can see that the ``edge'' of the $5\sigma$ detections occurs roughly when the resolving power is between 200,000 and 300,000. Hence, a spectrograph designed to detect anisotropic atmospheric escape should aim to have a resolving power of around 200,000 to 300,000. Our analysis done here is somewhat conservative, because it assumes that we are trying to detect anisotropic atmospheric escape with $5\sigma$ significance from just a single $t_\mathrm{exp}=360$ s exposure. In practice, many frames are taken and they are stacked, and so the effect of photon noise is reduced. We will take a resolving power of $\mathcal{R}$ = 300,000 as the requirement for VIPER.


\subsection{Observing Targets and Spectrograph Bandpass}

\subsubsection{Exoplanets}

Another key requirement for astronomical spectrographs is the bandpass, which also informs the choice of the disperser and the spectrograph design. To determine the bandpass, we need to determine the kinds of planets that VIPER will observe. From the NASA Exoplanet Archive, we selected confirmed planets that have the following properties:
\begin{itemize}
    \vspace{-4pt}
    \itemsep0em 
    \item Has been detected by transit
    \item Period between 0.9 and 10 days
    \item Radius between $3R_\oplus$ and $15R_\oplus$
    \item Mass between $0.01M_\mathrm{Jup}$ and $2M_\mathrm{Jup}$
    \item Declination above $-15^\circ$ so that it is observable from FLWO ($31.68^\circ$ latitude)
\end{itemize}
\begin{figure}[h]
    \centering
    \begin{subfigure}{0.325\textwidth}
        \centering
        \includegraphics[width=\textwidth]{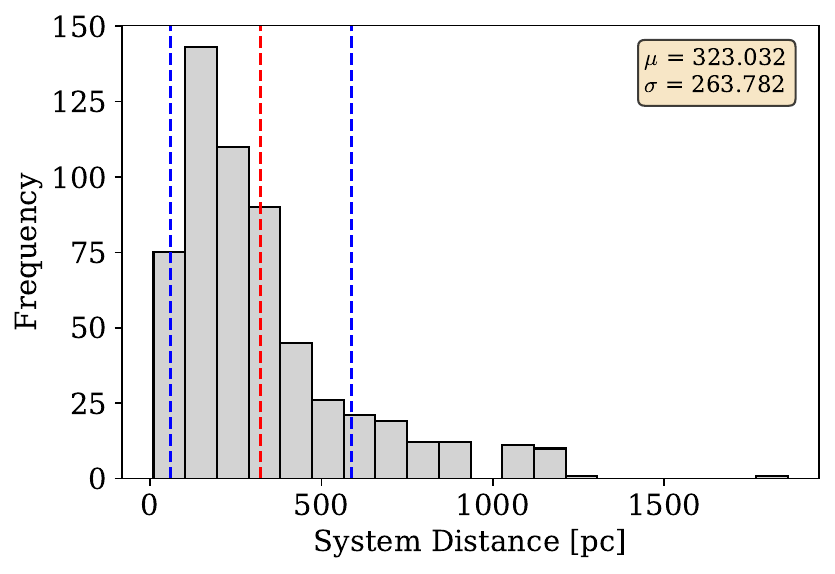}
        \caption{System distance}
    \end{subfigure}
    \begin{subfigure}{0.325\textwidth}
        \centering
        \includegraphics[width=\textwidth]{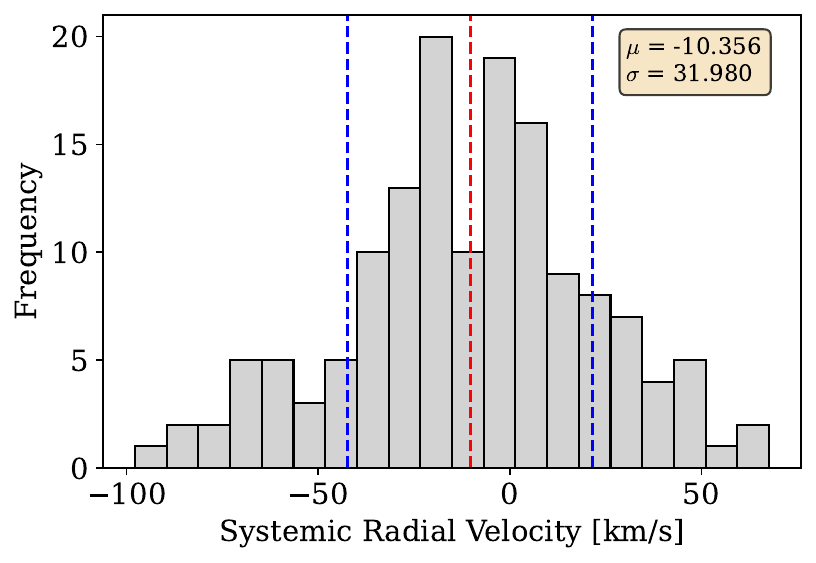}
        \caption{System radial velocity}
    \end{subfigure}
    \begin{subfigure}{0.325\textwidth}
        \centering
        \includegraphics[width=\textwidth]{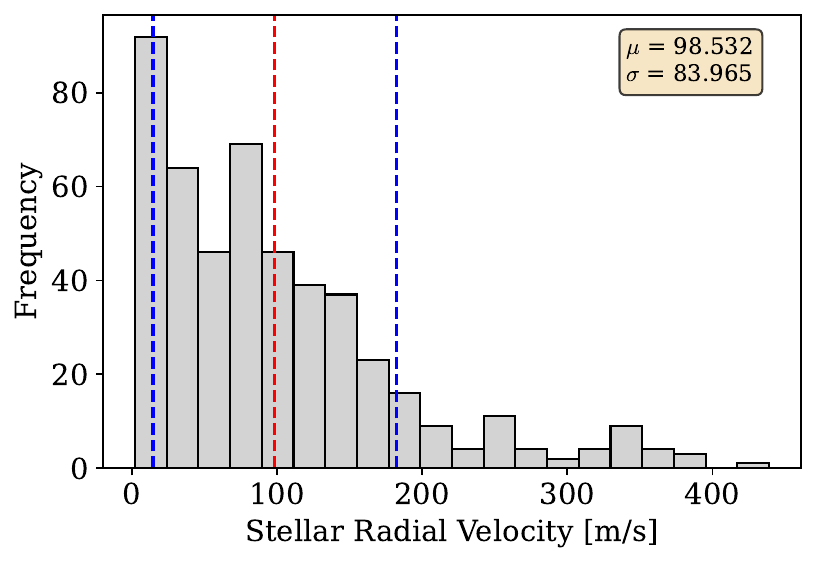}
        \caption{Stellar radial velocity}
    \end{subfigure}
        \begin{subfigure}{0.325\textwidth}
        \centering
        \includegraphics[width=\textwidth]{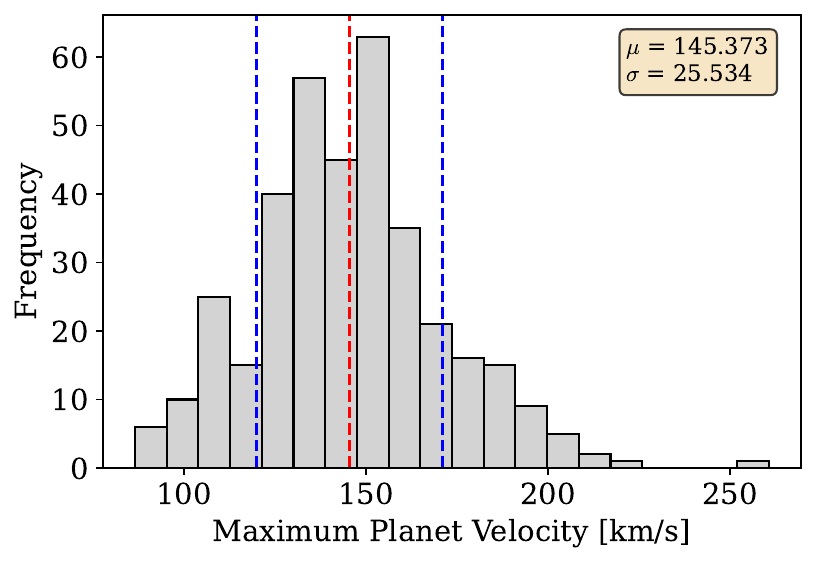}
        \caption{Maximum planet velocity}
    \end{subfigure}
    \begin{subfigure}{0.325\textwidth}
        \centering
        \includegraphics[width=\textwidth]{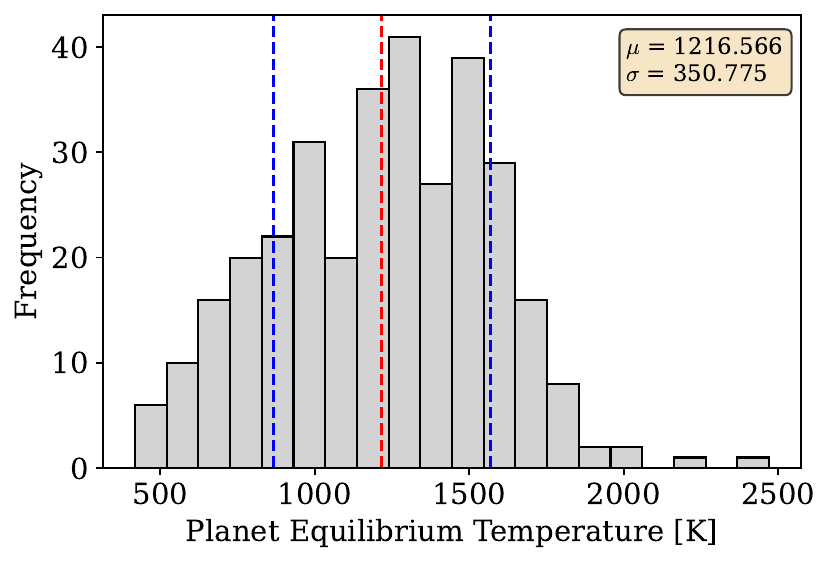}
        \caption{Planet equilibrium temperature}
        \label{fig:planet_eq}
    \end{subfigure}
    \begin{subfigure}{0.325\textwidth}
        \centering
        \includegraphics[width=\textwidth]{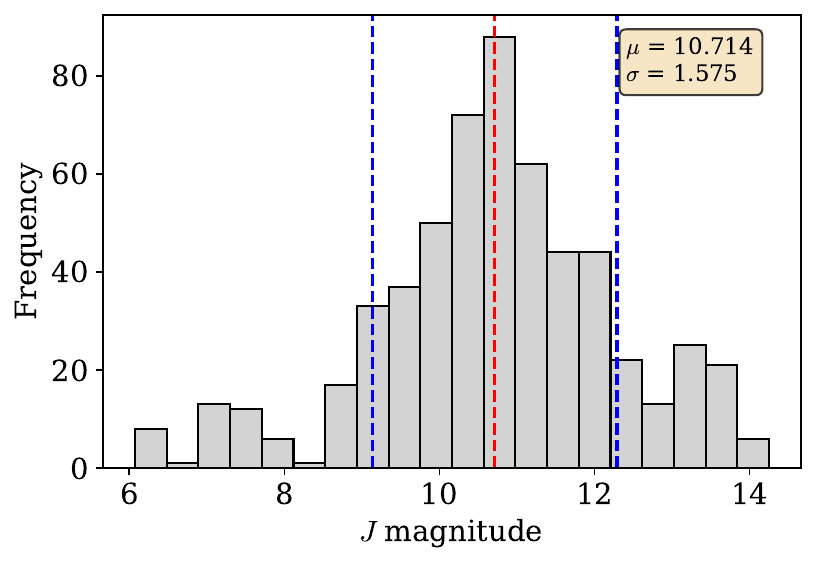}
        \caption{Stellar apparent magnitude}
    \end{subfigure}
    \caption{Properties of selected planets that may be probed by VIPER. These planets correspond to the red points in Figure \ref{fig:mass_radius_period}. ``$\mu$'' and ``$\sigma$'' denote the mean and standard deviations respectively of each property. The red dashed line denotes $\mu$ and the blue dashed lines denote $\mu\pm\sigma$.}
    \label{fig:sample_prop}
\end{figure}
This selection is shown by the red points in Figure \ref{fig:mass_radius_period}. Some properties of planets in this sample are shown in Figure \ref{fig:sample_prop}. All of the properties are from the NASA Exoplanet Archive, with the exception of the maximum planet velocity, which we compute by:
\begin{equation}
    v_\mathrm{max} = \sqrt{G(M_\star+M_\mathrm{p})\left(\frac{2}{a(1-e)}-\frac{1}{a}\right)}
\end{equation}
where $M_\star$ is the stellar mass, $M_\mathrm{p}$ is the planet mass, $G$ is the gravitational constant, $a$ is the semi-major axis of the orbit, and $e$ is the eccentricity of the orbit. This equation was obtained by substituting the distance between the planet and the star at periastron, $a(1-e)$, into Eq. (34) of Ref.~\citenum{SeagerBookCh2}. Table~\ref{tab:bandpass_planets} lists some estimated contributions to the line-of-sight velocity of the planets in the sample. The total sum of these contributions is 490 km/s, which only happens if all of the contributions are correlated. This gives a wavelength shift of 1.77 nm about 1083 nm, which implies a minimum bandpass of 3.54 nm about 1083 nm.

\begin{table}[H]
\centering
\begin{tabular}{ | m{0.5\textwidth} | m{0.3\textwidth}| } 
  \hline
  Contribution & Speed \\ 
  \hline
  \hline
  Maximum orbital speed of planet & 260 km/s \\
  Maximum absolute systemic radial velocity & 100 km/s \\
  Motion of Earth around Sun & 30 km/s \\
  Maximum speed of helium gas outflow from planet & $\lesssim100$ km/s \\
  \hline
  Total Sum & 490 km/s\\
  \hline
\end{tabular}
\caption{Velocity contributions along line-of-sight for exoplanets in Figure \ref{fig:sample_prop}.}\label{tab:bandpass_planets}
\end{table}


\subsubsection{Stellar Winds}

The helium 1083 nm triplet has also been used as a probe for stellar winds and outflows, in studies of stellar mass loss\cite{Dupree2009,Strader2015}. Table \ref{tab:bandpass_stars} lists some estimated contributions to the line-of-sight velocity of relevant stars. The total sum of these contributions is 530 km/s, which again, only happens if all of the contributions are correlated. This gives a wavelength shift of 1.91 nm about 1083 nm, which implies a minimum bandpass of 3.82 nm about 1083~nm.

\begin{table}[H]
\centering
\begin{tabular}{ | m{0.5\textwidth} | m{0.3\textwidth}| } 
  \hline
  Contribution & Speed \\ 
  \hline
  \hline
  Stellar outflows & $\sim 100$ km/s\cite{Strader2015} \\
  Stellar systemic velocity & $\lesssim400$ km/s\cite{Dupree2009} \\
  Motion of Earth around Sun & 30 km/s \\
  \hline
  Total Sum & 530 km/s\\
  \hline
\end{tabular}
\caption{Velocity contributions along line-of-sight for stars of interest.}\label{tab:bandpass_stars}
\end{table}


\subsection{Spectrograph Efficiency}

\begin{figure}[H]
    \centering
    \includegraphics[width=0.5\textwidth]{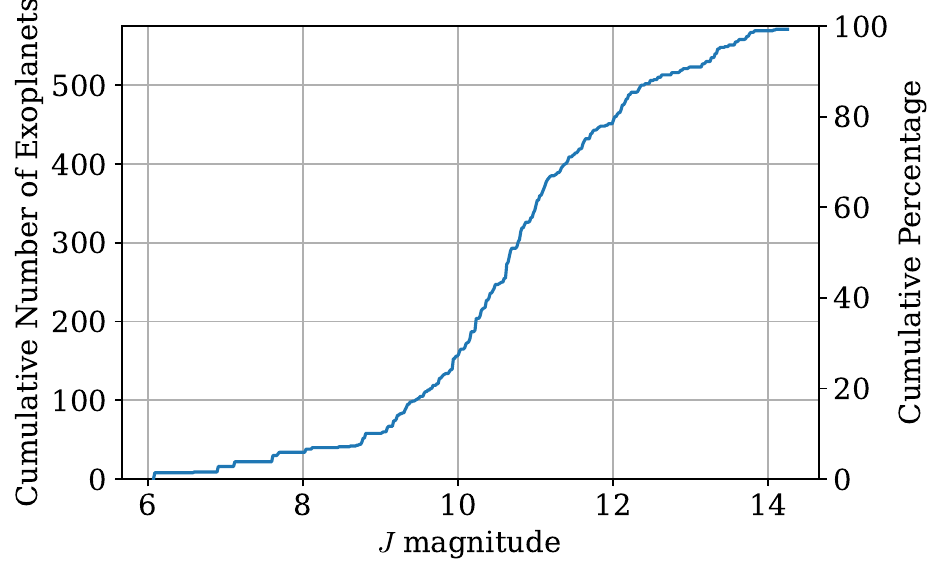}
    \caption{Cumulative histogram of $J$-band apparent magnitude, for the sample of exoplanets shown by the red points in Figure \ref{fig:mass_radius_period}.}
    \label{fig:J_mag_cum}
\end{figure}

The efficiency of the spectrograph will determine which targets can be observed. However, estimating the required efficiency is nontrivial because this would depend on the specific analysis methods used to detect anisotropic atmospheric escape, which will likely be more complicated than the procedure described in Section \ref{sec:science_hypo_test}. Nevertheless, as seen in Section \ref{sec:req_R}, an efficiency of $\epsilon=0.3$ allows for the detection of anisotropic atmospheric escape at $5\sigma$ under certain combinations of parameters, even for targets as faint as $m_J=10.5$. Based on this, we adopt an estimated minimum required efficiency of 0.3, and aim to target stars with $m_J\lesssim10$. This results in over 100 targets, as seen in Figure \ref{fig:J_mag_cum}. Another spectrograph specifically designed for atmospheric escape observations, called NIGHT\cite{FarretJentink2023}, with a lower resolving power of $\mathcal{R}$ = 70,000 and designed for a 2 m class telescope, also has a target list of around 100 planets, all with $m_J<10$, suggesting that our estimate is within a realistic and competitive range. Table \ref{tab:inst_req} summarizes the instrument requirements for VIPER discussed in this section.

\begin{table}[H]
\centering
\begin{tabular}{ | m{0.25\textwidth} | m{0.5\textwidth}| } 
  \hline
  Parameter & Requirement \\ 
  \hline
  \hline
  Resolving Power & 300,000 \\
  \hline
  Bandpass & $\geq3.54$ nm about $1083$ nm for exoplanets,\newline $\geq 3.82$ nm about $1083$ nm for stars \\
  \hline
  Efficiency & $\geq30\%$, with limiting $J$-band apparent magnitude of 10\\
  \hline
\end{tabular}
\caption{Instrument requirements for VIPER.}\label{tab:inst_req}
\end{table}


\subsection{Limitations of Grating-Based Spectrographs}

The instrument requirements outlined in Table \ref{tab:inst_req} are difficult to achieve with conventional astronomical spectrographs that use a diffraction grating as the primary disperser. Grating-based astronomical spectrographs suffer from a fundamental scaling relation which limits the maximum resolving power. If the grating is operated in the Littrow configuration, where the diffraction efficiency is highest, then maximum possible resolving power is:
\begin{equation}\label{eq:R_grat_spec}
    \mathcal{R}_\mathrm{max} = \frac{2D_\mathrm{coll}\tan{\theta_B}}{D_T \phi_T}
\end{equation}
where $D_T$ is the diameter of the telescope, $D_\mathrm{coll}$ is the diameter of the collimator, $\theta_B$ is the grating blaze angle, and $\phi_T$ is the angular width of the slit projected up to the sky. Since the resolving power scales with spectrograph size ($D_\mathrm{coll}$) and is inversely proportional to the telescope diameter, it is expensive and challenging to develop high-resolution grating-based spectrographs for large-aperture telescopes. An $\mathcal{R}$ = 300,000 grating-based spectrograph for the 6.5~m MMT would be impractical. In addition, conventional high-resolution astronomical spectrographs based on echelle gratings typically have throughputs of only 5--10\%, which is incompatible with our efficiency requirement. Since we do not require a broad bandpass, a VIPA would be an excellent choice as a primary disperser to satisfy our instrument requirements.
\section{VIPA Spectrograph Overview}\label{sec:VIPA_spec_overview}

\begin{figure}[H]
    \centering
    \includegraphics[width=0.9\textwidth]{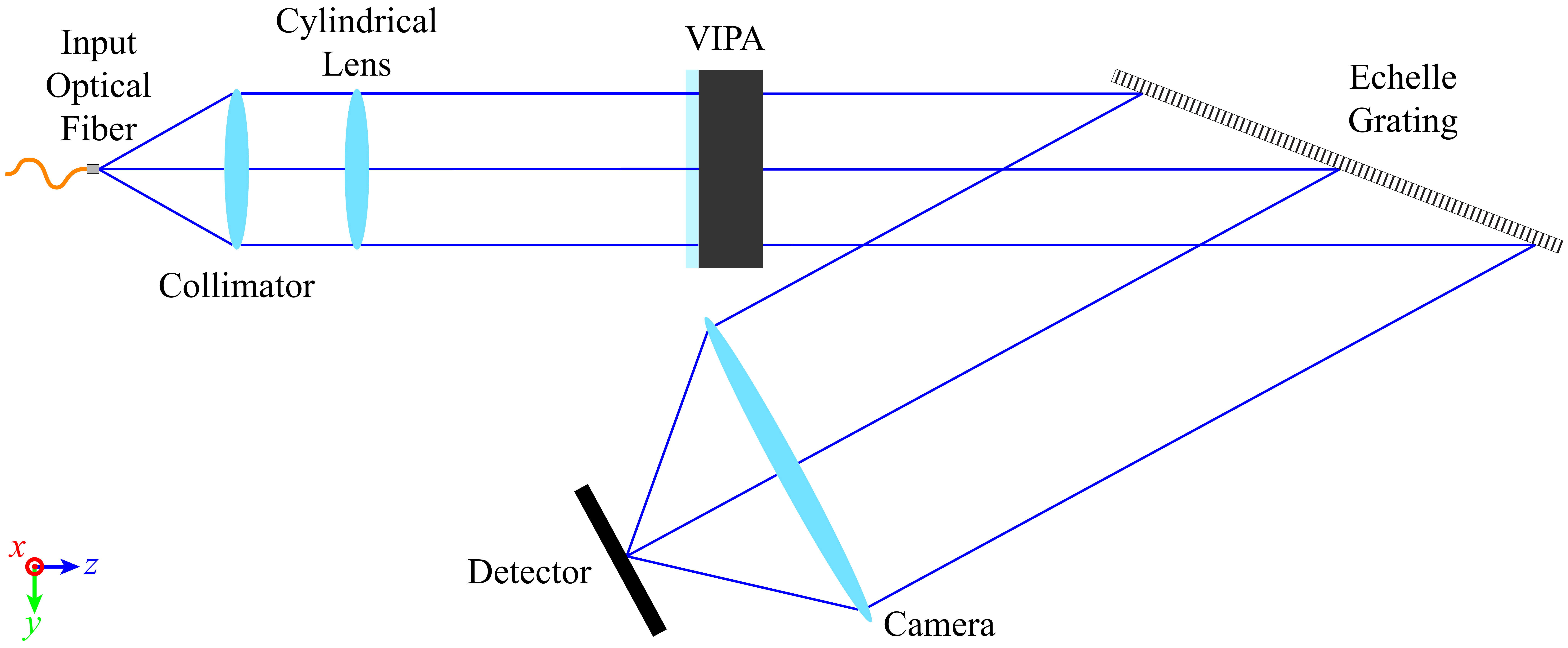}
    \caption{Illustration of a cross-dispersed VIPA spectrograph. Light from the input optical fiber is collimated by a collimator and is then line-focused by a cylindrical lens onto the back surface of the VIPA. The VIPA disperses light in the $x$-direction. An echelle grating then cross disperses the beam. Finally, a camera focuses the spectrum onto a detector. Unlike some designs in the literature, we do not employ an additional slit in our design.}
    \label{fig:VIPA_spec_top}
\end{figure}

\begin{figure}[H]
    \centering
    \includegraphics[width=0.9\textwidth]{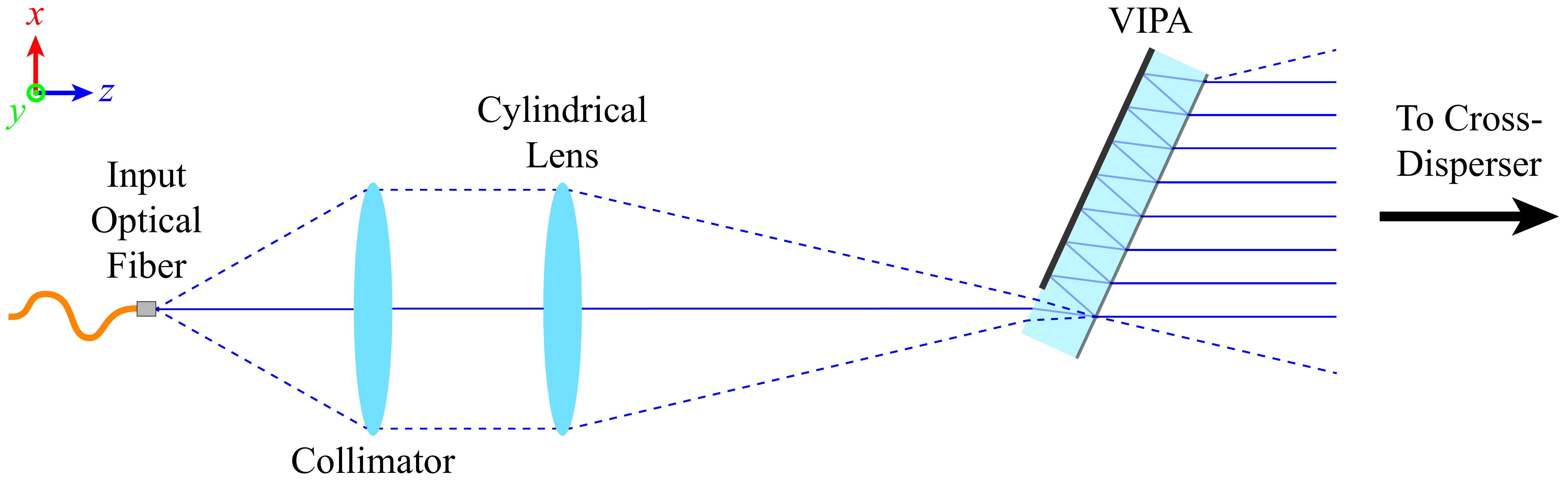}
    \caption{Side view of the cross-dispersed VIPA spectrograph shown in Figure \ref{fig:VIPA_spec_top}, from the input fiber to the VIPA.}
    \label{fig:VIPA_spec_side}
\end{figure}

\begin{figure}[H]
    \centering
    \includegraphics[width=0.9\textwidth]{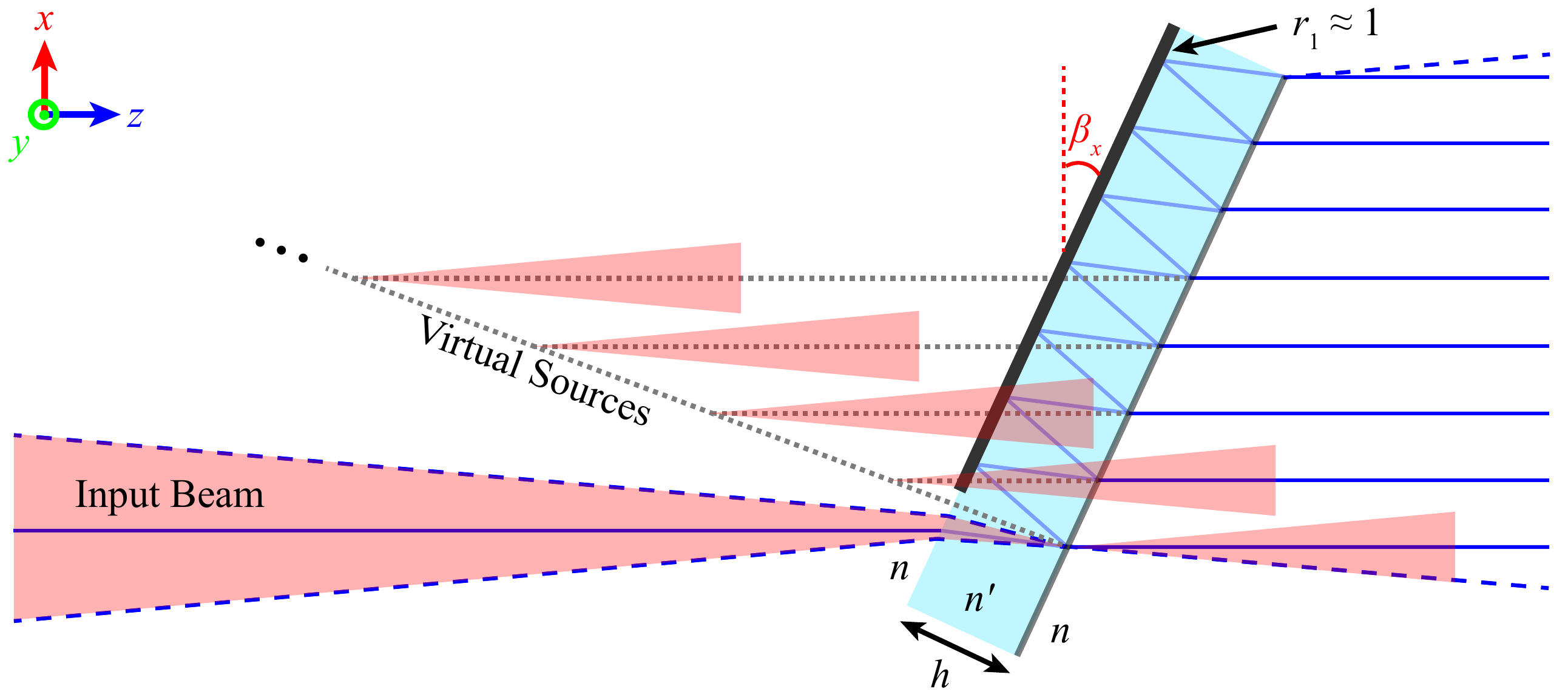}
    \caption{Illustration of a Virtually Imaged Phased Array (VIPA). Multiple reflections between the two internally reflective surfaces create a phased array of virtual sources that interfere to achieve high spectral resolution.}
    \label{fig:VIPA_ill}
\end{figure}

In this section, we provide a brief overview of how a cross-dispersed VIPA spectrograph functions. Figures \ref{fig:VIPA_spec_top} and \ref{fig:VIPA_spec_side} show the basic design of a cross-dispersed VIPA spectrograph. The coordinate conventions used in these figures will be used for the remainder of this paper. A multimode optical fiber is used as the input to the spectrograph, and light from the fiber is collimated by a collimator. The collimated beam is then line-focused by a cylindrical lens onto the back (exit) surface of the VIPA. The VIPA, illustrated in Figure \ref{fig:VIPA_ill}, is like a Fabry-Perot etalon, and disperses light in the $x$-direction. The beam leaving the VIPA is expanding in the $xz$ plane, but is collimated in the $yz$ plane. This beam is then cross-dispersed by an echelle grating, which disperses light in the $y$-direction. A set of camera optics then focuses the spectrum onto a detector, which can be rotated to better align with the VIPA diffraction orders.

\section{Virtually Imaged Phased Array (VIPA)}\label{sec:VIPA}

In this section, we discuss how a VIPA functions from the perspective of a Fabry-Perot etalon and we derive some equations which will be used in the spectrograph design process. Figure \ref{fig:VIPA_ill} shows an illustration of a VIPA, which consists of a dielectric plate with two surfaces that are partially reflective internally. A VIPA is basically a Fabry-Perot etalon where the input surface has an internal reflectivity near 100\%, with a small non-reflecting window allowing for light to enter. Multiple internal reflections between the two surfaces of the etalon create a phased array of virtual sources that interfere to achieve high spectral resolution.


\subsection{VIPA as a Fabry-Perot Etalon}

\begin{figure}[h]
    \centering
    \includegraphics[width=0.65\textwidth]{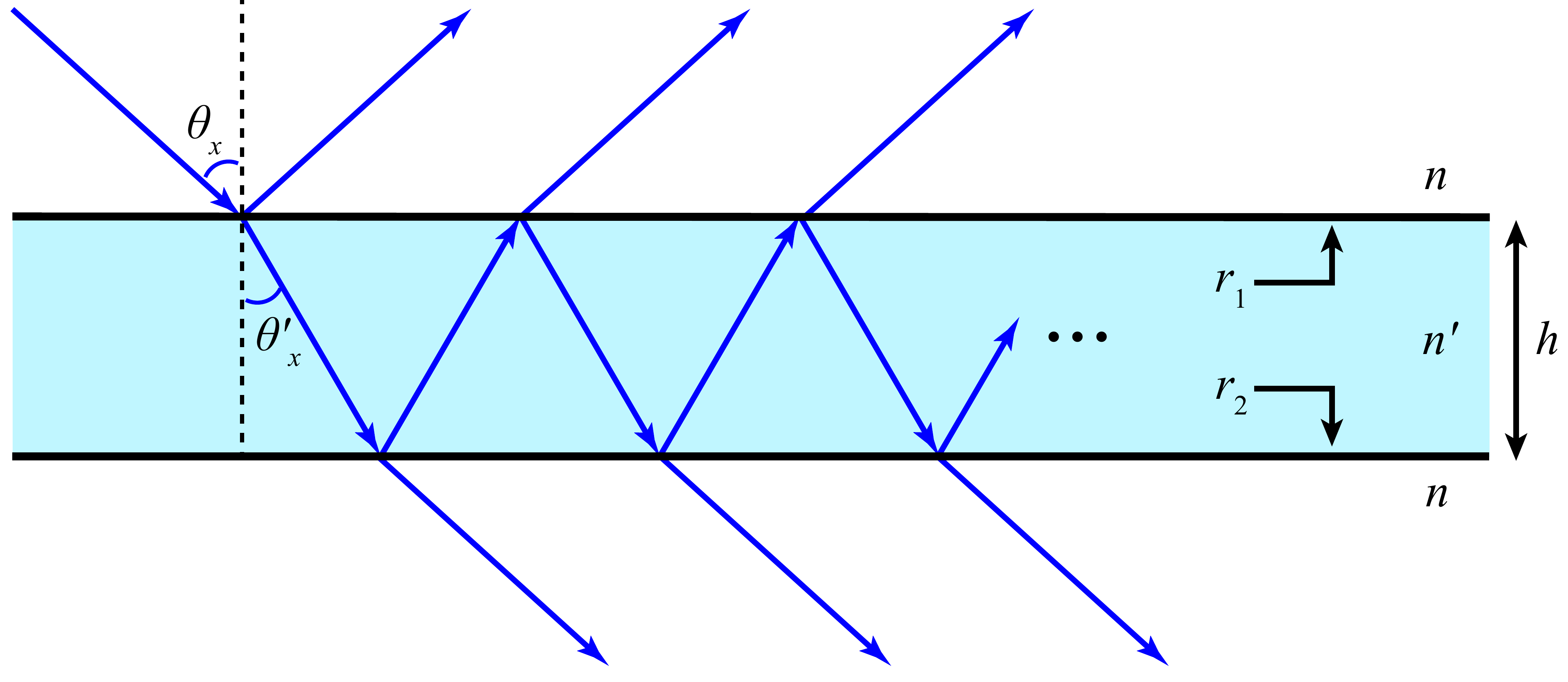}
    \caption{Illustration of a plane-parallel plate with two surfaces that are partially reflective internally.}
    \label{fig:plane_parallel_plate}
\end{figure}

A Fabry-Perot etalon can be thought of as a plane-parallel plate with partially reflective surfaces. Consider a plane-parallel plate, illustrated in Figure \ref{fig:plane_parallel_plate}, with thickness $h$ and refractive index $n'$, in a surrounding medium with refractive index $n$. Let $r_1$ and $r_2$ be the internal reflection coefficients of the front and back surfaces of the plate respectively. Let $t_2$ be the internal transmission coefficient of the back surface of the plate, where $|r_2|^2+|t_2|^2=1$. Suppose that a plane wave $E_\mathrm{in}=e^{-i(k_x x +k_y y)}$ impinges upon the plate, where $k_x=k\sin{\theta_x}$ and $k_y=k\sin{\theta_y}$ are the wavevector components and $k=2\pi n/\lambda$ is the wavenumber in the surrounding medium. If the plane wave makes one round trip inside the plate, then the optical phase accumulated is:
\begin{equation}
    \delta_\mathrm{FP} = 2hk_z' = 2h\sqrt{k'^2 - k_x'^2 - k_y'^2}
\end{equation}
where $k_x'=k'\sin{\theta_x'}$ and $k_y'=k'\sin{\theta_y'}$ are the wavevector components inside the etalon and \mbox{$k'=2\pi n'/\lambda$} is the wavenumber inside the etalon. Continuity of the transverse wavevector components across the interface (i.e., Snell's law) implies $k_x=k_x'$ and $k_y=k_y'$, and hence:
\begin{equation}\label{eq:delta_FP_theta}
    \delta_\mathrm{FP} = 2h\sqrt{k'^2 - k_x^2 - k_y^2} = 2h\sqrt{k'^2 - k^2\sin^2{\theta_x} - k^2 \sin^2{\theta_y}}
\end{equation}
The output from the plate can be obtained by taking a sum of plane waves exiting the plate\cite{BornWolf,YarivYeh}, each attenuated by an additional multiplicative factor of $r_1 r_2$ due to internal reflection off the front and back surfaces: 
\begin{align*}
    E_\mathrm{out} &= |E_\mathrm{in}| \left[ t_2 e^{-i\delta_\mathrm{FP}/2} + t_2 (r_1 r_2) e^{-3i\delta_\mathrm{FP}/2} + t_2 (r_1 r_2)^2 e^{-5i\delta_\mathrm{FP}/2} + \cdots\right] \nonumber\\
    &= |E_\mathrm{in}| t_2 e^{-i\delta_\mathrm{FP}/2} \left[1 + (r_1 r_2) e^{-i\delta_\mathrm{FP}} + (r_1 r_2)^2 e^{-2i\delta_\mathrm{FP}} + \cdots\right] \nonumber\\
    &= |E_\mathrm{in}| t_2 e^{-i\delta_\mathrm{FP}/2} \sum_{l=0}^\infty \left[ r_1 r_2 e^{-i\delta_\mathrm{FP}} \right]^l \nonumber\\
    &= |E_\mathrm{in}| \frac{t_2 e^{-i\delta_\mathrm{FP}/2}}{1-r_1 r_2 e^{-i\delta_\mathrm{FP}}}
\end{align*}
In the last step, we simplify the expression as a geometric series. Hence, the transfer function for a Fabry-Perot etalon is:
\begin{equation}\label{eq:H_FP}
    H_\mathrm{FP}(\theta_x,\theta_y)=\frac{t_2 e^{-i\delta_\mathrm{FP}/2}}{1-r_1 r_2 e^{-i\delta_\mathrm{FP}}}
\end{equation}
Note that this analysis ignores the external reflection coefficient at the front surface, since for a VIPA, the incident light enters through a non-reflecting window. The output intensity is the modulus squared of $H_\mathrm{FP}$:
\begin{equation}\label{eq:I_FP}
    I_\mathrm{FP} = |H_\mathrm{FP}|^2 = \frac{|t_2|^2}{(1-r_1r_2)^2 +4r_1r_2\sin^2{(\delta_\mathrm{FP}/2)}} = \frac{1-|r_2|^2}{(1-r_1r_2)^2} \frac{1}{1+F\sin^2{(\delta_\mathrm{FP}/2)}}
\end{equation}
where $F$ is a factor related to the etalon's finesse $\mathcal{F}=\pi\sqrt{F}/2$ and defined as\cite{BornWolf}:
\begin{equation}
    F\equiv \frac{4r_1r_2}{(1-r_1r_2)^2}
\end{equation}
Intensity peaks (resonances) in Eq.~(\ref{eq:I_FP}) occur wherever $\delta_\mathrm{FP}$ is an integer multiple of $2\pi$. If $\theta_y=0$, then the angles corresponding to the intensity peaks can be obtained by rearranging Eq.~(\ref{eq:delta_FP_theta}):
\begin{equation}\label{eq:theta_x_peak}
    \theta_{x,\mathrm{peak}} = \arcsin{\left(\frac{1}{k}\sqrt{k'^2 - \left(\frac{m\pi}{h}\right)^2} \right)} \:\:\: ,\:m \in \mathbb{Z}
\end{equation}

\begin{figure}[h]
    \centering
    \includegraphics[width=0.7\textwidth]{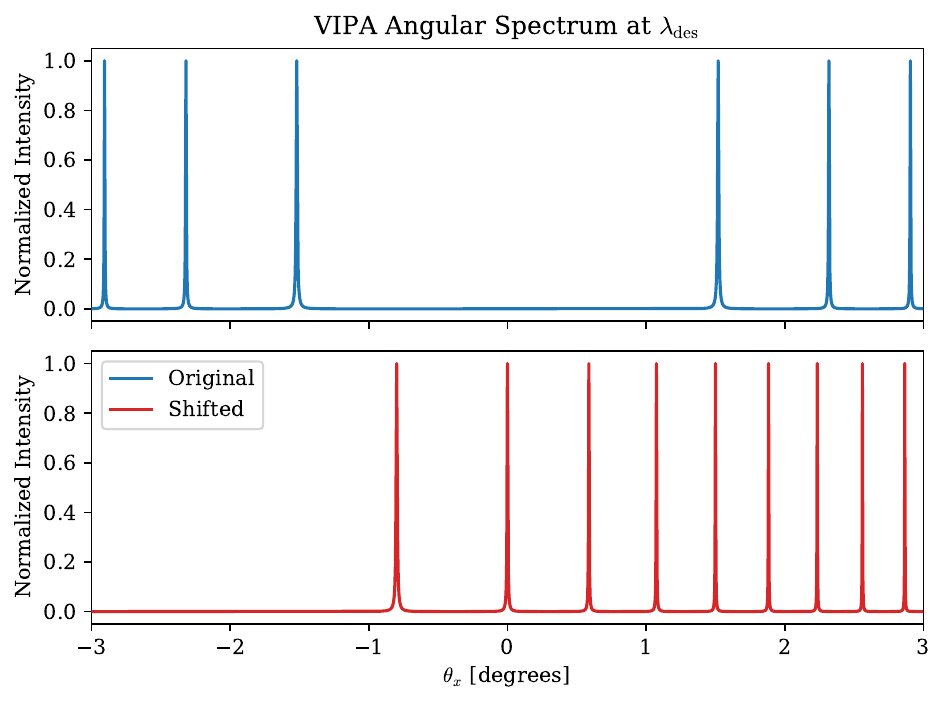}
    \caption{Angular spectrum of VIPA. The quantity being plotted is the normalized intensity, Eq.~(\ref{eq:I_FP}), with $\beta_x=0$ (blue) and $\beta_x=2.319^\circ$ (red), $\lambda_\mathrm{des}=1083$ nm, $n'=1.4494$, $n=1.0003$, and $h=1.68$ mm. The peaks correspond to different VIPA orders.}
    \label{fig:VIPA_ang_spec}
\end{figure}

A VIPA is a Fabry-Perot etalon where $r_1\approx 1$, and where the etalon is tilted at an angle $\beta_x$ (see Figure \ref{fig:VIPA_ill}) with respect to its normal so that its output angular spectrum is shifted away from the central intensity peak. This tilt is a modification of $\delta_\mathrm{FP}$ with the substitution $\theta_x\rightarrow\theta_x+\beta_x$:
\begin{equation}\label{eq:delta_FP}
    \delta_\mathrm{FP} = 2h\sqrt{k'^2 - k^2\sin^2{(\theta_x+\beta_x)} - k^2 \sin^2{\theta_y}}
\end{equation}
We select $\beta_x$ so that one of the side intensity peaks corresponding to some design wavelength $\lambda_\mathrm{des}$ is centered at $\theta_x=0$. In other words, we take $\beta_x$ to be one of the $\theta_{x,\mathrm{peak}}$ angles obtained from Eq.~(\ref{eq:theta_x_peak}), selecting a particular integer order $m$. This shift is illustrated in Figure \ref{fig:VIPA_ang_spec}.

\subsection{VIPA Resolving Power and Thickness}

Following the procedure in Ref.~\citenum{BornWolf}, the resolving power of the etalon can be found from the intensity function Eq.~(\ref{eq:I_FP}):
\begin{equation}\label{eq:R_FP}
    \mathcal{R}_\mathrm{FP} = \frac{\lambda}{\Delta\lambda} = \frac{\delta_\mathrm{FP}\sqrt{F}}{4.15} = \frac{2}{4.15} \sqrt{\frac{4r_1r_2}{(1-r_1r_2)^2}} h \sqrt{k'^2 - k^2\sin^2{(\theta_x+\beta_x)} - k^2 \sin^2{\theta_y}}
\end{equation}
where $\Delta \lambda$ is the wavelength resolution element. Given a desired design resolving power $\mathcal{R}_\mathrm{des}$, a minimum bound on the thickness can be found by rearranging Eq.~(\ref{eq:R_FP}) and taking $k_x=k_y=0$, since the etalon is usually operated in a regime where $k_x\ll k'$ and $k_y\ll k'$:
\begin{equation}\label{eq:h_min}
    h \geq h_\mathrm{min}= \frac{4.15\mathcal{R}_\mathrm{des}}{2\sqrt{F}k'}
\end{equation}


\subsection{VIPA Angular Dispersion and Camera Focal Length}

The angular dispersion of the etalon can be found by implicit differentiation of Eq.~(\ref{eq:delta_FP}) with respect to $\lambda$. Taking $\theta_y=0$ and $\frac{d\theta_y}{d\lambda}=0$, and rearranging, we obtain:
\begin{equation}
    \frac{d\theta_x}{d\lambda} = - \frac{2\left(\frac{1}{\lambda}\left[n'^2-n^2\sin^2{(\theta_x+\beta_x)}\right] - n'\frac{dn'}{d\lambda} -n\frac{dn}{d\lambda} \sin^2{(\theta_x+\beta_x)} \right)}{n^2 \sin{(2[\theta_x+\beta_x])}}
\end{equation}
If we assume that there is no refractive index dispersion, that is, $\frac{dn}{d\lambda}=\frac{dn'}{d\lambda}=0$, then we obtain the result as in Ref.~\citenum{Hu2015}:
\begin{equation}\label{eq:ang_disp_FP}
    \frac{d\theta_x}{d\lambda} = - \frac{2\left(n'^2-n^2\sin^2{(\theta_x+\beta_x)} \right)}{\lambda n^2 \sin{(2[\theta_x+\beta_x])}}
\end{equation}

The spectrum is focused by some camera optics of focal length $f_\mathrm{cam}$ onto a detector. The linear dispersion on the detector is:
\begin{equation}\label{eq:dxdlam}
    \frac{dx}{d\lambda} \approx \frac{\Delta x}{\Delta \lambda} = \frac{\Delta x}{\lambda/\mathcal{R}_\mathrm{des}}
\end{equation}
where $\Delta x$ is the spatial element on the detector that corresponds to $\Delta \lambda$. $\Delta x$ is taken to be the width of at least two pixels due to Nyquist sampling. The linear dispersion, for small angles, can also be written as:
\begin{equation}\label{eq:dxdlam_fcam}
    \left|\frac{dx}{d\lambda}\right| \approx f_\mathrm{cam} \left|\frac{d\theta_x}{d\lambda}\right|
\end{equation}
Hence, the focal length of the camera is:
\begin{equation}\label{eq:f_cam}
    f_\mathrm{cam} = \frac{\mathcal{R}_\mathrm{des} \Delta x}{\lambda} \left|\frac{d\theta_x}{d\lambda}\right|^{-1} =  \frac{\mathcal{R}_\mathrm{des} \Delta x\,n^2 \sin{(2[\theta_x+\beta_x])}}{2\left(n'^2-n^2\sin^2{(\theta_x+\beta_x)} \right)}
\end{equation}
where Eq.~(\ref{eq:ang_disp_FP}) was substituted. To find $f_\mathrm{cam}$, we evaluate Eq.~(\ref{eq:f_cam}) at $\theta_x=0$.


\subsection{VIPA Free Spectral Range and Constraints on Cross Dispersion}

Equation~(\ref{eq:delta_FP}) can be rewritten into another useful form:
\begin{equation}\label{eq:delta_FP_n}
    \delta_\mathrm{FP} = 2hk_0 \sqrt{n'^2-n^2\sin^2{(\theta_x+\beta_x)}-n^2\sin^2{\theta_y}}
\end{equation}
where $k_0=2\pi/\lambda=2\pi\nu/c$ is the vacuum wavenumber. Suppose that $\nu_m$ is the frequency corresponding to an intensity peak at integer order $m$. Hence, if we substitute $\delta_\mathrm{FP}=2\pi m$ and $k_0=2\pi\nu_m/c$ into Eq.~(\ref{eq:delta_FP_n}) and solve for $\nu_m$, we obtain:
\begin{equation}
    \nu_m = \frac{mc}{2h}\left[n'^2-n^2\sin^2{(\theta_x+\beta_x)}-n^2\sin^2{\theta_y}\right]^{-1/2}
\end{equation}
The free spectral range (FSR) of the etalon in frequency units is therefore:
\begin{equation}
    (\Delta\nu)_\mathrm{FSR} = \nu_{m+1}-\nu_m = \frac{c}{2h}\left[n'^2-n^2\sin^2{(\theta_x+\beta_x)}-n^2\sin^2{\theta_y}\right]^{-1/2}
\end{equation}
Noting that $\lambda \nu = c \Rightarrow |d\lambda|=\frac{\lambda^2}{c}|d\nu|$, the FSR of the etalon in wavelength units is then:
\begin{equation}
    (\Delta \lambda)_\mathrm{FSR} = \frac{\lambda^2}{2h\sqrt{n'^2-n^2\sin^2{(\theta_x+\beta_x)}-n^2\sin^2{\theta_y}}}
\end{equation}

The etalon's FSR determines the amount of cross dispersion required. The required resolving power in the cross dispersion direction is:
\begin{equation}\label{eq:R_CD}
    \mathcal{R}_\mathrm{CD} = \frac{\lambda}{(\Delta \lambda)_\mathrm{FSR}}
\end{equation}
Let $\Delta y_\mathrm{order}$ be the spatial separation between adjacent cross dispersed orders on the detector. Then, analogous to the $x$ direction (cf. Eqs. (\ref{eq:dxdlam}) and (\ref{eq:dxdlam_fcam})), we have, in the cross dispersion ($y$) direction:
\begin{align*}
    \frac{dy}{d\lambda} &\approx \frac{\Delta y_\mathrm{order}}{(\Delta \lambda)_\mathrm{FSR}} = \frac{\Delta y_\mathrm{order}}{\lambda/\mathcal{R}_\mathrm{CD}} \\
    \left|\frac{dy}{d\lambda}\right| &\approx f_\mathrm{cam} \left|\frac{d\theta_y}{d\lambda}\right|
\end{align*}
Combining these two equations, the angular dispersion required in the cross dispersion direction is:
\begin{equation}
    \left|\frac{d\theta_y}{d\lambda}\right| \approx \frac{\mathcal{R}_\mathrm{CD} \Delta y_\mathrm{order}}{\lambda f_\mathrm{cam}}
\end{equation}
This informs the choice of the cross disperser.


\subsection{VIPA Peak Transmission}

To calculate the peak transmission of a VIPA, one might naively think to take the peak transmission as the $\frac{1-|r_2|^2}{(1-r_1r_2)^2}$ term in $|H_\mathrm{FP}|^2$ (Eq.~(\ref{eq:I_FP})). However, as $|r_1|^2\rightarrow1$ or $|r_2|^2\rightarrow1$, $\frac{1-|r_2|^2}{(1-r_1r_2)^2}$ can be above 1, which results in an unphysical apparent paradox. The cause of this paradox is because our analysis ignored the external reflection coefficient at the front surface, and so energy is not conserved. $|H_\mathrm{FP}|^2$ basically assumes that the input is a source internally launched from the VIPA. Ref.~\citenum{Weiner2012} resolved this paradox by taking an average of the transmission over one round-trip phase shift $\delta_\mathrm{FP}$, deriving the peak transmission as:
\begin{equation}\label{eq:VIPA_transmission}
    T_\mathrm{VIPA} = \frac{1-|r_2|^2}{1-|r_1|^2|r_2|^2}
\end{equation}
An alternate method, which also arrives at the same result, is to consider the attenuation of intensity by the reflectivities $|r_1|^2$ and $|r_2|^2$ in each round trip; then the peak transmission can be found by evaluating the geometric series $T_\mathrm{VIPA} = (1-|r_2|^2) \sum_{l=0}^\infty (|r_1|^2|r_2|^2)^l = \frac{1-|r_2|^2}{1-|r_1|^2|r_2|^2}$.
\section{VIPA Injection}\label{sec:VIPA_inject}

In this section, we model the injection of light into the VIPA when a multimode optical fiber is used as the input to the spectrograph. In order to use a Fabry-Perot etalon for spectral dispersion, the incident light must have a non-negligible angular spread along the dispersion direction. The transmission of an etalon is angle-dependent (see Eq.~(\ref{eq:I_FP}) and note the angular dependence in $\delta_\mathrm{FP}$), and so the angular spectrum of the input beam modulates the output angular spectrum of the VIPA. The input beam cannot be perfectly collimated along the $x$-direction, because this would result in a delta-function-like angular spectrum and therefore no angular dispersion. Hence, a cylindrical lens is used to inject light into the VIPA. The angular spectrum of the input beam is crucial to the VIPA's operation as a spectral disperser and controls the VIPA's diffraction efficiency.

\begin{figure}[h]
    \centering
    \includegraphics[width=0.7\textwidth]{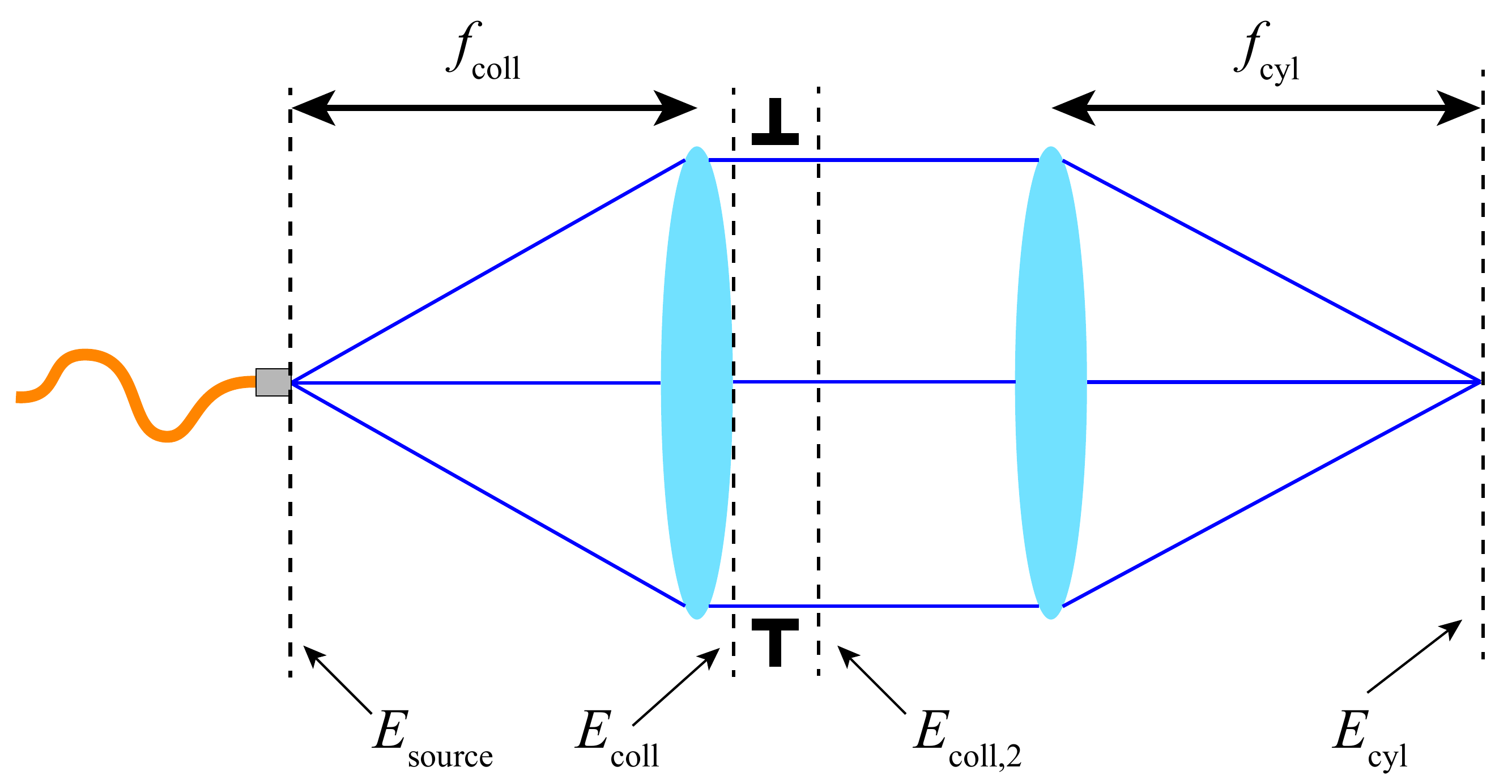}
    \caption{Illustration of injection optics for VIPA.}
    \label{fig:VIPA_inject2lenses}
\end{figure}

A number of papers in the literature often assume that a Gaussian beam is injected into the VIPA\cite{Xiao2004,Xiao2005,Hu2015}. This description works well when a single mode optical fiber is fed into the system, as it has been used in this situation in the literature. However, a Gaussian beam is not a suitable description when the input is a multimode optical fiber, because the output of a multimode optical fiber is not a Gaussian beam. Instead, here we discuss a model where the beam injected into the VIPA is from a relay system which reimages a circular input fiber of diameter $s$ onto the back surface of the VIPA. This relay system is illustrated in Figure \ref{fig:VIPA_inject2lenses}, consisting of a collimator with focal length $f_\mathrm{coll}$ and a cylindrical lens with focal length $f_\mathrm{cyl}$. 

A simple crude model for the output electric field of a multimode optical fiber of diameter $s$ is the circle function:
\begin{equation}\label{eq:fiber_circle}
    E_\mathrm{source}(x,y) = 
    \begin{cases} 
      1 & \mathrm{if}\:\:\:x^2+y^2\leq(s/2)^2 \\
      0 & \mathrm{otherwise}
   \end{cases}
\end{equation}
The corresponding angular spectrum $\tilde{E}_\mathrm{cyl}(\theta_x,\theta_y)$ at the focal plane of the cylindrical lens can be derived by principles of Fourier optics. A full derivation is provided in Appendix \ref{sec:app_VIPA_input}. With some approximations, we obtain:
\begin{equation}\label{eq:cyl_ang_spec}
    \tilde{E}_\mathrm{cyl}(\theta_x,\theta_y) = 2k C_1(\theta_y)\, \sinc{\left(C_1(\theta_y)\, k f_\mathrm{cyl} \tan{\theta_x}\right)}
\end{equation}
where $\sinc{x}\equiv \frac{\sin{x}}{x}$ and where we define:
\begin{equation}\label{eq:C1}
    C_1(\theta_y) \equiv \arctan{\left(\frac{1}{f_\mathrm{coll}} \sqrt{\left(\frac{s}{2}\right)^2-(f_\mathrm{cyl}\tan{\theta_y})^2}\right)}
\end{equation}

\begin{figure}[h]
    \centering
    \begin{subfigure}{0.49\textwidth}
        \centering
        \includegraphics[width=\textwidth]{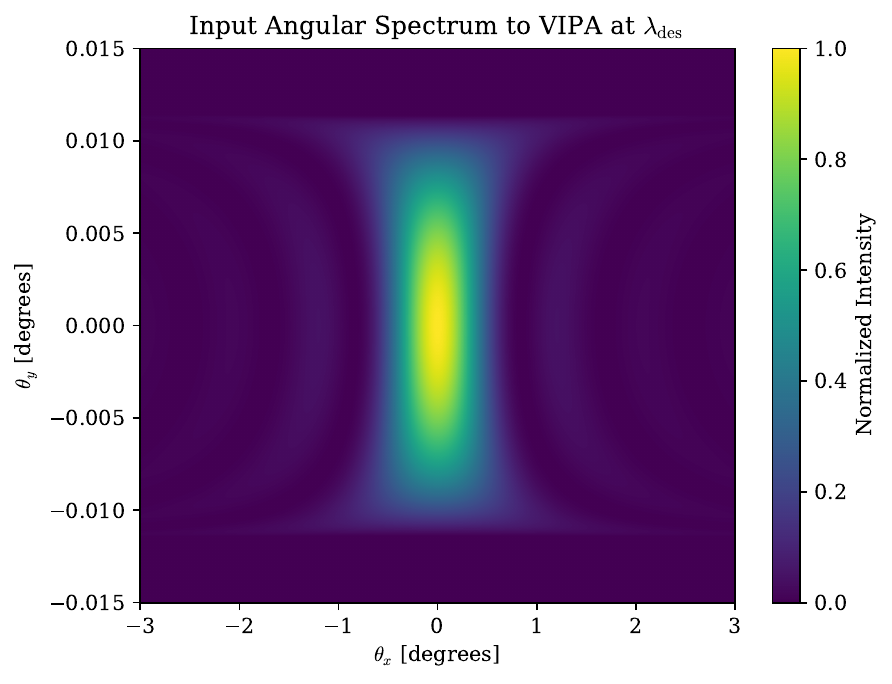}
        \caption{Input angular spectrum to VIPA}
        \label{fig:VIPA_in_ang_spec}
    \end{subfigure}
    \begin{subfigure}{0.49\textwidth}
        \centering
        \includegraphics[width=\textwidth]{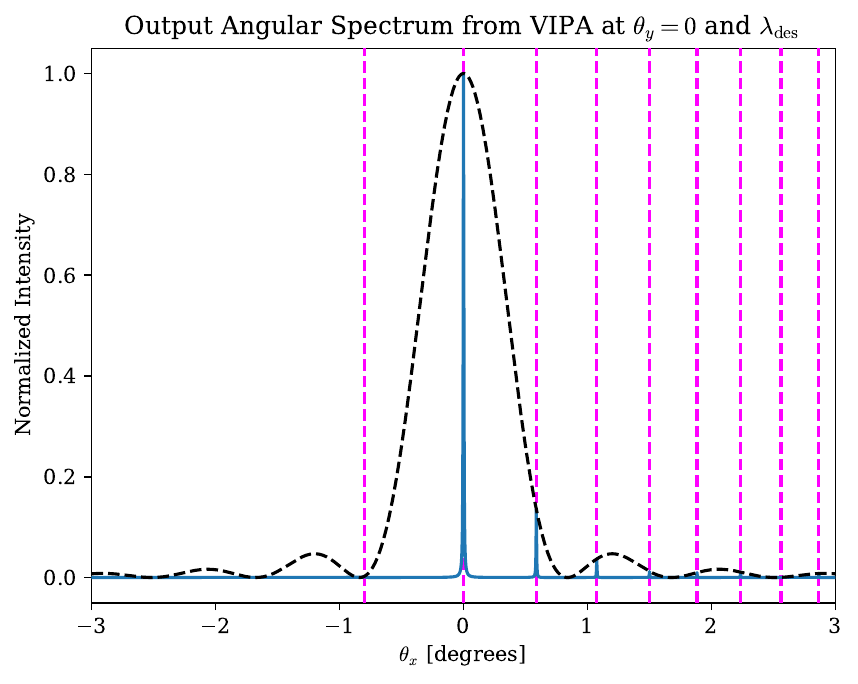}
        \caption{VIPA output angular spectrum}
        \label{fig:VIPA_out_ang_spec}
    \end{subfigure}
    \caption{The left figure shows the input angular spectrum to the VIPA; this a plot of Eq.~(\ref{eq:cyl_ang_spec}) with $f_\mathrm{coll}=126$ mm, $f_\mathrm{cyl}=186.333$ mm, $s=50$~\textmu m, $k=2\pi n/\lambda_\mathrm{des}$. The right figure shows the output angular spectrum from the VIPA at $\theta_y=0$ and $\lambda_\mathrm{des}$. The output angular spectrum is modulated by the black dashed curve, which is Eq.~(\ref{eq:cyl_ang_spec}). The magenta dashed lines denote the intensity peaks, which are the same as those in the lower subplot of Figure \ref{fig:VIPA_ang_spec}.}
\end{figure}

$\tilde{E}_\mathrm{cyl}(\theta_x,\theta_y)$ modulates the output angular spectrum of the VIPA and is referred to as the ``diffraction envelope'' in some papers\cite{Zhu2023}. If we want the VIPA orders to have high diffraction efficiency, then we want light to be concentrated a few orders of the VIPA near $\theta_x=0$. Let $\theta_\mathrm{FAR}$ be the free angular range (FAR\cite{Metz2013}) of the etalon, which is the angular separation between two adjacent intensity peaks at a fixed wavelength. We want to set the width of intensity function $|\tilde{E}_\mathrm{cyl}(\theta_x,\theta_y)|^2$ such that it envelops a certain number of $\theta_\mathrm{FSR}$'s. Ref.~\citenum{Bourdarot2018} takes the width such that the modulus squared of the diffraction envelope intersects the VIPA intensity peak adjacent to the center VIPA intensity peak at $1/e^2$ of the maximum intensity. Numerically, $|\sinc{x}|^2$ has value $1/e^2$ at $x\approx\pm2.199$. Hence, taking $\theta_y=0$ and using the small angle approximation in Eqs.~(\ref{eq:C1}) and (\ref{eq:cyl_ang_spec}), we obtain a constraint for the relationship between $f_\mathrm{cyl}$ and $f_\mathrm{coll}$:
\begin{equation}\label{eq:f_cyl_f_coll}
    \frac{f_\mathrm{cyl}}{f_\mathrm{coll}} \approx \frac{2(2.199)}{\theta_\mathrm{FAR}ks}
\end{equation}
Figures \ref{fig:VIPA_in_ang_spec} and \ref{fig:VIPA_out_ang_spec} show the VIPA input angular spectrum and output angular spectrum. In Figure \ref{fig:VIPA_out_ang_spec}, we can see that with an appropriate selection of $f_\mathrm{cyl}/f_\mathrm{coll}$, the VIPA orders adjacent to the center order are suppressed, resulting in a high diffraction efficiency in the center order.

\section{Cross-Disperser}\label{sec:cross_disp}

\subsection{Blazed Echelle Diffraction Grating}

In this section, we discuss the operation of the cross-disperser, which is an echelle grating. Cross dispersion is along the $y$-direction. Consider a blazed diffraction grating with a groove separation of $d$ and a grating normal parallel to the $z$-axis. Suppose that the grating grooves are initially aligned parallel to the $x$-axis, but that the grating may be rotated by an angle $\gamma$ about the $z$-axis, effectively changing the alignment of the grooves. If an incident electric field with angular spectrum $\tilde{E}_\mathrm{G,in}(\theta_x,\theta_y)$ illuminates the grating, then the output angular spectrum of the grating is:
\begin{equation}\label{eq:Etilde_G}
    \tilde{E}_\mathrm{G}(\theta_x,\theta_y) \propto \sum_{m\in\mathbb{Z}} \frac{b \sinc{\left(\delta_{\mathrm{g,ce}}(\theta_x,\xi(\theta_x,\theta_y,m),\theta_y)\right)} \,|\cos{\theta_y}|}{|\cos{(\xi(\theta_x,\theta_y,m))}|} \,\tilde{E}_\mathrm{G,in}(\theta_x,\xi(\theta_x,\theta_y,m))
\end{equation}
A full derivation of Eq.~(\ref{eq:Etilde_G}) can be found in Appendix \ref{sec:app_grating}. In Eq.~(\ref{eq:Etilde_G}), $m$ is the diffraction order of the grating, $b$ is the effective groove separation given by Eq.~(\ref{eq:b}), $\delta_\mathrm{g,ce}$ is given by Eq.~(\ref{eq:deltagce_mod}), and $\xi$ is a function defined as:
\begin{equation}\label{eq:xi_fcn}
    \xi(\theta_x,\theta_y,m) \equiv \arcsin{\left[\frac{2\pi m}{kd\cos{(\gamma-\theta_x)}}-\sin{\theta_y}\right]}
\end{equation}
Eq.~(\ref{eq:Etilde_G}) takes into account the effect of conical diffraction\cite{Yang2016}, which is important to model because this is the underlying cause of grating smile distortion when there are off-axis field angles\cite{Leung2022}, which is the case here because light is dispersed in the $x$-direction by the VIPA before illuminating the grating. See Appendix \ref{sec:app_grating} for a more detailed discussion.

While echelle gratings are usually operated in multiple orders, here we use it as a cross-disperser and only operate it in one order $m_\mathrm{cen}$, which is informed by the blaze angle $\theta_B$ and design wavelength $\lambda_\mathrm{des}$:
\begin{equation}\label{eq:m_cen}
    m_\mathrm{cen} = \mathrm{round}\left\{\frac{2d\sin{\theta_B}}{\lambda_\mathrm{des}}\right\}
\end{equation}
Then we can write Eq.~(\ref{eq:Etilde_G}) as:
\begin{equation}\label{eq:Etilde_G_onem}
    \tilde{E}_\mathrm{G}(\theta_x,\theta_y) \propto \frac{b \sinc{\left(\delta_{\mathrm{g,ce}}(\theta_x,\xi(\theta_x,\theta_y,m_\mathrm{cen}),\theta_y)\right)} \,|\cos{\theta_y}|}{|\cos{(\xi(\theta_x,\theta_y,m_\mathrm{cen}))}|} \,\tilde{E}_\mathrm{G,in}(\theta_x,\xi(\theta_x,\theta_y,m_\mathrm{cen}))
\end{equation}


\subsection{Collimator Focal Length}

The properties of the grating inform the choice of the collimator focal length. Suppose that the beam illuminating the grating makes an angle $\alpha_0$ with the grating normal. Then the diffraction angle corresponding to $\alpha_0$ is:
\begin{equation}\label{eq:beta_0}
    \beta_0\equiv\xi(0,\alpha_0,m_\mathrm{cen})
\end{equation}
Then the collimator focal length is:
\begin{equation}\label{eq:f_coll}
    f_\mathrm{coll} = \frac{f_\mathrm{cam} s}{\Delta y_\mathrm{res}} \left|\frac{\cos{\alpha_0}}{\cos{\beta_0}} \right|
\end{equation}
where $\frac{\cos{\alpha_0}}{\cos{\beta_0}}$ is the anamorphic magnification and $\Delta y_\mathrm{res}$ is the desired spatial width of one VIPA order on the detector. This is the size of the image of the fiber on the detector. $f_\mathrm{coll}$ is a quantity that should be maximized so that the size of the image of the fiber on the detector is small. However, in practice, $f_\mathrm{coll}$ cannot be arbitrarily large because the collimated beam size is limited by one of the optical elements downstream. Oftentimes, the aperture stop is the VIPA itself. Suppose that the input fiber has focal ratio $(f/\#)_\mathrm{fiber}$ and the VIPA has a diameter (clear aperture) of $D_\mathrm{VIPA}$. Then $f_\mathrm{coll}$ is:
\begin{equation}\label{eq:f_coll_limit}
    f_\mathrm{coll} = (f/\#)_\mathrm{fiber} D_\mathrm{VIPA}
\end{equation}


\subsection{Grating Diffraction Efficiency}

\begin{figure}[h]
    \centering
    \begin{subfigure}{0.49\textwidth}
        \centering
        \includegraphics[width=\textwidth]{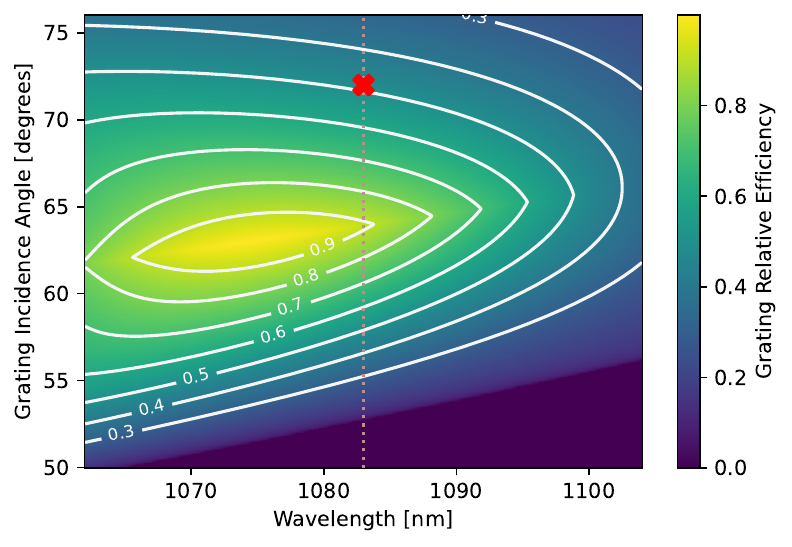}
        \caption{Grating relative efficiency}
        \label{fig:grating_eff}
    \end{subfigure}
    \begin{subfigure}{0.49\textwidth}
        \centering
        \includegraphics[width=\textwidth]{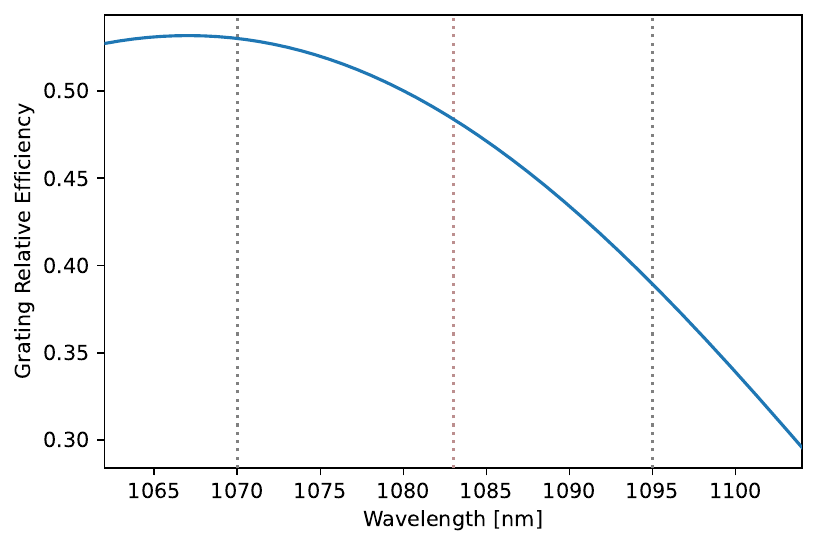}
        \caption{Grating relative efficiency at $\alpha=72^\circ$}
        \label{fig:grating_eff_slice}
    \end{subfigure}
    \caption{Relative efficiency of the echelle grating used in VIPER's design, computed using Eq.~(\ref{eq:grating_eff}). The grating parameters are $d=1/79$~mm and $\theta_B=63^\circ$, and we operate the grating in order $m_\mathrm{cen}=21$. The red ``X'' in the left figure marks the incidence angle we selected, which is $72^\circ$. This incidence angle does not result in the highest diffraction efficiency, but we had to pick this due to technical constraints, which are discussed in Section \ref{sec:spec_design}.}
    \label{fig:grating_eff_fig}
\end{figure}

The relative efficiency of a blazed diffraction grating in order $m$ is given by Bottema\cite{Bottema1981,Bottema1981SPIE} as:
\begin{equation}\label{eq:grating_eff}
    E_{r,m} = 
    \begin{cases} 
      \frac{\cos{\alpha}}{\cos{\beta_m}} \sinc^2{\left[\frac{1}{2}kd\cos{\alpha} \left(\sin{(\alpha-\theta_B)}+\sin{(\beta_m-\theta_B)}\right)\right]} & \mathrm{if}\:\:\: \alpha>\beta_m \\
      \frac{\cos{\beta_m}}{\cos{\alpha}} \sinc^2{\left[\frac{1}{2}kd\cos{\beta_m} \left(\sin{(\alpha-\theta_B)}+\sin{(\beta_m-\theta_B)}\right)\right]} & \mathrm{if}\:\:\: \alpha<\beta_m
   \end{cases}
\end{equation}
where $\alpha$ is the incidence angle and $\beta_m$ is the diffraction angle corresponding to $\alpha$ at order $m$. Figure \ref{fig:grating_eff_fig} shows the relative efficiency of the echelle grating used in VIPER's design, at diffraction order $m=21$.

\section{Spectrograph Model}\label{sec:spec_model}

Combining the results from the previous sections, an end-to-end analytic model of the cross-dispersed VIPA spectrograph shown in Figures \ref{fig:VIPA_spec_top} and \ref{fig:VIPA_spec_side} can be described by the following equations:
\begin{align}
    \tilde{E}_\mathrm{cyl}(\theta_x,\theta_y) &\propto 2k C_1(\theta_y)\, \sinc{\left(C_1(\theta_y)\, k f_\mathrm{cyl} \tan{\theta_x}\right)} \tag{\ref{eq:cyl_ang_spec}} \\
    H_\mathrm{FP}(\theta_x,\theta_y) &= \frac{t_2 e^{-i\delta_\mathrm{FP}/2}}{1-r_1 r_2 e^{-i\delta_\mathrm{FP}}} \tag{\ref{eq:H_FP}} \\
    \tilde{E}_\mathrm{FP,out}(\theta_x,\theta_y) &= \tilde{E}_\mathrm{cyl}(\theta_x,\theta_y) \, H_\mathrm{FP}(\theta_x,\theta_y) \label{eq:Etilde_FPout} \\
    \tilde{E}_\mathrm{G,in}(\theta_x,\theta_y) &= \tilde{E}_\mathrm{FP,out}(\theta_x,\theta_y-\alpha_0) \label{eq:Etilde_Gin} \\
    \tilde{E}_\mathrm{G}(\theta_x,\theta_y) &\propto \frac{b \sinc{\left(\delta_{\mathrm{g,ce}}(\theta_x,\xi(\theta_x,\theta_y,m_\mathrm{cen}),\theta_y)\right)} \,|\cos{\theta_y}|}{|\cos{(\xi(\theta_x,\theta_y,m_\mathrm{cen}))}|} \,\tilde{E}_\mathrm{G,in}(\theta_x,\xi(\theta_x,\theta_y,m_\mathrm{cen}))\tag{\ref{eq:Etilde_G_onem}} \\
    \tilde{E}_\mathrm{G,out}(\theta_x,\theta_y) &= \tilde{E}_\mathrm{G}(\theta_x,\theta_y+\beta_0) \label{eq:Etilde_M1} \\
    \tilde{E}_\mathrm{rot}(\theta_x,\theta_y) &= \tilde{E}_\mathrm{M1}(\theta_x\cos{(-\kappa})-\theta_y\sin{(-\kappa)}, \theta_x\sin{(-\kappa})+\theta_y\cos{(-\kappa)}) \label{eq:Etilde_rot} \\
    E_\mathrm{D}(x,y) &\propto \tilde{E}_\mathrm{rot}(\arctan{(x/f_\mathrm{cam})}, \arctan{(y/f_\mathrm{cam})}) \label{eq:E_D}
\end{align}
Equations that have appeared in the previous sections are referenced again. It is important to note that there is an implicit $k$ dependence in each equation. Equation~(\ref{eq:Etilde_FPout}) is the output angular spectrum of the VIPA modulated by the angular spectrum of the injection beam. Equation~(\ref{eq:Etilde_Gin}) rotates the optical axis to the reference frame of the grating. Equation~(\ref{eq:Etilde_M1}) rotates the optical axis so that it coincides with the diffracted beam from the grating corresponding to $\theta_x=0$ and $\lambda_\mathrm{des}$. Equation~(\ref{eq:Etilde_rot}) describes an optional rotation of the detector by angle $\kappa$ so that the detector rows align better with the VIPA orders. Equation~(\ref{eq:E_D}) is the field amplitude at the detector. The intensity on the detector, when monochromatic light with wavenumber $k$ is inputted into the system, is the modulus squared of Eq.~(\ref{eq:E_D}):
\begin{equation}\label{eq:I_D}
    I_\mathrm{D}(x,y,k)=|E_\mathrm{D}(x,y,k)|^2
\end{equation}
If there is a broadband source with a flat spectrum between wavenumbers $k_\mathrm{min}$ and $k_\mathrm{max}$, then the intensity is:
\begin{equation}\label{eq:I_DBB}
    I_\mathrm{D,BB}(x,y)=\left| \int_{k_\mathrm{min}}^{k_\mathrm{max}} E_\mathrm{D}(x,y,k) \,dk \right|^2
\end{equation}
which can be approximated numerically by taking a sum instead of an integral.
\section{Spectrograph Design}\label{sec:spec_design}

In this section, we discuss the optical design of VIPER. First, in Section \ref{sec:spec_design_proc}, we discuss a systematic design procedure for cross-dispersed VIPA-based spectrographs and demonstrate the design process for VIPER. The resulting design parameters for VIPER are presented in Table~\ref{tab:design}. Then in Section \ref{sec:spec_VIPER_sim}, we demonstrate an end-to-end model of VIPER using the parameters in Table~\ref{tab:design} and the equations from Section \ref{sec:spec_model}, showing the spectrograph footprint and intensity distribution at the detector. In Section \ref{sec:spec_design_Zemax}, we present an optical design of VIPER modeled in Ansys Zemax OpticStudio. In Section \ref{sec:spec_design_tele_interface}, we discuss preliminary efforts to develop VIPER's interface with the Tillinghast Telescope and MMT. In Section, \ref{sec:spec_VIPER_throughput}, we discuss VIPER's estimated throughput. Finally, in Section \ref{sec:spec_design_challenges}, we highlight some challenges in the design process of cross-dispersed VIPA spectrographs.

\subsection{VIPA Spectrograph Design Procedure}\label{sec:spec_design_proc}

\subsubsection{VIPA Properties}

We started by selecting a design resolving power $\mathcal{R}_\mathrm{des}=3\times10^5$ and a design wavelength $\lambda_\mathrm{des}=1083$~nm. We assumed that the VIPA's front and back surfaces have internal reflectivities of $|r_1|^2=99.5\%$ and $|r_2|^2=95\%$ respectively, which are comparable to the reflectivities of VIPAs in the literature\cite{Carlotti2022,Hu2015,Metz2013,Yang2020}. We assumed that the VIPA is operating in air, which has refractive index $n=1.0003$, and that the VIPA is made of fused silica, which has refractive index $n'=1.4494$ and is also common in the literature\cite{Bourdarot2018,Zhu2020,Zhu2023,Yang2020}. From these parameters, we used Eq.~(\ref{eq:h_min}) to determine that the minimum VIPA thickness is $h_\mathrm{min}=1.042$~mm. From this, we selected a VIPA thickness of $h=1.68$~mm, which is a thickness offering from a company, LightMachinery Inc., that makes VIPAs. We then used Eq.~(\ref{eq:theta_x_peak}) to select a VIPA tilt angle of $\beta_x=2.319^\circ$, shifting the VIPA angular spectrum so that there is a peak at $\theta_x=0$ when $\lambda=\lambda_\mathrm{des}$. With the properties and tilt angle of the VIPA confirmed, we then used Eq.~(\ref{eq:f_cam}) to compute the camera focal length, which is $f_\mathrm{cam}=86.718$~mm. As part of this calculation, we assumed a detector pixel pitch of $p_D=5$ \textmu m and $\Delta x = 3p_D$. Note that it is ideal to select a detector with a smaller pixel pitch because generally $f_\mathrm{cam}$ is a quantity which we would like to minimize (more on this later in Section~\ref{sec:spec_design_challenges}).

\subsubsection{Cross-Disperser}

The next step is to select the cross-disperser. The choice of the cross-disperser depends on the FSR of the VIPA and also the size of the input optical fiber. Using Eq.~(\ref{eq:R_CD}), we found that the resolving power required to resolve two overlapping VIPA orders at $\theta_x=0$, separated by one FSR, is $\mathcal{R}_\mathrm{CD}=4495$. We used a $s=50$~\textmu m diameter optical fiber as the input, with light coming out at a focal ratio of $f/7$ (more on this later in Section~\ref{sec:spec_design_tele_interface}). Since the fiber is an extended source, it will have a finite size on the detector. Thus, in practice, we need a cross-disperser with high angular dispersion so that the orders, which have a finite width, do not overlap with each other. Hence, we selected an echelle grating as the cross-disperser. The echelle grating we selected has a groove density of $1/d=79$ lines/mm and a blaze angle of $\theta_B=63^\circ$.

\begin{figure}[h]
    \centering
    \includegraphics[width=\textwidth]{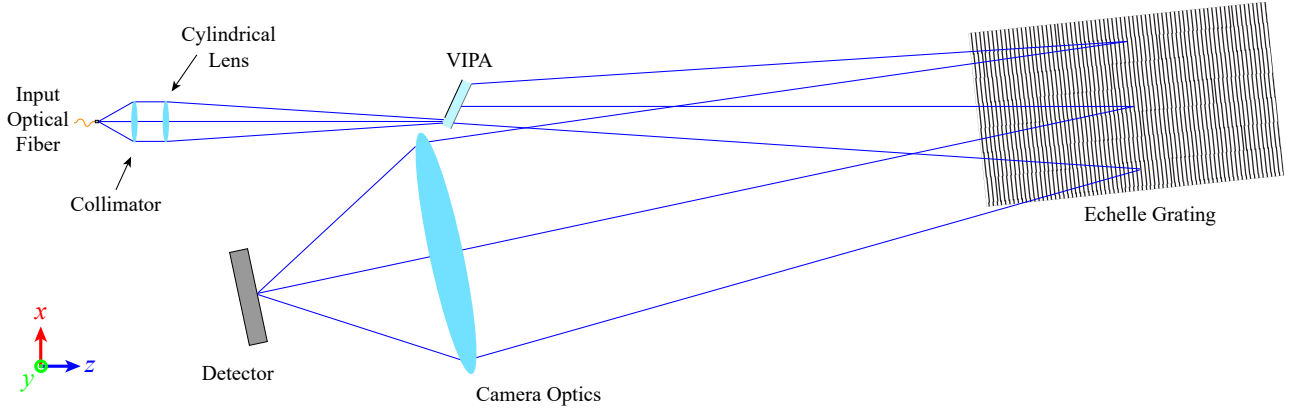}
    \caption{Illustration of a cross-dispersed VIPA spectrograph with the echelle grating in quasi-Littrow configuration. We attempted but abandoned this configuration. This configuration is difficult to implement practically because the grating needs to be placed far away so that the reflected beam can clear the VIPA. The camera optics would be impractically large. Compare this with the design illustrated in Figure~\ref{fig:VIPA_spec_top}, which is the design we actually use.}
    \label{fig:VIPA_spec_quasiLittrow}
\end{figure}

Initially, we wanted to operate the grating in the quasi-Littrow configuration, in order to maximize diffraction efficiency \cite{Schroeder1980}. This configuration is illustrated in Figure \ref{fig:VIPA_spec_quasiLittrow}. In the quasi-Littrow configuration, we would set $\alpha_0=\theta_B$, and then tilt the grating by an angle $\gamma$. However, we found that this kind of configuration was difficult to implement in practice, because the diverging beam exiting from the VIPA would force the grating to be placed far away, so that there is enough room for the diffracted beam, which is reflected, to clear the VIPA. By the time the reflected diverging beam clears the VIPA, it would be very enlarged. The resulting camera optics would have to be large and fast, and hence would be expensive and impractical. In Figure \ref{fig:VIPA_spec_quasiLittrow}, we see that the incident and diffracted beams are both mostly in the $xz$ plane, which is the same plane in which the VIPA output beam is diverging. This is in contrast to Figure \ref{fig:VIPA_spec_top}, in which the incident and diffracted beams are in the $yz$ plane, which is the same plane in which the VIPA output beam is collimated. We therefore abandoned the quasi-Littrow configuration for now, and decided to operate the echelle in a configuration where $\alpha_0>\theta_B$ and $\gamma=0$, which is common in astronomical spectrographs\cite{Schroeder1980}.

In order for the components to be arranged in a configuration that is practically possible, we selected an incidence angle of $\alpha_0=72^\circ$, which is the same incidence angle as in Hectoechelle\cite{Szentgyorgyi2011}, an optical echelle spectrograph for the MMT. Using Eq.~(\ref{eq:m_cen}), we determined that the echelle will be operated in grating order $m_\mathrm{cen}=21$. The corresponding diffraction angle was computed by Eq.~(\ref{eq:beta_0}) as $\beta_0=57.683^\circ$.

\begin{table}[b]
\centering
\begin{tabular}{ | m{0.4\textwidth} | m{0.1\textwidth}| m{0.12\textwidth}| m{0.08\textwidth}| m{0.16\textwidth} | } 
  \hline
  Variable & Symbol & Equation & Choice & Value \\ 
  \hline
  \hline
  Design resolving power & $\mathcal{R}_\mathrm{des}$ & -- & S & $3\times10^5$ \\ 
  Design wavelength & $\lambda_\mathrm{des}$ & -- & S & $1083$ nm \\
  VIPA front surface internal reflectivity & $|r_1|^2$ & -- & S & $99.5\%$ \\
  VIPA back surface internal reflectivity & $|r_2|^2$ & -- & S & $95\%$ \\
  VIPA refractive index & $n'$ & -- & S & $1.4494$ \\
  Surrounding refractive index & $n$ & -- & S & $1.0003$ \\
  VIPA clear aperture & $D_\mathrm{VIPA}$ & -- & S & $18$ mm \\
  Input optical fiber diameter & $s$ & -- & S & $50$ \textmu m \\
  Input optical fiber focal ratio & $(f/\#)_\mathrm{fiber}$ & -- & S & $f/7$ \\
  Detector pixel pitch & $p_D$ & -- & S & $5$ \textmu m \\
  \hline
  Minimum VIPA thickness & $h_\mathrm{min}$ & Eq.~(\ref{eq:h_min}) & C & $1.042$ mm \\
  VIPA thickness & $h$ & -- & S & $1.68$ mm \\
  VIPA input angle & $\beta_x$ & Eq.~(\ref{eq:theta_x_peak}) & C & $2.319^\circ$\\
  Camera focal length & $f_\mathrm{cam}$ & Eq.~(\ref{eq:f_cam}) & C & $86.718$ mm\\
  Required cross-dispersion resolving power & $\mathcal{R}_\mathrm{CD}$ & Eq.~(\ref{eq:R_CD}) & C & $4495$ \\
  \hline
  Grating groove spacing & $d$ & -- & S & $1/79$ mm\\
  Grating blaze angle & $\theta_B$ & -- & S & $63^\circ$ \\
  Grating tilt angle & $\gamma$ & -- & S & $0^\circ$ \\
  Grating incidence angle & $\alpha_0$ & -- & S & $72^\circ$ \\
  Grating order & $m_\mathrm{cen}$ & Eq.~(\ref{eq:m_cen}) & C & $21$ \\
  Diffraction angle corresponding to $\alpha_0$ & 
  $\beta_0$ & Eq.~(\ref{eq:beta_0}) & C & $57.683^\circ$ \\
  \hline
  Detector tilt angle & $\kappa$ & -- & C & $3.892^\circ$ \\
  Collimator focal length & $f_\mathrm{coll}$ & Eqs.~(\ref{eq:f_coll})/(\ref{eq:f_coll_limit}) & C & $126.000$ mm \\
  Cylindrical lens focal length & $f_\mathrm{cyl}$ & Eq.~(\ref{eq:f_cyl_f_coll}) & C & $186.333$ mm \\
  \hline
\end{tabular}
\caption{Design parameters of VIPER. In the ``Choice'' column, ``S'' refers to parameters which we select, while ``C'' refers to parameters we compute.}\label{tab:design}
\end{table}

\subsubsection{Collimator, Cylindrical Lens, and Detector Rotation}

With the grating configuration confirmed, the focal length of the collimator $f_\mathrm{coll}$ can be computed using Eq.~(\ref{eq:f_coll}). However, in practice, $f_\mathrm{coll}$ is often limited by the size of one of the optics downstream. We took the aperture stop to be the VIPA itself, and assumed that the VIPA's clear aperture is $D_\mathrm{VIPA}=18$~mm, which is the clear aperture of off-the-shelf VIPAs manufactured by LightMachinery Inc. Limited by this aperture, we used Eq.~(\ref{eq:f_coll_limit}) to compute the collimator focal length as $f_\mathrm{coll}=126$~mm. Then using Eq.~(\ref{eq:f_cyl_f_coll}), we computed the cylindrical lens focal length as $f_\mathrm{cyl}=186.333$~mm.

At this point, we have selected all the parameters required for us to create a model of VIPER, using the equations in Section \ref{sec:spec_model}. Since the spectrum is cross dispersed, the VIPA orders will be tilted. This is shown in Figure \ref{fig:spectrograph_ang_spec_BB}, which is a plot of the modulus squared of Eq.~(\ref{eq:Etilde_M1}) integrated over many wavelengths. In order to align the VIPA orders with the pixel rows, we decided to rotate the detector by angle $\kappa=3.892^\circ$. The design is now complete, and the parameters of VIPER are presented in Table \ref{tab:design}.


\subsection{VIPER Simulation}\label{sec:spec_VIPER_sim}

In this section, we show VIPER's footprint and intensity distribution at the detector, computed using the end-to-end model in Section \ref{sec:spec_model} and the parameters in Table \ref{tab:design}. 

\subsubsection{Angular Spectrum before Detector Rotation}

\begin{figure}[H]
    \centering
    \includegraphics[width=0.6\textwidth]{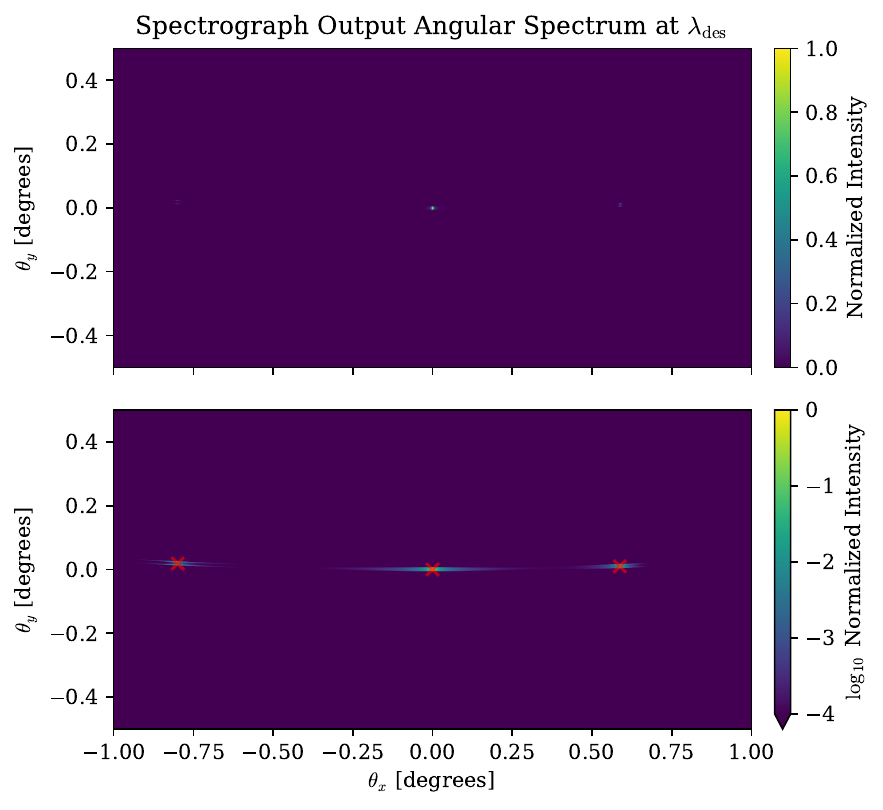}
    \caption{Plot of VIPER's output angular spectrum at the design wavelength $\lambda_\mathrm{des}$, prior to detector rotation. The quantity being plotted is $|\tilde{E}_\mathrm{G,out}(\theta_x,\theta_y,2\pi n/\lambda_\mathrm{des})|^2$, the modulus squared of Eq.~(\ref{eq:Etilde_M1}) evaluated at $\lambda_\mathrm{des}$. The top plot uses a linear scale while the bottom plot uses a logarithmic scale. The red X's mark the expected location of intensity peaks computed using ray optics, using the grating equation (Eq.~(\ref{eq:grating_eq_gamma})) and Eq.~(\ref{eq:theta_x_peak}). Notice how the intensity of the peaks adjacent to the center peak are suppressed, by design. Compare this with Figure \ref{fig:VIPA_out_ang_spec}. }
    \label{fig:spectrograph_output_ang_spec_wavdes}
\end{figure}

\begin{figure}[H]
    \centering
    \includegraphics[width=0.7\textwidth]{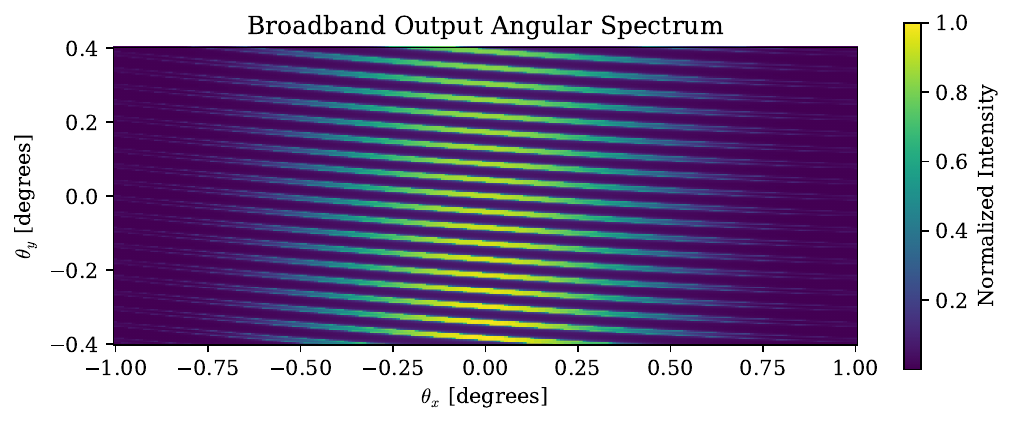}
    \caption{Plot of VIPER's broadband output angular spectrum, prior to detector rotation. The quantity being plotted is $\left|\int \tilde{E}_\mathrm{G,out}(\theta_x,\theta_y,k')\, dk'\right|^2$, where $\tilde{E}_\mathrm{G,out}(\theta_x,\theta_y,k)$ is Eq.~(\ref{eq:Etilde_M1}). Notice that the orders are tilted, and so we rotate the detector. Also note that the intensity falls off from the center vertical line -- this is due to the input angular spectrum to the VIPA (the black dashed curve in Figure \ref{fig:VIPA_out_ang_spec}).}
    \label{fig:spectrograph_ang_spec_BB}
\end{figure}

Figure \ref{fig:spectrograph_output_ang_spec_wavdes} shows VIPER's output angular spectrum before the detector rotation, for a monochromatic input at the design wavelength $\lambda_\mathrm{des}$. In the logarithmically scaled plot, we can see three intensity peaks, which correspond to three different VIPA orders. The $\theta_x$ locations of these intensity peaks are the same as those in the lower subplot of Figure \ref{fig:VIPA_ang_spec} and those marked by the magenta dashed lines in Figure \ref{fig:VIPA_out_ang_spec}. VIPER's output angular spectrum is being modulated by the angular spectrum of the injected electric field (Eq.~(\ref{eq:cyl_ang_spec})), which causes orders away from the center to be suppressed, by design. Figure \ref{fig:spectrograph_ang_spec_BB} shows VIPER's output angular spectrum for a broadband input. The modulation effect of the injection angular spectrum is clearly seen. In this figure, we can also see that the orders are tilted. In order to align the orders with the pixel rows, we rotated the detector.

\subsubsection{Spectrum following Detector Rotation}

\begin{figure}[H]
    \centering
    \includegraphics[width=0.77\textwidth]{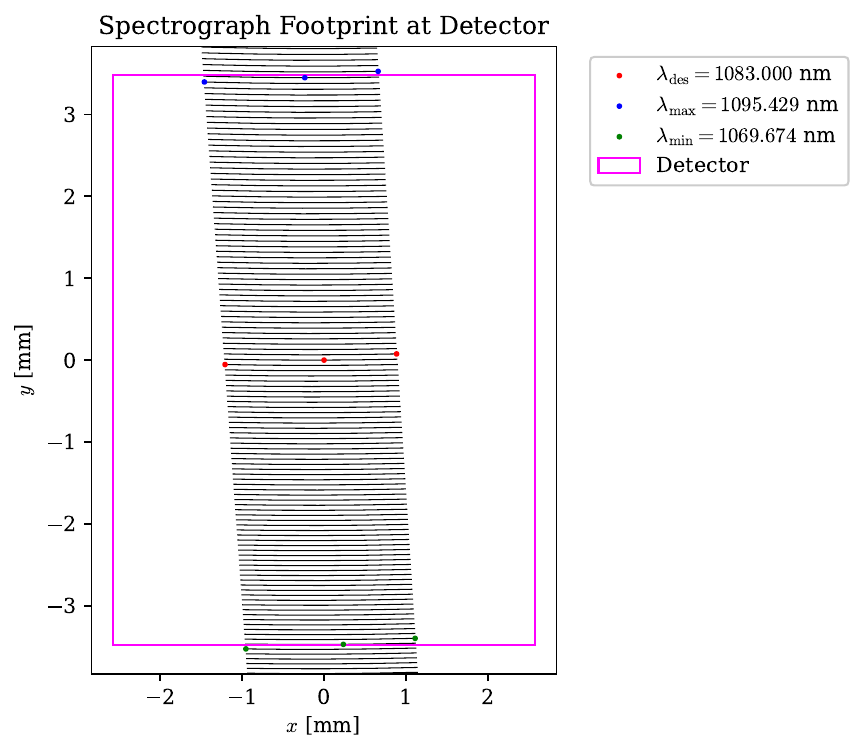}
    \caption{Plot of VIPER's footprint on the detector. Each black curve represents one VIPA order. The detector is rotated by $\kappa=3.892^\circ$ to align the orders with the pixel rows. The magenta box marks the extent of the detector, which has a 5 \textmu m pixel pitch with $1032 \times 1392$ pixels.}
    \label{fig:footprint_detector}
\end{figure}

\begin{figure}[H]
    \centering
    \includegraphics[width=0.55\textwidth]{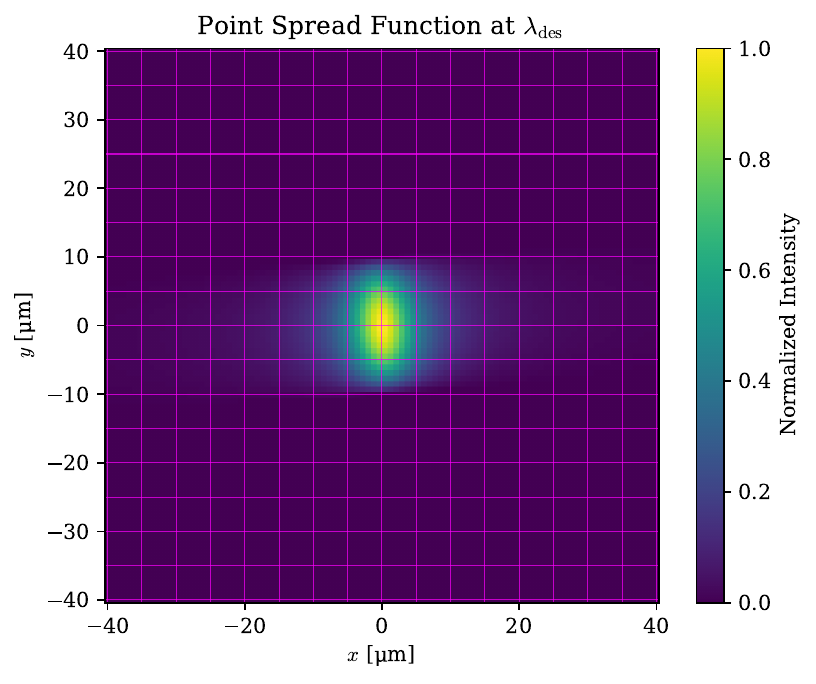}
    \caption{Plot of VIPER's monochromatic spectrum on the detector at $\lambda_\mathrm{des}$. The quantity being plotted is Eq.~(\ref{eq:I_D}) evaluated at $k=2\pi n/\lambda_\mathrm{des}$. This is the point spread function at $\lambda_\mathrm{des}$, in the center VIPA order at $\theta_x=0$ and $\theta_y=0$. The magenta lines mark the pixel boundaries.}
    \label{fig:spectrum_wavdes}
\end{figure}

\begin{figure}[H]
    \centering
    \includegraphics[width=0.7\textwidth]{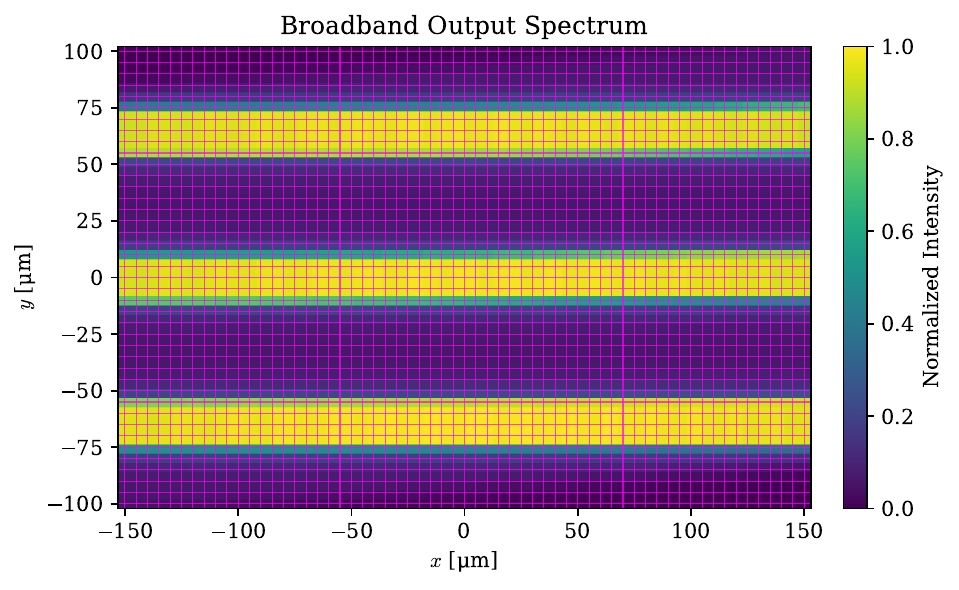}
    \caption{Plot of VIPER's broadband spectrum on the detector. The quantity being plotted is Eq.~(\ref{eq:I_DBB}). The magenta lines mark the pixel boundaries. The orders are well separated from each other.}
    \label{fig:spectrum_BB}
\end{figure}

Figure \ref{fig:footprint_detector} shows VIPER's footprint at the detector. Each black curve represents the extent of one VIPA order. From the footprint, we can see that VIPER's total wavelength range is around 25~nm, with a minimum wavelength of 1069.674~nm and a maxmimum wavelength of 1095.429~nm, for our particular assumed choice of a detector. This 25~nm wavelength range significantly exceeds our requirement listed in Table~\ref{tab:inst_req}. This extended wavelength range was not deliberately designed for, but emerged naturally as a consequence of the selected optical design parameters. Figures \ref{fig:spectrum_wavdes} and \ref{fig:spectrum_BB} respectively show the point spread function and broadband intensity at the detector.

\subsection{Zemax Optical Design}\label{sec:spec_design_Zemax}

\begin{figure}[H]
    \centering
    \includegraphics[width=\textwidth]{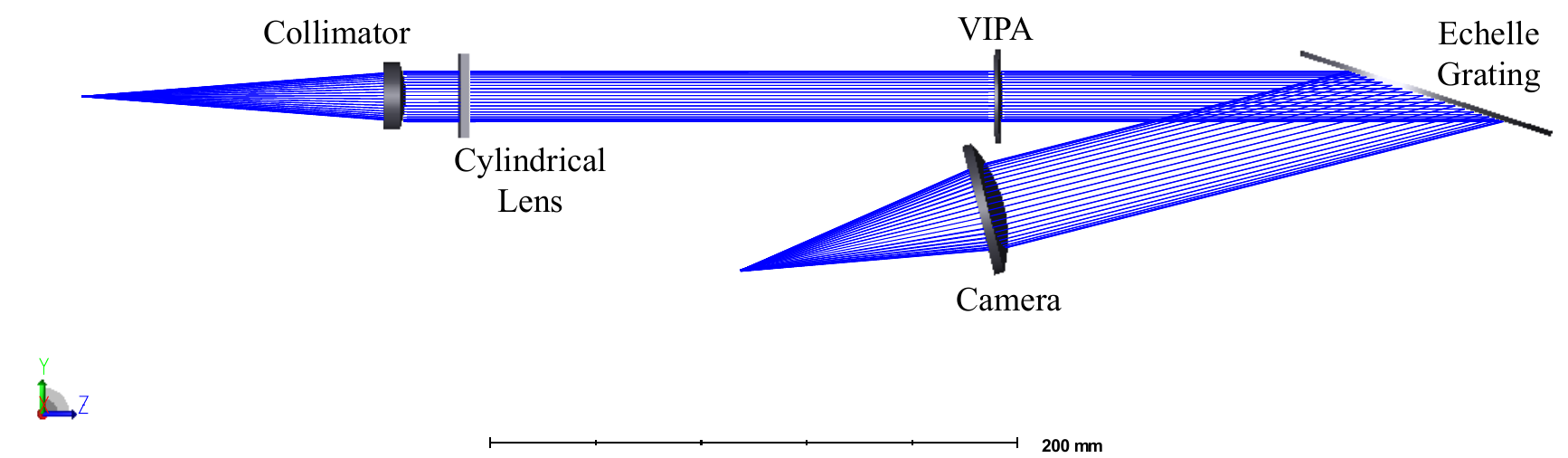}
    \caption{Ansys Zemax OpticStudio model of VIPER, using mixed mode. This is analogous to Figure \ref{fig:VIPA_spec_top}.}
    \label{fig:Zemax_shaded_model}
\end{figure}

In this section, we present a real-world optical design of VIPER modeled in Ansys Zemax OpticStudio. We picked off-the-shelf lenses with focal lengths close to those reported in Table \ref{tab:design}. For the collimator, we selected a Thorlabs AC254-125-C achromatic doublet with a focal length of 125 mm. For the cylindrical lens, we selected a Thorlabs LJ1653L1-C plano-convex cylindrical lens with a focal length of 200 mm. For the camera optics, we selected a Thorlabs AL50100-C aspheric lens with a focal length of 100 mm. VIPER was modeled in Zemax's mixed mode using these components, illustrated in Figure \ref{fig:Zemax_shaded_model}.

\begin{figure}[H]
    \centering
    \includegraphics[width=0.8\textwidth]{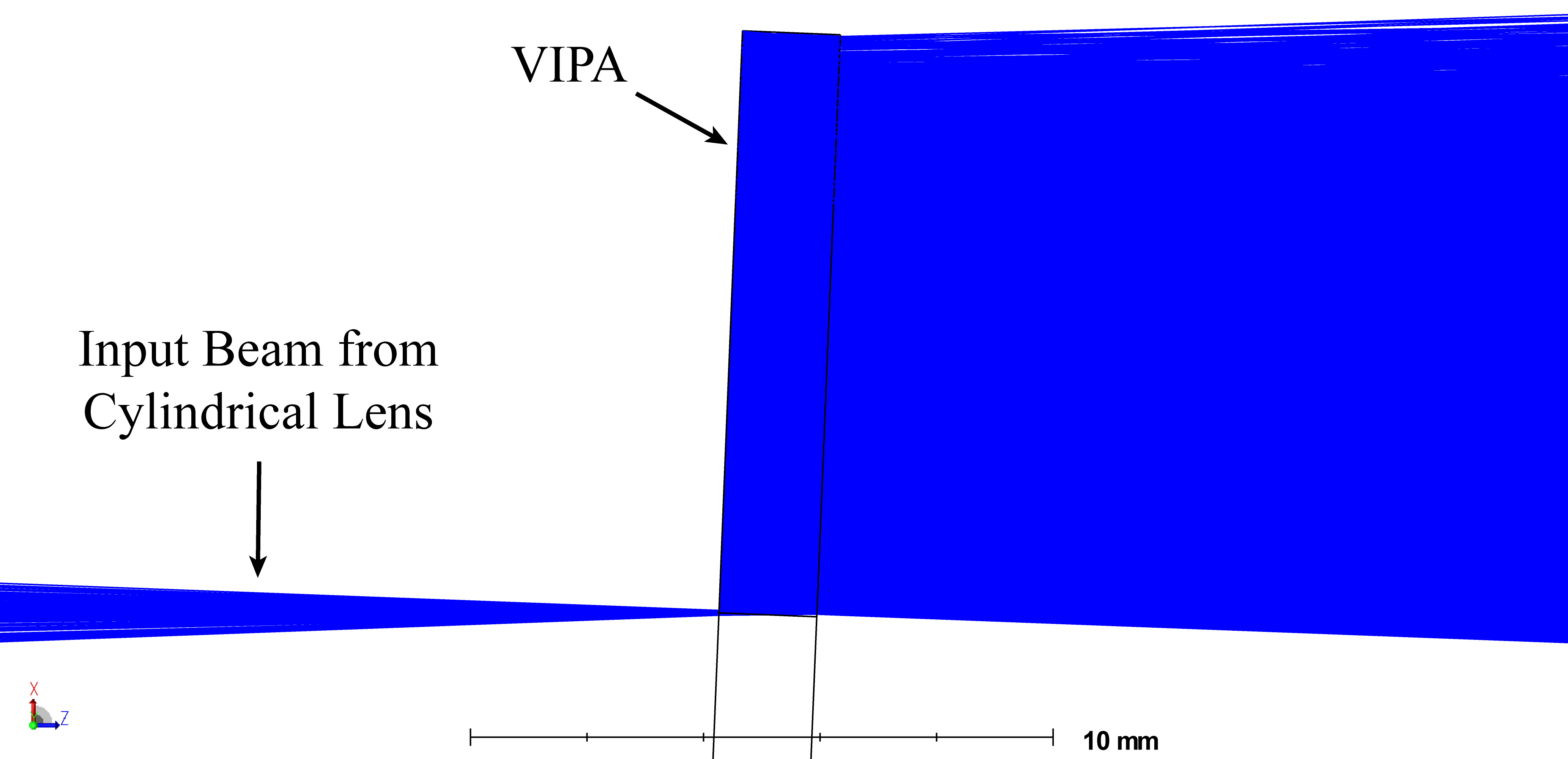}
    \caption{Ansys Zemax OpticStudio model of VIPA in VIPER, using non-sequential mode.}
    \label{fig:Zemax_VIPA}
\end{figure}

\begin{figure}[H]
    \centering
    \includegraphics[width=0.8\textwidth]{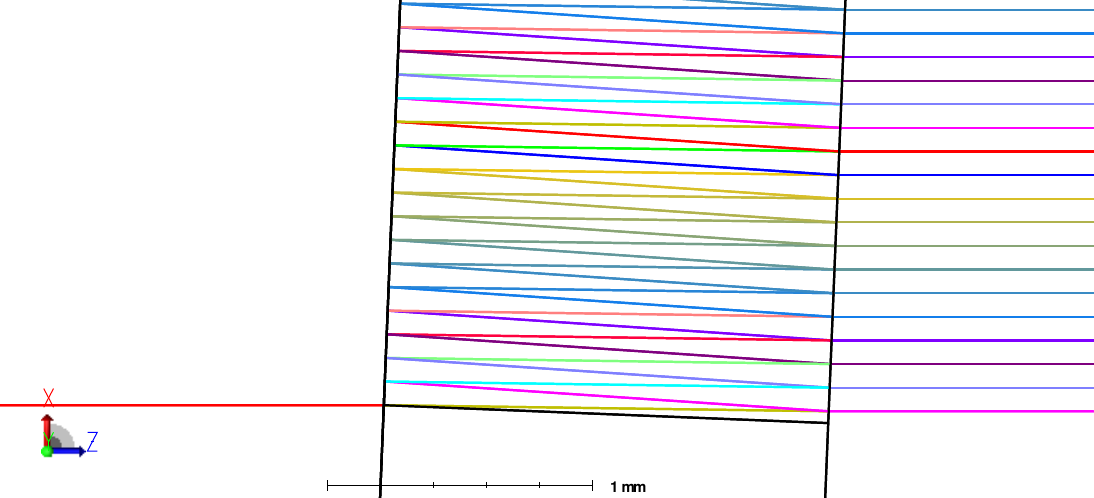}
    \caption{On-axis single ray trace in Ansys Zemax OpticStudio model of VIPA in VIPER, using non-sequential mode. The colors denote different ray segments. The non-sequential ray is split at the front and back internally reflective surfaces of the VIPA.}
    \label{fig:Zemax_VIPA_SRT}
\end{figure}

The VIPA was modeled as a non-sequential component within mixed mode, illustrated in Figure \ref{fig:Zemax_VIPA}. The VIPA itself can be modeled as a rectangular volume or cylinder volume. Using Zemax's nesting capability\cite{Yang2020} in non-sequential mode, we placed a mirror at the back surface of the VIPA, leaving an opening for the beam from the cylindrical lens to enter. Light then reflects internally between the two VIPA surfaces. Figure~\ref{fig:Zemax_VIPA_SRT} shows an on-axis single ray trace through the same VIPA, also in non-sequential mode. We can see that the non-sequential rays are split at the two surfaces, behaving the same way as in an actual VIPA.

This Zemax model demonstrates that our choice of parameters in Table \ref{tab:design} is physically realizable. Unlike most conventional high-resolution astronomical spectrographs with $\mathcal{R}\gtrsim10^5$, VIPER is compact, on the order of 50~cm in length as seen in Figure \ref{fig:Zemax_shaded_model}. This reduces costs and technical complexity significantly.

\subsection{Telescope Interface}\label{sec:spec_design_tele_interface}

In this section, we discuss preliminary efforts to develop VIPER's interface with the Tillinghast Telescope and MMT, both at FLWO on Mount Hopkins, Arizona, USA.

\subsubsection{Tillinghast Telescope}\label{sec:spec_design_Tilling_interface}

The Tillinghast Telescope is a 1.5 m (60 inch) diameter telescope at the FLWO Ridge, at 7800 ft above sea level and 800 ft below the peak of Mount Hopkins. The Tillinghast is used exclusively for spectroscopy. At its native $f/10$ focal ratio, the Tillinghast has a plate scale of 72.72~\textmu m/arcsec. If we convert the $f/10$ beam to $f/7$ to feed an optical fiber, the linear distance on the focal plane corresponding to 1 arcsec is $72.72\times \frac{7}{10}=50.90$~\textmu m. If we assume that the seeing at the Tillinghast is the same as that of the nearby MMT, which has a reported median seeing of 0.87 arcsec\cite{Pickering2018}, then the minimum optical fiber diameter required for a $f/7$ feed is $50.90\times0.87=44.28$~\textmu m. This is why we selected a $s=50$~\textmu m diameter fiber and $f/7$ in Table \ref{tab:design}.

\begin{figure}[h]
    \centering
    \includegraphics[width=0.5\textwidth]{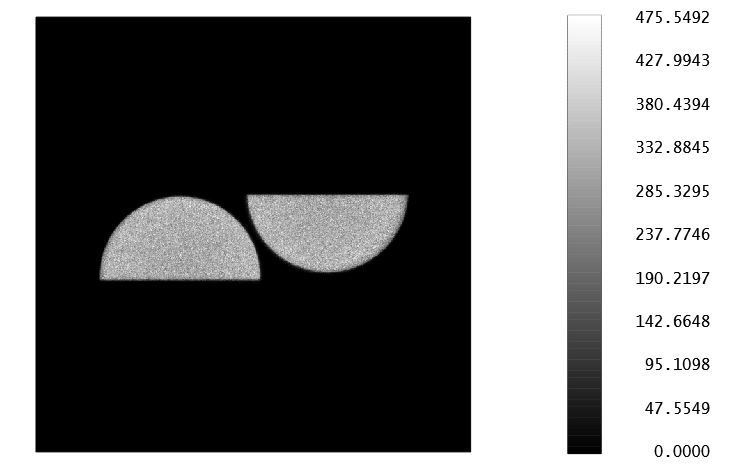}
    \caption{Ansys Zemax OpticStudio geometric image analysis simulation of a Bowen-Walraven pupil slicer currently in development by some members of our team, for an upgrade to the EXPRES spectrograph. A similar slicer used in EXPRES achieved a throughput and injection efficiency of over 90\%\cite{Jurgenson2016}.}
    \label{fig:EXPRES_slicer}
\end{figure}

However, a recent internal report in our team found that the Tillinghast Telescope performance has degraded as a result of guiding and instrumental problems, with seeing varying between 1.3~arcsec and 1.9~arcsec, with an average of 1.6~arcsec. At the worst-case seeing of 1.9 arcsec with a $f/7$ feed, a 96.71~\textmu m diameter optical fiber would be required. To address this, we are planning to feed the $f/7$ beam into a 100 \textmu m diameter circular fiber, slice the output light using a mirror Bowen-Walraven pupil slicer\cite{Avila2012,Tala2017}, and then feed this into a rectangular 50~\textmu m by 200~\textmu m optical fiber, where the longer edge of the rectangle is parallel with the $x$-axis in Figure \ref{fig:VIPA_spec_side}. Some members of our team are currently developing Bowen-Walraven pupil slicers for G-CLEF\cite{Szentgyorgyi2024_OG} and EXPRES\cite{Jurgenson2016}, both high resolution spectrographs, and the development of VIPER's slicer will happen in tandem with these projects. Figure \ref{fig:EXPRES_slicer} shows a Zemax geometric image analysis simulation of a Bowen-Walraven pupil slicer in development by our team for an upgrade to EXPRES. Such a slicer has a throughput and injection efficiency of over 90\%.

Going from a $50$~\textmu m diameter circular fiber to a 50~\textmu m by 200~\textmu m rectangular optical fiber would require slight changes in our design presented in Table \ref{tab:design}. The main change is that with the factor of 4 increase in the optical fiber size along the $x$-direction, the cylindrical lens focal length would have to decrease accordingly by a factor of 4, according to Eq.~(\ref{eq:f_cyl_f_coll}). The focal lengths of the collimator and camera can remain unchanged. Note that the inherent resolving power of the VIPA does not depend on the fiber size, as seen in Eq.~(\ref{eq:R_FP}). In the end-to-end spectrograph model presented in Section \ref{sec:spec_model}, Eq.~(\ref{eq:cyl_ang_spec}) would have to be replaced with an appropriate equation describing a rectangular optical fiber as the input, but the other equations can remain the same and still be used. 


\subsubsection{MMT}

We are also exploring the possibility of placing VIPER on the 6.5~m MMT. At its native $f/5$ focal ratio, the MMT has a plate scale of 157.56~\textmu m/arcsec. If we convert the $f/5$ beam to $f/7$ to feed an optical fiber, and assume the same 0.87 arcsec seeing, then the minimum optical fiber diameter required is 191.91~\textmu m. Basically, a 200~\textmu m diameter optical fiber is required, which is 4 times as large as the one in our current design. It would be difficult to design a cross-dispersed VIPA spectrograph with our configuration in Figure \ref{fig:VIPA_spec_top} using a 200~\textmu m diameter optical fiber, for reasons which we will discuss in Section \ref{sec:spec_design_challenges}. One simple solution is to employ a slit orthogonal to the cross-dispersion direction, which is what is done in the literature\cite{Zhu2023}. However, we would like to avoid this because we want to maximize throughput.

One possibility is to feed VIPER using the new MMT adaptive optics (AO) system, called MAPS\cite{Morzinski2020,Anugu2020,Morzinski2024}. MAPS is an AO upgrade which is designed to feed two instruments at the MMT: AIRES and MMT-Pol. However, other instruments can also access the AO-corrected focal plane, and MAPS provides a $f/15$ feed. MAPS is currently online, but performance refinements are being made. The expected Strehl ratio in the $J$-band is around 0.4\cite{Anugu2020}. The AO-corrected seeing disk may be on the order of 0.1 arcsec at $\lambda=1$ \textmu m, which implies that a roughly 22~\textmu m diameter fiber would suffice at a $f/7$ feed. Our current design would work with this. However, our target selection in Figure \ref{fig:mass_radius_period} may be affected, because MAPS uses natural guide star AO, and so a relatively bright star is required near the target.


\subsection{VIPER Throughput}\label{sec:spec_VIPER_throughput}
 
\begin{table}[H]
\centering
\begin{tabular}{ | m{0.45\textwidth} | m{0.07\textwidth}| m{0.4\textwidth} | } 
  \hline
  Component & $T$ & Note/Reference\\ 
  \hline
  \hline
  Pupil slicer throughput and injection efficiency & $>90\%$ & See Section \ref{sec:spec_design_Tilling_interface} and Ref. \citenum{Jurgenson2016} for an example\\
  Collimator transmission & $99.2\%$ & Thorlabs AC254-125-C achromatic doublet\\
  Cylindrical lens transmission & $99.5$ & Thorlabs LJ1653L1-C cylindrical lens\\
  VIPA theoretical transmission & $91.3\%$ & See Eq.~(\ref{eq:VIPA_transmission}) \\
  Echelle grating relative diffraction efficiency & $48.3\%$ & See Figure \ref{fig:grating_eff_slice} \\
  Echelle grating substrate reflectivity & $92.5\%$ & Bare aluminum\\
  Camera transmission & $99.5\%$ & Thorlabs AL50100-C aspheric lens \\
  \hline
  Total & $36\%$ & \\
  \hline
\end{tabular}
\caption{VIPER's estimated throughput at design wavelength $\lambda_\mathrm{des}=1083$ nm.}\label{tab:throughput}
\end{table}

Table \ref{tab:throughput} shows the estimated throughput of VIPER at the 1083 nm design wavelength. The total throughput at this wavelength is currently 36\%, which is higher than most conventional high-resolution astronomical spectrographs, which have throughputs of 5--10\%. The component that limits the throughput in our design is the echelle grating, which has a relative diffraction efficiency of 48.3\% at 1083~nm. We are looking for ways to improve the throughput, and investigating if alternative cross-dispersers could be used.


\subsection{Design Challenges}\label{sec:spec_design_challenges}

In this section, we highlight the challenges in designing multimode fiber-fed cross-dispersed VIPA spectrographs. One of the greatest challenges is the optical fiber size. We want the optical fiber to be large so that more light can enter the spectrograph and so that the injection requirements at the telescope are less stringent. However, most VIPA-based spectrographs use a single mode fiber as the input because a larger optical fiber input is more challenging, for several reasons.

Firstly, a larger optical fiber has a narrower angular spectrum, which means that the cylindrical lens has to have a shorter focal length to compensate, as seen in Eq.~(\ref{eq:f_cyl_f_coll}), where $f_\mathrm{cyl}\propto s^{-1}$. If the cylindrical lens has a shorter focal length, then the beam exiting the VIPA will be more spread out along the $x$-direction, meaning that the optics downstream will have to be larger. This particularly affects the size of the camera optics, which in practice, cannot be made arbitrarily large.

A larger optical fiber also means that the cross-disperser has to provide greater linear dispersion on the detector because the individual orders will be wider in the cross dispersion direction. We experienced this problem and had to use an echelle as the cross-disperser, which is not ideal because echelle gratings have relatively poor diffraction efficiency. Due to the expanding beam exiting from the VIPA, we could not find a practical way to operate the echelle in quasi-Litrrow configuration (see Figure \ref{fig:VIPA_spec_quasiLittrow}), because the camera optics would be impractically large. Instead, we had to rely on a more inefficient configuration where $\alpha_0>\theta_B$.

The root problem here is that the image of the fiber on the detector along the cross-dispersion direction is too large. Along the cross-dispersion direction, we are basically experiencing the same design challenges as in a conventional grating-based astronomical spectrograph (see Eq.~(\ref{eq:R_grat_spec})). As in a conventional grating-based astronomical spectrograph, the solution is to make the demagnification between the collimator and camera larger; or in other words, to increase the collimator focal length and to decrease the camera focal length. However, we cannot arbitrarily decrease the camera focal length because there is a lower limit imposed by the VIPA, in Eq.~(\ref{eq:f_cam}). We also cannot arbitrarily increase the collimator focal length because this is constrained by the VIPA clear aperture, in Eq.~(\ref{eq:f_coll_limit}).

In order to address this root problem, some multimode fiber-fed cross-dispersed VIPA spectrographs have employed an additional slit, for example in Ref.~\citenum{Zhu2023}. The input optical fiber is reimaged onto a slit which runs orthogonal to the cross-dispersion direction. With this slit, the effective fiber size along the cross-dispersion direction is reduced, and so the image of the fiber on the detector along the cross-dispersion direction is small. However, this is something which we did not want to do, because a lot of light is lost with the slit. With a careful selection of parameters, we were able to avoid the use of a slit in our design.

As alluded to in the previous paragraphs, the size of the camera optics is a challenge. The camera focal length is set by Eq.~(\ref{eq:f_cam}). However, the specific geometry of the other components and their focal lengths will influence the size of the camera optics. In our current design, the camera needs to be fast, with a focal ratio of around $\lesssim f/2$. The spread of the output angular spectrum after the cross disperser would require the camera to have a half field angle on the order of $2.5^\circ$. These two requirements on aperture and field size respectively may be difficult to achieve with off-the-shelf components. According to Ref.~\citenum{SPIE_FG27}, microscope objectives and single aspheric lenses may not work well with the $2.5^\circ$ half field angle and $f/2$ aperture requirements. For the camera, we may need to use a double Gauss lens or a Petzval lens with a field flattener. This is something that we are investigating. If the camera is made even larger, then the focal ratio will be much smaller than $f/2$. In this case, with the $2.5^\circ$ half field angle, there might be no suitable lens designs that can image the spectrum with acceptable aberrations.
\newpage
\section{Conclusions and Next Steps}

We are developing VIPER, a high-resolution, narrowband, cross-dispersed VIPA spectrograph designed specifically to observe the helium 1083 nm triplet absorption line. The primary science goal of VIPER is to detect anisotropic atmospheric escape in gaseous planets. To maximize throughput, VIPER is fed by a multimode optical fiber and operates without an additional slit, unlike other VIPA spectrograph designs in the literature.

The optical design for VIPER follows a design methodology we developed for VIPA spectrographs aimed at maximizing throughput and diffraction efficiency. VIPER's design parameters are listed in Table \ref{tab:design}. The current design of VIPER will achieve a resolving power of $\mathcal{R}$ = 300,000 over a bandpass of 25 nm, with an estimated throughput of 36\% at the 1083 nm design wavelength. We developed a wave-optics-based end-to-end analytic model for cross-dispersed VIPA spectrographs, summarized in Section \ref{sec:spec_model}, and we used this model to demonstrate VIPER's performance in Section \ref{sec:spec_design}.

There are several design challenges, as discussed in Section \ref{sec:spec_design_challenges}. In particular, the current echelle cross-disperser has relatively low diffraction efficiency, and we are actively searching for potential alternatives with enough angular dispersion. We are also developing a mirror Bowen-Walraven pupil slicer, which we will use to feed a rectangular fiber. We aim for VIPER to be on-sky by 2027--2028.
\appendix
\numberwithin{equation}{section}
\section{Derivation of Input Angular Spectrum to VIPA}\label{sec:app_VIPA_input}

The angular spectrum $\tilde{E}_\mathrm{cyl}$ at the focal plane of the cylindrical lens can be derived by principles of Fourier optics. The angular spectrum $\tilde{U}(k_x,k_y)$ of scalar electric field $U(x,y)$ obeys\cite{Goodman}:
\begin{subequations}
\begin{align}
    \tilde{U}(k_x,k_y) &= \int_{-\infty}^\infty \int_{-\infty}^\infty U(x,y)\, e^{-i(k_x x + k_y y)} \,dx\,dy \\
    U(x,y) &= \int_{-\infty}^\infty \int_{-\infty}^\infty \tilde{U}(k_x,k_y)\, e^{i(k_x x + k_y y)} \,dk_x\,dk_y \label{eq:Utilde_k_to_U}
\end{align}
\end{subequations}
We can make a change of variables to obtain the angular spectrum in terms of angles $\theta_x$ and $\theta_y$ instead of the wavevector components $k_x=k\sin{\theta_x}$ and $k_y=k\sin{\theta_y}$. The Jacobian of the transformation is $\left|\frac{\partial(k_x,k_y)}{\partial(\theta_x,\theta_y)}\right| = k^2\cos{\theta_x}\cos{\theta_y}$. Substituting $dk_x\,dk_y = \left|\frac{\partial(k_x,k_y)}{\partial(\theta_x,\theta_y)}\right| d\theta_x\,d\theta_y$ in Eq.~(\ref{eq:Utilde_k_to_U}), we obtain:
\begin{subequations}
\begin{align}
    \tilde{U}(\theta_x,\theta_y) &= \int_{-\infty}^\infty \int_{-\infty}^\infty U(x,y) \,e^{-ik(x\sin{\theta_x}+y\sin{\theta_y})} \,dx\,dy \\
     U(x,y) &= \iint\limits_{\sin^2{\theta_x}+\sin^2{\theta_y}\leq1} \tilde{U}(\theta_x,\theta_y) \,k\cos{\theta_x}\cos{\theta_y}\,e^{ik(x\sin{\theta_x}+y\sin{\theta_y})} \,d\theta_x\,d\theta_y \label{eq:Utilde_theta_to_U}
\end{align}
\end{subequations}
Consider Figure \ref{fig:VIPA_inject2lenses}. By the principles of Fourier optics, ignoring some phase factors, we have:
\begin{subequations}
\begin{align}
    E_\mathrm{coll}(x,y) &\propto \tilde{E}_\mathrm{source}(\arctan{(x/f_\mathrm{coll})}, \arctan{(y/f_\mathrm{coll})}) \\
    \tilde{E}_\mathrm{coll}(\theta_x,\theta_y) &\propto E_\mathrm{source}(f_\mathrm{coll}\tan{\theta_x}, f_\mathrm{coll}\tan{\theta_y}) \\
    E_\mathrm{coll,2}(x,y) &\propto E_\mathrm{coll}(x,y) \,T(x,y) \\
    \tilde{E}_\mathrm{coll,2}(\theta_x,\theta_y) &\propto \tilde{E}_\mathrm{coll}(\theta_x,\theta_y) * \tilde{T}(\theta_x,\theta_y) \\
    \tilde{E}_\mathrm{cyl}(\theta_x,\theta_y) &\propto \int \tilde{E}_\mathrm{coll,2}(\varphi_x,\theta_y) \,k\cos{\varphi_x} \, e^{ik\sin{\varphi_x}(f_\mathrm{cyl}\tan{\theta_x})}  \, d\varphi_x \label{eq:Etildecyl_integral}
\end{align}
\end{subequations}
Note that Eq.~(\ref{eq:Etildecyl_integral}) is an analog of Eq.~(\ref{eq:Utilde_theta_to_U}). The cylindrical lens only focuses light along the $x$ direction and so there is only a single integral.

If we assume that the collimator and cylindrical lens are infinitely large, then $T(x,y)=1$ and $\tilde{T}(\theta_x,\theta_y)=\delta(\theta_x)\delta(\theta_y)$. This simplifies the calculations greatly since then $\tilde{E}_\mathrm{coll,2}\propto \tilde{E}_\mathrm{coll}\propto E_\mathrm{source}$, where $E_\mathrm{source}$ is the circle function from Eq.~(\ref{eq:fiber_circle}). Substituting this into Eq.~(\ref{eq:Etildecyl_integral}), the angular spectrum of the electric field after the cylindrical lens is:
\begin{align*}\label{eq:cyl_ang_spec}
    \tilde{E}_\mathrm{cyl}(\theta_x,\theta_y) &\propto \int \tilde{E}_\mathrm{coll,2}(\varphi_x,\theta_y) \,k\cos{\varphi_x} \, e^{ik\sin{\varphi_x}(f_\mathrm{cyl}\tan{\theta_x})}  \, d\varphi_x \nonumber\\
    &= \int E_\mathrm{source}(f_\mathrm{coll}\tan{\varphi_x}, f_\mathrm{coll}\tan{\theta_y}) \,k\cos{\varphi_x} \, e^{ik\sin{\varphi_x}(f_\mathrm{cyl}\tan{\theta_x})}  \, d\varphi_x \nonumber \\
    &= \int_{-C_1(\theta_y)}^{C_1(\theta_y)} k\cos{\varphi_x} \, e^{ik\sin{\varphi_x}(f_\mathrm{cyl}\tan{\theta_x})}  \, d\varphi_x \nonumber
\end{align*}
where we define:
\begin{equation}
    C_1(\theta_y) \equiv \arctan{\left(\frac{1}{f_\mathrm{coll}} \sqrt{\left(\frac{s}{2}\right)^2-(f_\mathrm{cyl}\tan{\theta_y})^2}\right)} \tag{\ref{eq:C1}}
\end{equation}
Then noting that $C_1\ll1$ and hence $\varphi_x\ll1$, we can approximate $\cos{\varphi_x}\approx1$ and $\sin{\varphi_x}\approx \varphi_x$, so that:
\begin{align}
    \tilde{E}_\mathrm{cyl}(\theta_x,\theta_y) &\propto \int_{-C_1(\theta_y)}^{C_1(\theta_y)} k\cos{\varphi_x} \, e^{ik\sin{\varphi_x}(f_\mathrm{cyl}\tan{\theta_x})}  \, d\varphi_x \nonumber \\
    &\approx \int_{-C_1(\theta_y)}^{C_1(\theta_y)} k  e^{ik\varphi_x(f_\mathrm{cyl}\tan{\theta_x})}  \, d\varphi_x \nonumber\\
    &= 2k C_1(\theta_y)\, \sinc{\left(C_1(\theta_y)\, k f_\mathrm{cyl} \tan{\theta_x}\right)} \tag{\ref{eq:cyl_ang_spec}}
\end{align}
\section{Output Angular Spectrum of a Blazed Grating}\label{sec:app_grating}

\subsection{Blazed Diffraction Grating}

Consider a blazed diffraction grating. Suppose that the grating normal is the $z$-axis, that the grating grooves are parallel to the $x$-axis, and that the incident light is on a plane that is perpendicular to the grating grooves (i.e, on the $yz$ plane). In this case, the incident and diffracted rays lie on the same plane of incidence, which is the $yz$ plane. Let the incident and diffracted wavevector $y$-components in the plane of incidence be $k_{y0}=k\sin{\theta_{y0}}$ and $k_y=k\sin{\theta_y}$ respectively, where $\theta_{y0}$ and $\theta_y$ are respectively the incidence and diffracted angles in the plane of incidence. Let $d$ be the groove separation and let $m$ be the integer-valued diffraction order. Then the effect of the grating can be written as a statement of conservation of wavevector:
\begin{subequations}
\begin{align}
    k_y&=\frac{2\pi m}{d} - k_{y0} \label{eq:grating_eq_k}\\
    \frac{2\pi m}{kd} &= \sin{\theta_{y0}}+\sin{\theta_y} \label{eq:grating_eq}
\end{align}
\end{subequations}
Equation~(\ref{eq:grating_eq}) is the commonly-known grating equation. From this equation, we can obtain the phase difference between the centers of adjacent grating grooves, which is:
\begin{equation}\label{eq:deltagcc_og}
    \delta_\mathrm{g,cc} = kd(\sin{\theta_{y0}}+\sin{\theta_y})
\end{equation}
Ref.~\citenum{Schroeder} gives the phase difference between the center and edge of one groove as:
\begin{equation}\label{eq:deltagce_og}
    \delta_{\mathrm{g,ce}} = \frac{kb}{2} \left[\sin{(\theta_{y0}-\theta_B)} + \sin{(\theta_y-\theta_B)}\right]
\end{equation}
where $\theta_B$ is the blaze angle of the grating, and $b$ is the effective groove separation, which is:
\begin{equation}\label{eq:b}
    b =
    \begin{cases} 
      d\cos{\theta_B} & \mathrm{if}\:\:\: \theta_{y0}>\theta_B \\
      \frac{d\cos{\theta_{y0}}}{\cos{(\theta_{y0}-\theta_B)}} & \mathrm{if}\:\:\: \theta_{y0}\leq\theta_B
   \end{cases}
\end{equation}
With the phase differences $\delta_\mathrm{g,cc}$ and $\delta_\mathrm{g,ce}$, we can derive the output angular spectrum of a diffraction grating $\tilde{E}_\mathrm{G}(\theta_x,\theta_y)$ given some input angular spectrum $\tilde{E}_\mathrm{G,in}(\theta_x,\theta_y)$.

The output of a diffraction grating can be thought of as the interference of light diffracted through a series of $N_\mathrm{il}$ evenly spaced identical apertures, with each aperture being the illuminated blazed groove. Each aperture is a wave source with equal amplitude (determined from $\delta_{\mathrm{g,ce}}$) and equal phase difference $\delta_\mathrm{g,cc}$. Following this paradigm, we can obtain the transfer function of a diffraction grating\cite{BornWolf}:
\begin{equation}
    H_\mathrm{G}(\theta_{y0},\theta_y,k) \propto b\sinc{(\delta_{\mathrm{g,ce}})} \, \frac{1-e^{-iN_\mathrm{il}\delta_\mathrm{g,cc}}}{N_\mathrm{il}(1-e^{-i\delta_\mathrm{g,cc}})}
\end{equation}
where $N_\mathrm{il}$ is the number of illuminated grooves on the grating. The $\frac{1-e^{-iN_\mathrm{il}\delta_\mathrm{g,cc}}}{N_\mathrm{il}(1-e^{-i\delta_\mathrm{g,cc}})}$ term comes from the evaluation of a geometric series\cite{SalehTeich}, which arises from the summation of $N_\mathrm{il}$ sources with equal amplitude and equal phase difference $\delta_\mathrm{g,cc}$. Note that this term acts like a Dirac comb, peaking wherever $\delta_\mathrm{g,cc}$ is an integer multiple of $2\pi$. So, we can write:
\begin{equation*}
    \lim_{N_\mathrm{il}\rightarrow\infty} \frac{1-e^{-iN_\mathrm{il}\delta_\mathrm{g,cc}}}{N_\mathrm{il}(1-e^{-i\delta_\mathrm{g,cc}})} \rightarrow \sum_{m\in\mathbb{Z}} \delta(\delta_\mathrm{g,cc}-2m\pi)
\end{equation*}
Hence, $H_\mathrm{G}(\theta_{y0},\theta_y,k)$ could be written as:
\begin{equation}\label{eq:grating_transfer_1D_og}
    H_\mathrm{G}(\theta_{y0},\theta_y) \propto b\sinc{(\delta_{\mathrm{g,ce}})} \sum_{m\in\mathbb{Z}} \delta(\delta_\mathrm{g,cc}-2m\pi)
\end{equation}
Note that $H_\mathrm{G}(\theta_{y0},\theta_y,k)$ is not shift-invariant in angle space.


\subsection{Conical Diffraction}

Now, suppose that the incident light instead makes an angle $\gamma$ with respect to the plane perpendicular to the grating grooves (i.e., the $yz$ plane). In this case, the diffracted light will lie on a cone, rather than on a single plane of incidence (rather than on the $yz$ plane); this effect is known as conical diffraction, and arises due to conservation of wavevector, since the incident and output wavevector $x$-component is constant. The axis of the cone is the $x$-axis, parallel to the grating grooves. In this situation, the $\gamma$ angle requires Eq.~(\ref{eq:grating_eq_k}) to be modified, as this equation describes the wavevector components on the plane perpendicular to the grating grooves. The required modification is a $\cos{\gamma}$ multiplicative factor which quantifies the projection of wavevector components onto the $yz$ plane\cite{Yang2016}:
\begin{subequations}
\begin{align}
    k_y\cos{\gamma} &= \frac{2\pi m}{d} - k_{y0} \cos{\gamma} \\
    \frac{2\pi m}{kd} &= (\sin{\theta_{y0}}+\sin{\theta_y})\cos{\gamma} \label{eq:grating_eq_gamma}
\end{align}
\end{subequations}
The inclusion of this $\cos{\gamma}$ projection factor is crucial to accurately model off-axis field angles. This factor is the cause of smile distortion seen in long-slit grating spectrographs\cite{Leung2022}.

For highest diffraction efficiency, the grating is ideally operated in the Littrow configuration, where the incidence angle and the diffraction angle are both equal to blaze angle, i.e., $\theta_{y0}=\theta_y=\theta_B$, at a particular wavelength known as the blaze wavelength. However, in practice, a pure Littrow configuration is difficult to implement because the diffracted beam would go back onto the incident beam. Instead, the grating is rotated in the $z$-axis by an angle $\gamma$ in what is called the quasi-Littrow configuration. This configuration has the highest diffraction efficiency\cite{Schroeder1980}. The action of rotating the grating by $\gamma$ about the $z$-axis is the same as making the incident light come in at an angle $\gamma$ with respect to the plane perpendicular to the grating grooves. Regardless of whether or not we rotate the grating, we also need to account for the nonzero ``field angles'' in the $x$-direction, from dispersion by the VIPA. If the incident ``field angle'' is $\theta_{x0}$, and the grating is rotated by angle $\gamma$ about the $z$-axis, then we should modify the grating equation with a $\cos{(\gamma-\theta_{x0})}$ projection factor:
\begin{equation}
    \frac{2\pi m}{kd} = (\sin{\theta_{y0}}+\sin{\theta_y})\cos{(\gamma-\theta_{x0})}
\end{equation}
Similarly, the phase differences in Eq.~(\ref{eq:deltagcc_og}) and (\ref{eq:deltagce_og}) are modified accordingly:
\begin{align}
    \delta_\mathrm{g,cc}(\theta_{x0},\theta_{y0},\theta_y) &= kd(\sin{\theta_{y0}}+\sin{\theta_y})\cos{(\gamma-\theta_{x0})} \\
    \delta_{\mathrm{g,ce}}(\theta_{x0},\theta_{y0},\theta_y) &= \frac{kb}{2} \left[\sin{(\theta_{y0}-\theta_B)} + \sin{(\theta_y-\theta_B)}\right]\cos{(\gamma-\theta_{x0})} \label{eq:deltagce_mod}
\end{align}
The transfer function in Eq.~(\ref{eq:grating_transfer_1D_og}) should then be modified to include dependence on the incident and output angles in the $x$-direction, which are $\theta_{x0}$ and $\theta_x$ respectively. The reflection grating acts as a mirror along the $x$-direction, and hence:
\begin{align}
    H&_\mathrm{G}(\theta_{x0},\theta_{y0},\theta_x,\theta_y) \nonumber\\
    &\propto b\sinc{(\delta_{\mathrm{g,ce}})} \sum_{m\in\mathbb{Z}} \delta(\theta_x-\theta_{x0}) \,\delta(\delta_\mathrm{g,cc}-2m\pi) \nonumber \\
    &= b\sinc{(\delta_{\mathrm{g,ce}}(\theta_{x0},\theta_{y0},\theta_y))} \sum_{m\in\mathbb{Z}}  \delta(\theta_x-\theta_{x0}) \, \delta\left(\theta_y-\arcsin{\left[\frac{2\pi m}{kd\cos{(\gamma-\theta_{x0})}}-\sin{\theta_{y0}}\right]}\right)
\end{align}

We can now find the output angular spectrum of a diffraction grating $\tilde{E}_\mathrm{G}(\theta_x,\theta_y)$ given some input angular spectrum $\tilde{E}_\mathrm{G,in}(\theta_x,\theta_y)$. Since $H_\mathrm{G}(\theta_{x0},\theta_{y0},\theta_x,\theta_y)$ is not shift-invariant in angle space, a double integral must be done to obtain the output angular spectrum:
\begin{equation}
    \tilde{E}_\mathrm{G}(\theta_x,\theta_y) = \iint H_\mathrm{G}(\theta_{x0},\theta_{y0},\theta_x,\theta_y) \, \tilde{E}_\mathrm{G,in}(\theta_{x0},\theta_{y0}) \, d\theta_{x0}\,d\theta_{y0}
\end{equation}
We can immediately integrate out the $\theta_{x0}$ coordinate to write:
\begin{equation}\label{eq:Etilde_G_intermediate}
    \tilde{E}_\mathrm{G}(\theta_x,\theta_y) \propto \int b\sinc{\left(\delta_{\mathrm{g,ce}} (\theta_x,\theta_{y0},\theta_y)\right)} \tilde{E}_\mathrm{G,in}(\theta_x,\theta_{y0})\sum_{m\in\mathbb{Z}} \delta\left(\theta_y-f(\theta_{y0})\right) d\theta_{y0}
\end{equation}
where we define:
\begin{equation}\label{eq:f_func}
    f(\theta_{y0}) \equiv \arcsin{\left[\frac{2\pi m}{kd\cos{(\gamma-\theta_x)}}-\sin{\theta_{y0}}\right]}
\end{equation}
Now, note the following property of the Dirac delta function:
\begin{align*}
    \delta(x-f(u))=\sum_i\frac{\delta(u-\hat{u}_i)}{|f'(\hat{u}_i)|} && \textrm{where}\:\: \hat{u}_i=f^{-1}(x) 
\end{align*}
Hence, for arbitrary function $g$:
\begin{align}\label{eq:delta_comp_int}
    \int_{-\infty}^{\infty}\delta(x-f(u)) \, g(u) \,du = \sum_i \frac{g(\hat{u}_i)}{|f'(\hat{u}_i)|} && \textrm{where}\:\: \hat{u}_i=f^{-1}(x)
\end{align}
The $u$ and $x$ variables in Eq.~(\ref{eq:delta_comp_int}) are analogous to $\theta_{y0}$ and $\theta_y$ respectively. Let us define:
\begin{equation}
    \xi(\theta_x,\theta_y,m) \equiv \hat{\theta}_{y0} =f^{-1}(\theta_y) = \arcsin{\left[\frac{2\pi m}{kd\cos{(\gamma-\theta_x)}}-\sin{\theta_y}\right]} \tag{\ref{eq:xi_fcn}}
\end{equation}
Note that there is only one unique value of $\hat{\theta}_{y0}$ given $\theta_x$, $\theta_y$, and $m$. Next, note that:
\begin{equation}\label{eq:fprime_eval}
    f'(\hat{\theta}_{y0}) = \left.\frac{df}{d\theta_{y0}} \right\vert_{\theta_{y0}=\hat{\theta}_{y0}} = -\frac{\cos{\hat{\theta}_{y0}}}{\cos{\theta_y}}
\end{equation}
This is the anamorphic magnification. Then using Eqs.~(\ref{eq:f_func})--(\ref{eq:fprime_eval}), Eq.~(\ref{eq:Etilde_G_intermediate}) can be simplified to:
\begin{equation}
    \tilde{E}_\mathrm{G}(\theta_x,\theta_y) \propto \sum_{m\in\mathbb{Z}} \frac{b \sinc{\left(\delta_{\mathrm{g,ce}}(\theta_x,\xi(\theta_x,\theta_y,m),\theta_y)\right)} \,|\cos{\theta_y}|}{|\cos{(\xi(\theta_x,\theta_y,m))}|} \,\tilde{E}_\mathrm{G,in}(\theta_x,\xi(\theta_x,\theta_y,m)) \tag{\ref{eq:Etilde_G}}
\end{equation}

\newpage
\acknowledgments 
 
The authors thank Andrea Cordaro, James Owen, Dimitar Sasselov, Ashley Villar, Vigneshwaran Krishnamurthy, Phillip Cargile, Samuel Yee, Juliana García-Mejía, Emily Pass, Suresh Sivanandam, Katie Morzinski, Shaojie Chen, and Ralf Konietzka for helpful discussions. M.C.H.~Leung gratefully acknowledges the support of the Natural Sciences and Engineering Research Council of Canada (NSERC) through an NSERC Postgraduate Scholarship – Doctoral (PGS D). M.C.H.~Leung thanks SPIE for providing a conference fee waiver and travel grant to attend SPIE Optics + Photonics 2025.

\bibliography{report} 

\begin{thebibliography}{10}

\bibitem{Fulton2017}
Fulton, B.~J., Petigura, E.~A., Howard, A.~W., Isaacson, H., Marcy, G.~W., Cargile, P.~A., Hebb, L., Weiss, L.~M., Johnson, J.~A., Morton, T.~D., Sinukoff, E., Crossfield, I. J.~M., and Hirsch, L.~A., ``{The California-Kepler Survey. III. A Gap in the Radius Distribution of Small Planets*},'' {\em The Astronomical Journal}~{\bf 154}(3),  109 (2017).

\bibitem{Owen2019}
Owen, J.~E., ``{Atmospheric Escape and the Evolution of Close-In Exoplanets},'' {\em Annual Review of Earth and Planetary Sciences}~{\bf 47}(1),  67–90 (2019).

\bibitem{Oklopi2018}
Oklopčić, A. and Hirata, C.~M., ``{A New Window into Escaping Exoplanet Atmospheres: 10830 {\AA} Line of Helium},'' {\em The Astrophysical Journal}~{\bf 855}(1),  L11 (2018).

\bibitem{Zhang2023}
Zhang, Z., Morley, C.~V., Gully-Santiago, M., MacLeod, M., Oklopčić, A., Luna, J., Tran, Q.~H., Ninan, J.~P., Mahadevan, S., Krolikowski, D.~M., Cochran, W.~D., Bowler, B.~P., Endl, M., Stefánsson, G., Tofflemire, B.~M., Vanderburg, A., and Zeimann, G.~R., ``{Giant tidal tails of helium escaping the hot Jupiter HAT-P-32 b},'' {\em Science Advances}~{\bf 9}(23) (2023).

\bibitem{MacLeod2022}
MacLeod, M. and Oklopčić, A., ``{Stellar Wind Confinement of Evaporating Exoplanet Atmospheres and Its Signatures in 1083 nm Observations},'' {\em The Astrophysical Journal}~{\bf 926}(2),  226 (2022).

\bibitem{Schreyer2023}
Schreyer, E., Owen, J.~E., Spake, J.~J., Bahroloom, Z., and Di~Giampasquale, S., ``{Using helium 10830 {\AA} transits to constrain planetary magnetic fields},'' {\em Monthly Notices of the Royal Astronomical Society}~{\bf 527}(3),  5117–5130 (2023).

\bibitem{Nail2024}
Nail, F., Oklopčić, A., and MacLeod, M., ``{Effects of planetary day-night temperature gradients on He 1083 nm transit spectra},'' {\em Astronomy \& Astrophysics}~{\bf 684},  A20 (2024).

\bibitem{Nail2025}
Nail, F., MacLeod, M., Oklopčić, A., Gully-Santiago, M., Morley, C.~V., and Zhang, Z., ``Cold dayside winds shape large leading streams in evaporating exoplanet atmospheres,'' {\em Astronomy \& Astrophysics}~{\bf 695},  A186 (2025).

\bibitem{Owen2023}
Owen, J.~E. and Schlichting, H.~E., ``Mapping out the parameter space for photoevaporation and core-powered mass-loss,'' {\em Monthly Notices of the Royal Astronomical Society}~{\bf 528}(2),  1615–1629 (2023).

\bibitem{Rukdee2023}
Rukdee, S., Ben-Ami, S., López-Morales, M., Szentgyorgyi, A., Charbonneau, D., García-Mejía, J., and Buchner, J., ``{First on-sky results of a Fabry–Perot Instrument for Oxygen Searches (FIOS) prototype},'' {\em Astronomy \& Astrophysics}~{\bf 678},  A114 (2023).

\bibitem{Shirasaki1996}
Shirasaki, M., ``{Large angular dispersion by a virtually imaged phased array and its application to a wavelength demultiplexer},'' {\em Optics Letters}~{\bf 21}(5),  366 (1996).

\bibitem{Bourdarot2017}
Bourdarot, G., Le~Coarer, E., Alecian, E., Bonfils, X., and Rabou, P., ``{Nano-VIPA: a miniaturized high-resolution echelle spectrometer, for the monitoring of young stars from a 6u cubesat},'' in [{\em International Conference on Space Optics — ICSO 2016}{\nolinebreak\hspace{0.1em}]},  {\em Proc. SPIE} {\bf 10562},  239 (2017).

\bibitem{Bourdarot2018}
Bourdarot, G., Le~Coarer, E., Mouillet, D., Jocou, L., Rabou, P., Correia, J.-J., Bonfils, X., Stadler, E., Carlotti, A., Forveille, T., Vigan, A., Artigau, E., Doyon, R., Vallée, P., and Magnard, Y., ``{Experimental test of a 40 cm-long R=100 000 spectrometer for exoplanet characterisation},'' in [{\em Ground-based and Airborne Instrumentation for Astronomy VII}{\nolinebreak\hspace{0.1em}]},  {\em Proc. SPIE} {\bf 10702},  107025Y (2018).

\bibitem{Carlotti2022}
Carlotti, A., Bidot, A., Mouillet, D., Correia, J.-J., Jocou, L., Curaba, S., Delboulbé, A., Le~Coarer, E., Rabou, P., Bourdarot, G., Forveille, T., Bonfils, X., Vasisht, G., Mawet, D., Burruss, R.~S., Oppenheimer, R., Doyon, R., Artigau, E., and Vallée, P., ``{On-sky demonstration at Palomar Observatory of the near-IR, high-resolution VIPA spectrometer},'' in [{\em Ground-based and Airborne Instrumentation for Astronomy IX}{\nolinebreak\hspace{0.1em}]},  {\em Proc. SPIE} {\bf 12184},  52 (2022).

\bibitem{Zhu2020}
Zhu, X., Lin, D., Hao, Z., Wang, L., and He, J., ``{A VIPA Spectrograph with Ultra-high Resolution and Wavelength Calibration for Astronomical Applications},'' {\em The Astronomical Journal}~{\bf 160}(3),  135 (2020).

\bibitem{Zhu2023}
Zhu, X., Lin, D., Zhang, Z., Xie, X., and He, J., ``{Dispersion Characteristics of the Multi-mode Fiber-fed VIPA Spectrograph},'' {\em The Astronomical Journal}~{\bf 165}(6),  228 (2023).

\bibitem{Rukdee2024}
Rukdee, S., ``{Conceptual Design of VIOLA: Vipa Instrument for Oxygen Loaded Atmospheres},'' in [{\em Ground-based and Airborne Instrumentation for Astronomy X}{\nolinebreak\hspace{0.1em}]},  {\em Proc. SPIE} {\bf 13096},  334 (2024).

\bibitem{Strader2015}
Strader, J., Dupree, A.~K., and Smith, G.~H., ``{THE 10830 {\AA} HELIUM LINE AMONG EVOLVED STARS IN THE GLOBULAR CLUSTER M4},'' {\em The Astrophysical Journal}~{\bf 808}(2),  124 (2015).

\bibitem{Dupree1986}
Dupree, A.~K., ``{Mass Loss from Cool Stars},'' {\em Annual Review of Astronomy and Astrophysics}~{\bf 24}(1),  377–420 (1986).

\bibitem{Cooke2022}
Cooke, R.~J., Noterdaeme, P., Johnson, J.~W., Pettini, M., Welsh, L., Peroux, C., Murphy, M.~T., and Weinberg, D.~H., ``{Primordial Helium-3 Redux: The Helium Isotope Ratio of the Orion Nebula*},'' {\em The Astrophysical Journal}~{\bf 932}(1),  60 (2022).

\bibitem{NASAExoArchive}
{NASA Exoplanet Science Institute}, ``{Planetary Systems Table}.'' DOI: 10.26133/NEA12, Version: 2025-07-11 12:03 (2025).

\bibitem{CastroGonzlez2024}
Castro-González, A., Bourrier, V., Lillo-Box, J., Delisle, J.-B., Armstrong, D.~J., Barrado, D., and Correia, A. C.~M., ``{Mapping the exo-Neptunian landscape: A ridge between the desert and savanna},'' {\em Astronomy \& Astrophysics}~{\bf 689},  A250 (2024).

\bibitem{Mazeh2016}
Mazeh, T., Holczer, T., and Faigler, S., ``{Dearth of short-period Neptunian exoplanets: A desert in period-mass and period-radius planes},'' {\em Astronomy \& Astrophysics}~{\bf 589},  A75 (2016).

\bibitem{Owen2022}
Owen, J.~E., Murray-Clay, R.~A., Schreyer, E., Schlichting, H.~E., Ardila, D., Gupta, A., Loyd, R. O.~P., Shkolnik, E.~L., Sing, D.~K., and Swain, M.~R., ``{The fundamentals of Lyman $\alpha$ exoplanet transits},'' {\em Monthly Notices of the Royal Astronomical Society}~{\bf 518}(3),  4357–4371 (2022).

\bibitem{Owen2017}
Owen, J.~E. and Wu, Y., ``{The Evaporation Valley in the Kepler Planets},'' {\em The Astrophysical Journal}~{\bf 847}(1),  29 (2017).

\bibitem{Rogers2021}
Rogers, J.~G. and Owen, J.~E., ``Unveiling the planet population at birth,'' {\em Monthly Notices of the Royal Astronomical Society}~{\bf 503}(1),  1526–1542 (2021).

\bibitem{Gupta2020}
Gupta, A. and Schlichting, H.~E., ``{Signatures of the core-powered mass-loss mechanism in the exoplanet population: dependence on stellar properties and observational predictions},'' {\em Monthly Notices of the Royal Astronomical Society}~{\bf 493}(1),  792–806 (2020).

\bibitem{Ginzburg2018}
Ginzburg, S., Schlichting, H.~E., and Sari, R., ``Core-powered mass-loss and the radius distribution of small exoplanets,'' {\em Monthly Notices of the Royal Astronomical Society}~{\bf 476}(1),  759–765 (2018).

\bibitem{Ionov2018}
Ionov, D.~E., Pavlyuchenkov, Y.~N., and Shematovich, V.~I., ``{Survival of a planet in short-period Neptunian desert under effect of photoevaporation},'' {\em Monthly Notices of the Royal Astronomical Society}~{\bf 476}(4),  5639–5644 (2018).

\bibitem{Ginzburg2016}
Ginzburg, S., Schlichting, H.~E., and Sari, R., ``{SUPER-EARTH ATMOSPHERES: SELF-CONSISTENT GAS ACCRETION AND RETENTION},'' {\em The Astrophysical Journal}~{\bf 825}(1),  29 (2016).

\bibitem{Krishnamurthy2024}
Krishnamurthy, V. and Cowan, N.~B., ``{Helium in Exoplanet Exospheres: Orbital and Stellar Influences},'' {\em The Astronomical Journal}~{\bf 168}(1),  30 (2024).

\bibitem{Lee2022}
Lee, E.~J., Karalis, A., and Thorngren, D.~P., ``{Creating the Radius Gap without Mass Loss},'' {\em The Astrophysical Journal}~{\bf 941}(2),  186 (2022).

\bibitem{Luque2022}
Luque, R. and Pallé, E., ``Density, not radius, separates rocky and water-rich small planets orbiting m dwarf stars,'' {\em Science}~{\bf 377}(6611),  1211–1214 (2022).

\bibitem{Owen2018}
Owen, J.~E. and Lai, D., ``Photoevaporation and high-eccentricity migration created the sub-jovian desert,'' {\em Monthly Notices of the Royal Astronomical Society}~{\bf 479}(4),  5012–5021 (2018).

\bibitem{Vissapragada2022}
Vissapragada, S., Knutson, H.~A., Greklek-McKeon, M., Oklopčić, A., Dai, F., dos Santos, L.~A., Jovanovic, N., Mawet, D., Millar-Blanchaer, M.~A., Paragas, K., Spake, J.~J., Tinyanont, S., and Vasisht, G., ``{The Upper Edge of the Neptune Desert Is Stable Against Photoevaporation},'' {\em The Astronomical Journal}~{\bf 164}(6),  234 (2022).

\bibitem{Matsakos2016}
Matsakos, T. and K\"{o}nigl, A., ``{ON THE ORIGIN OF THE SUB-JOVIAN DESERT IN THE ORBITAL-PERIOD–PLANETARY-MASS PLANE},'' {\em The Astrophysical Journal Letters}~{\bf 820}(1),  L8 (2016).

\bibitem{Cauley2018}
Cauley, P.~W., Kuckein, C., Redfield, S., Shkolnik, E.~L., Denker, C., Llama, J., and Verma, M., ``{The Effects of Stellar Activity on Optical High-resolution Exoplanet Transmission Spectra},'' {\em The Astronomical Journal}~{\bf 156}(5),  189 (2018).

\bibitem{Seager2000}
Seager, S. and Sasselov, D.~D., ``{Theoretical Transmission Spectra during Extrasolar Giant Planet Transits},'' {\em The Astrophysical Journal}~{\bf 537}(2),  916–921 (2000).

\bibitem{Drake1971}
Drake, G. W.~F., ``{Theory of Relativistic Magnetic Dipole Transitions: Lifetime of the Metastable 2 $^3S$ State of the Heliumlike Ions},'' {\em Physical Review A}~{\bf 3}(3),  908–915 (1971).

\bibitem{Corsaro1973}
Corsaro, F.~A. and Hameka, H.~F., ``{Fine structure of the 2 $^3P$ state of the helium atom},'' {\em The Journal of Chemical Physics}~{\bf 58}(7),  2851–2854 (1973).

\bibitem{NIST_ASD}
Kramida, A., {Yu.~Ralchenko}, Reader, J., and {NIST ASD Team}, ``{NIST Atomic Spectra Database}.'' { (ver. 5.12), {https://physics.nist.gov/asd}. National Institute of Standards and Technology} (2024).

\bibitem{Dupree2009}
Dupree, A.~K., Smith, G.~H., and Strader, J., ``{FAST WINDS AND MASS LOSS FROM METAL-POOR FIELD GIANTS},'' {\em The Astronomical Journal}~{\bf 138}(5),  1485–1501 (2009).

\bibitem{Spake2018}
Spake, J.~J., Sing, D.~K., Evans, T.~M., Oklopčić, A., Bourrier, V., Kreidberg, L., Rackham, B.~V., Irwin, J., Ehrenreich, D., Wyttenbach, A., Wakeford, H.~R., Zhou, Y., Chubb, K.~L., Nikolov, N., Goyal, J.~M., Henry, G.~W., Williamson, M.~H., Blumenthal, S., Anderson, D.~R., Hellier, C., Charbonneau, D., Udry, S., and Madhusudhan, N., ``{Helium in the eroding atmosphere of an exoplanet},'' {\em Nature}~{\bf 557}(7703),  68–70 (2018).

\bibitem{Allart2023}
Allart, R., Lemée-Joliecoeur, P.-B., Jaziri, A.~Y., Lafrenière, D., Artigau, E., Cook, N., Darveau-Bernier, A., Dang, L., Cadieux, C., Boucher, A., Bourrier, V., Deibert, E.~K., Pelletier, S., Radica, M., Benneke, B., Carmona, A., Cloutier, R., Cowan, N.~B., Delfosse, X., Donati, J.-F., Doyon, R., Figueira, P., Forveille, T., Fouqué, P., Gaidos, E., Gu, P.-G., Hébrard, G., Kiefer, F., Kóspál, A., Jayawardhana, R., Martioli, E., Dos~Santos, L.~A., Shang, H., Turner, J.~D., and Vidotto, A.~A., ``{Homogeneous search for helium in the atmosphere of 11 gas giant exoplanets with SPIRou},'' {\em Astronomy \& Astrophysics}~{\bf 677},  A164 (2023).

\bibitem{Guilluy2024}
Guilluy, G., D’Arpa, M.~C., Bonomo, A.~S., Spinelli, R., Biassoni, F., Fossati, L., Maggio, A., Giacobbe, P., Lanza, A.~F., Sozzetti, A., Borsa, F., Rainer, M., Micela, G., Affer, L., Andreuzzi, G., Bignamini, A., Boschin, W., Carleo, I., Cecconi, M., Desidera, S., Fardella, V., Ghedina, A., Mantovan, G., Mancini, L., Nascimbeni, V., Knapic, C., Pedani, M., Petralia, A., Pino, L., Scandariato, G., Sicilia, D., Stangret, M., and Zingales, T., ``{The GAPS Programme at TNG: LIV. A He I survey of close-in giant planets hosted by M-K dwarf stars with GIANO-B},'' {\em Astronomy \& Astrophysics}~{\bf 686},  A83 (2024).

\bibitem{Masson2024}
Masson, A., Vinatier, S., Bézard, B., López-Puertas, M., Lampón, M., Debras, F., Carmona, A., Klein, B., Artigau, E., Dethier, W., Pelletier, S., Hood, T., Allart, R., Bourrier, V., Cadieux, C., Charnay, B., Cowan, N.~B., Cook, N.~J., Delfosse, X., Donati, J.-F., Gu, P.-G., Hébrard, G., Martioli, E., Moutou, C., Venot, O., and Wyttenbach, A., ``{Probing atmospheric escape through metastable He I triplet lines in 15 exoplanets observed with SPIRou},'' {\em Astronomy \& Astrophysics}~{\bf 688},  A179 (2024).

\bibitem{GullySantiago2024}
Gully-Santiago, M., Morley, C.~V., Luna, J., MacLeod, M., Oklopčić, A., Ganesh, A., Tran, Q.~H., Zhang, Z., Bowler, B.~P., Cochran, W.~D., Krolikowski, D.~M., Mahadevan, S., Ninan, J.~P., Stefánsson, G., Vanderburg, A., Zalesky, J.~A., and Zeimann, G.~R., ``{A Large and Variable Leading Tail of Helium in a Hot Saturn Undergoing Runaway Inflation},'' {\em The Astronomical Journal}~{\bf 167}(4),  142 (2024).

\bibitem{Cohen2003}
Cohen, M., Wheaton, W.~A., and Megeath, S.~T., ``{Spectral Irradiance Calibration in the Infrared. XIV. The Absolute Calibration of 2MASS},'' {\em The Astronomical Journal}~{\bf 126}(2),  1090–1096 (2003).

\bibitem{SeagerBookCh2}
Murray, C.~D. and Correia, A. C.~M., ``{Keplerian Orbits and Dynamics of Exoplanets},'' in [{\em {Exoplanets}}{\nolinebreak\hspace{0.1em}]},  Seager, S., ed., University of Arizona Press (2011).

\bibitem{FarretJentink2023}
Farret~Jentink, C., Bourrier, V., Lovis, C., Allart, R., Chazelas, B., Lendl, M., Dumusque, X., and Pepe, F., ``Night: A compact, near-infrared, high-resolution spectrograph to survey helium in exoplanet systems,'' {\em Monthly Notices of the Royal Astronomical Society}~{\bf 527}(3),  4467–4482 (2023).

\bibitem{BornWolf}
Born, M. and Wolf, E.,  [{\em {Principles of Optics}}{\nolinebreak\hspace{0.1em}]}, Cambridge University Press, 7~ed. (2019).

\bibitem{YarivYeh}
Yariv, A. and Yeh, P.,  [{\em {Photonics}}{\nolinebreak\hspace{0.1em}]}, Oxford University Press, 7~ed. (2007).

\bibitem{Hu2015}
Hu, X., Sun, Q., Li, J., Li, C., Liu, Y., and Zhang, J., ``Spectral dispersion modeling of virtually imaged phased array by using angular spectrum of plane waves,'' {\em Optics Express}~{\bf 23}(1),  1 (2015).

\bibitem{Weiner2012}
Weiner, A.~M., ``Reply to comment on “generalized grating equation for virtually-imaged phased-array spectral dispersions”,'' {\em Applied Optics}~{\bf 51}(34),  8187 (2012).

\bibitem{Xiao2004}
Xiao, S., Weiner, A., and Lin, C., ``{A Dispersion Law for Virtually Imaged Phased-Array Spectral Dispersers Based on Paraxial Wave Theory},'' {\em IEEE Journal of Quantum Electronics}~{\bf 40}(4),  420–426 (2004).

\bibitem{Xiao2005}
Xiao, S. and Weiner, A.~M., ``{An eight-channel hyperfine wavelength demultiplexer using a virtually imaged phased-array (VIPA)},'' {\em IEEE Photonics Technology Letters}~{\bf 17}(2),  372–374 (2005).

\bibitem{Metz2013}
Metz, P., Block, H., Behnke, C., Krantz, M., Gerken, M., and Adam, J., ``Tunable elastomer-based virtually imaged phased array,'' {\em Optics Express}~{\bf 21}(3),  3324 (2013).

\bibitem{Yang2016}
Yang, Q., ``Compact high-resolution littrow conical diffraction spectrometer,'' {\em Applied Optics}~{\bf 55}(18),  4801 (2016).

\bibitem{Leung2022}
Leung, M. C.~H., Chen, S., and Jurgenson, C., ``Accurately measuring hyperspectral imaging distortion in grating spectrographs using a clustering algorithm,'' in [{\em Advances in Optical and Mechanical Technologies for Telescopes and Instrumentation V}{\nolinebreak\hspace{0.1em}]},  {\em Proc. SPIE} {\bf 12188},  121883W (2022).

\bibitem{Bottema1981}
Bottema, M., ``Echelle efficiencies: theory and experiment; comment,'' {\em Applied Optics}~{\bf 20}(4),  528 (1981).

\bibitem{Bottema1981SPIE}
Bottema, M., ``{Echelle Efficiency And Blaze Characteristics},'' in [{\em Periodic Structures, Gratings, Moire Patterns, and Diffraction Phenomena I}{\nolinebreak\hspace{0.1em}]},  {\em Proc. SPIE} {\bf 0240},  171–177 (1981).

\bibitem{Yang2020}
Yang, D., Liu, J.-t., Fan, X.-k., Zhu, W.-h., Wang, S., and Song, X.-q., ``Simulation analysis and implementation of spectral dispersion system based on virtually imaged phased array,'' {\em Optoelectronics Letters}~{\bf 16},  268–271 (July 2020).

\bibitem{Schroeder1980}
Schroeder, D.~J. and Hilliard, R.~L., ``Echelle efficiencies: theory and experiment,'' {\em Applied Optics}~{\bf 19}(16),  2833 (1980).

\bibitem{Szentgyorgyi2011}
Szentgyorgyi, A., Furesz, G., Cheimets, P., Conroy, M., Eng, R., Fabricant, D., Fata, R., Gauron, T., Geary, J., McLeod, B., Zajac, J., Amato, S., Bergner, H., Caldwell, N., Dupree, A., Goddard, R., Johnston, E., Meibom, S., Mink, D., Pieri, M., Roll, J., Tokarz, S., Wyatt, W., Epps, H., Hartmann, L., and Meszaros, S., ``{Hectochelle: A Multiobject Optical Echelle Spectrograph for the MMT},'' {\em Publications of the Astronomical Society of the Pacific}~{\bf 123}(908),  1188–1209 (2011).

\bibitem{Pickering2018}
Pickering, T.~E., ``{Seeing statistics and characteristics at the MMT Observatory 2003-2018},'' in [{\em Ground-based and Airborne Telescopes VII}{\nolinebreak\hspace{0.1em}]},  {\em Proc. SPIE} {\bf 10700},  196, SPIE (2018).

\bibitem{Avila2012}
Avila, G., Guirao, C., and Baader, T., ``High efficiency inexpensive 2-slices image slicers,'' in [{\em Ground-based and Airborne Instrumentation for Astronomy IV}{\nolinebreak\hspace{0.1em}]},  {\em Proc. SPIE} {\bf 8446},  84469M, SPIE (2012).

\bibitem{Tala2017}
Tala, M., Vanzi, L., Avila, G., Guirao, C., Pecchioli, E., Zapata, A., and Pieralli, F., ``Two simple image slicers for high resolution spectroscopy,'' {\em Experimental Astronomy}~{\bf 43}(2),  167–176 (2017).

\bibitem{Szentgyorgyi2024_OG}
Szentgyorgyi, A., Ben-Ami, S., Oh, J.~S., Park, C., Baldwin, D., Bean, J., Bichkovsky, A., Brennan, P., Catropa, D., Chou, C.-Y., Chun, M.-Y., Close, L., Crane, J.~D., Doherty, P., Durusky, D., Eastman, J., Epps, H., Evans, I.~N., Hartmann, V., Hershko, O., Jang, B.-H., Jang, J.-G., Jeong, U., Jordan, A., Jurgenson, C., Kansky, J., Kim, J., Kim, K.-M., Kim, S., Kim, Y., Lee, S., Leung, M., Ling, H.-H., Lupinari, H., Males, J.~R., McCracken, K., de~Oliveira, C.~M., Millan-Gabet, R., Mueller, M., Oh, H., Onyuksel, C., Park, B.-G., Park, S.-J., Park, W., Podgorski, W., Seifahrt, A., Smith, M., Sofer-Rimalt, Y., Wang, S.-Y., Unger, A., Uomoto, A., and Yu, Y.-S., ``{Innovations in the design and construction of the GMT-Consortium Large Earth Finder (G-CLEF), a first-light instrument for the Giant Magellan Telescope (GMT)},'' in [{\em Ground-based and Airborne Instrumentation for Astronomy X}{\nolinebreak\hspace{0.1em}]},  {\em Proc. SPIE} {\bf 13096},  130960Z (2024).

\bibitem{Jurgenson2016}
Jurgenson, C., Fischer, D., McCracken, T., Sawyer, D., Szymkowiak, A., Davis, A., Muller, G., and Santoro, F., ``{EXPRES: a next generation RV spectrograph in the search for earth-like worlds},'' in [{\em Ground-based and Airborne Instrumentation for Astronomy VI}{\nolinebreak\hspace{0.1em}]},  {\em Proc. SPIE} {\bf 9908},  99086T (2016).

\bibitem{Morzinski2020}
Morzinski, K.~M., Montoya, M., Fellows, C., Durney, O., Ford, J., West, G., Gardner, A., Vaz, A., Anugu, N., Mailhot, E., Carlson, J., Harrison, L., Gacon, F., Downey, E., Hinz, P.~M., Jones, T., Patience, J., Sivanandam, S., Chen, S., Lamb, M.~P., Butko, A., Liu, S., Hardy, T., and Jannuzi, B., ``{Development and status of MAPS, the MMT AO exoPlanet characterization system},'' in [{\em Adaptive Optics Systems VII}{\nolinebreak\hspace{0.1em}]},  {\em Proc. SPIE} {\bf 11448},  114481L (2020).

\bibitem{Anugu2020}
Anugu, N., Durney, O., M.~Morzinski, K., Hinz, P., Sivanandam, S., Males, J., Gardner, A.~K., Fellows, C., Montoya, M., West, G., Vaz, A., Mailhot, E., Carlson, J., Chen, S., Lamb, M., Butko, A., Downey, E., Tyler, J., and Jannuzi, B., ``{Design and development of a high-speed visible pyramid wavefront sensor for the MMT AO system},'' in [{\em Adaptive Optics Systems VII}{\nolinebreak\hspace{0.1em}]},  {\em Proc. SPIE} {\bf 11448},  114485J (2020).

\bibitem{Morzinski2024}
Morzinski, K.~M., Montoya, M., Patience, J., et~al., ``{Commissioning MAPS, the MMT AO exoPlanet characterization System},'' in [{\em Adaptive Optics Systems IX}{\nolinebreak\hspace{0.1em}]},  {\em Proc. SPIE} {\bf 13097},  130970D (2024).

\bibitem{SPIE_FG27}
Bentley, J. and Olson, C.,  [{\em {Field Guide to Lens Design}}{\nolinebreak\hspace{0.1em}]}, SPIE (2012).

\bibitem{Goodman}
Goodman, J.~W.,  [{\em {Introduction to Fourier Optics}}{\nolinebreak\hspace{0.1em}]}, The McGraw-Hill Companies, 2~ed. (1992).

\bibitem{Schroeder}
Schroeder, D.~J.,  [{\em {Astronomical Optics}}{\nolinebreak\hspace{0.1em}]}, Academic Press, 2~ed. (2000).

\bibitem{SalehTeich}
Saleh, B. E.~A. and Teich, M.~C.,  [{\em {Fundamentals of Photonics}}{\nolinebreak\hspace{0.1em}]}, John Wiley \& Sons, Inc. (1991).

\end{thebibliography}
\bibliographystyle{spiebib} 

\end{document}